\documentclass[prl, aps, twocolumn, floats, showpacs, superscriptaddress]{revtex4-2}
\usepackage[latin3]{inputenc}
\usepackage[makeroom]{cancel}
\usepackage{graphicx}
\usepackage{amsmath}
\usepackage{physics}
\usepackage{dsfont}
\usepackage{epstopdf}
\usepackage{amsfonts}
\usepackage{amssymb}
\usepackage{xcolor}
\usepackage{graphics} 
\usepackage{dcolumn}
\usepackage{bm}
\usepackage{array}
\usepackage{appendix}
\usepackage{nicefrac}
\usepackage{cellspace} 
\usepackage[left=2cm,right=2cm,top=2cm,bottom=2cm]{geometry}
\usepackage[normalem]{ulem}
\RequirePackage[
   hyperindex,colorlinks,bookmarksnumbered,
   plainpages=true,pdfstartview=FitH]{hyperref}
\hypersetup{linkcolor=blue,urlcolor=blue,citecolor=blue}

\begin{document}
\title{
Universal relations between thermoelectrics and noise in mesoscopic transport across a tunnel junction
 }
\author{Andrei I. Pavlov}
\email{andrei.pavlov@kit.edu}
\affiliation{IQMT, Karlsruhe Institute of Technology, 76131 Karlsruhe, Germany}
\author{Mikhail N. Kiselev}
\affiliation{The Abdus Salam International Centre for Theoretical Physics, Strada Costiera 11, I-34151 Trieste, Italy}
\begin{abstract}
We develop a unified theory of weakly probed differential observables for currents and noise in transport experiments. Our findings uncover a set of universal transport relations between thermoelectric and noise properties of a system probed through a tunnel contact, with the Wiedemann-Franz law being just one example of such universality between charge and heat currents. We apply this theory to various quantum systems, including multichannel Kondo, quantum Hall and Sachdev-Ye-Kitaev quantum dots, resonant impurity and two-stage Kondo models, and demonstrate that each of the microscopic theories is characterized by a set of universal relations connecting conductance and thermoelectrics with noise. Violations of these relations indicate additional energy scales emerging in a system.
\end{abstract}
\maketitle
Quantum transport probes are widely used for obtaining information about mesoscopic systems. 
Thermoelectric and shot noise measurements are employed for detecting signatures of Kondo physics \cite{Scheibner2005, Yamauchi2011, Borzenets2020} and characterization of quantum dots \cite{Godijin1999, Barthold2006}, serve as direct quantum information probes \cite{Hartman2018}. In heavy fermion materials and strange metals, they provide experimental \cite{Dong2013, Chen2023} and theoretical \cite{Wang2024, Gleis2025, Mravlje2016} insights into the nature of the charge carriers and interactions dominating in different regimes. In holographic systems, the transport observables are related to thermodynamics of black holes \cite{Sachdev2015, Davison2017, Inkof2020}. In addition to widely used thermoelectric and shot noise measurements \cite{Martin2005}, delta-T noise, arising purely due to temperature bias, has been recently measured experimentally \cite{Lumbroso2018, Sivre2019, Larocque2020}, which opens possibilities for utilizing it as an experimental probe.
Nevertheless, it is often experimentally challenging to probe differential transport observables, as this procedure requires a simultaneous control over different biases applied to a system \cite{Scheibner2005}. Furthermore, such probes may give ambiguous results about the microscopic properties of the system. For instance, violations of the Wiedemann-Franz (WF) law may have various underlying reasons \cite{Vavilov2005, Tanatar2007, Kubala2008}. 

In this Letter, we develop a linear response theory for transport through a tunnel contact that treats all currents and noise on the same footing. This theory provides a set of universal relations between different transport observables, leading to one-to-one connections between various Fano factors, and establishes regimes of equivalence between thermoelectric and noise measurements. The obtained universality allows for experimental flexibility regarding a choice of observables for obtaining the same information about the probed system. At the same time, violations of the established universal relations provide a more nuanced information about the microscopic properties of the system comparing to existing analysis. After presenting the general theory, we illustrate its application to a wide variety of systems, including multichannel Kondo devices, quantum Hall systems, holographic systems (Sachdev-Ye-Kitaev (SYK) model and its generalizations \cite{Tikhanovskaya2021, Chowdhury2022}), the resonant impurity model, and the two-stage charge Kondo (2CK) model. Generalizing the concept of the Lorenz number and the Lorenz ratio, we show that each of the microscopic theories results into a specific set of universal constants relating different transport observables to each other. Violations of these relations point to additional energy scales that effectively emerge within the system.
\paragraph*{Model ---}
We consider a quantum system ($S$) weakly coupled through a tunnel junction to a metallic lead ($L$). This is a typical setup for tunneling spectroscopy and transport probes widely used in mesoscopic and nanoscopic experiments \cite{vanderVaart1995, GoldhaberGordon1998, GoldhaberGordon1998b,  Godijin1999, Nauen2002, Nauen2004, Slot2004, Scheibner2005, Heersche2006, Barthold2006,  Huard2007, Osorio2008, Yamauchi2011, Pruser2011, Perrin2013, Burzuri2014, Iftikhar2015, Iftikhar2018, Gehring2019, Borzenets2020, Guo2021, Dani2022, Pouse2023, Piquard2023, Anderson2024}. The temperature of the system is $T$, while the temperature of the lead is $T+\Delta T$. There is a voltage bias $\Delta V$ between them; both $\Delta V$ and $\Delta T$ are assumed to be small enough so the linear response theory can be justified. For now, we do not specify the nature of the probed system. It can be another metallic lead, a quantum dot of various types (e.g., a multiterminal Kondo device, an SYK quantum dot), or some extended non-Fermi-liquid system \cite{Suppl}. 
Throughout the Letter, we put $\hbar=e=k_B=1$.

In the following, we consider transport of charge and heat across the tunnel junction, induced due to the voltage and temperature bias. Within the considered setup, there are two types of current - charge current $I_c$ and heat current $I_h$. For these two currents, there are three possible types of current-current correlations that constitute noise: charge noise $S_c$, heat noise $S_h$ and mixed noise $S_m$ (that accounts for correlations between charge and heat currents) \cite{Sanchez2013, Battista2014, Crepieux2014}. This consideration can easily be extended to other types of transport (e.g., spin transport \cite{Jacquod2012}), multiterminal setups \cite{Mazza2014, Benenti2017}, and higher moments of the corresponding currents.

\paragraph*{Noise coefficients---}
A steady state tunneling current and zero frequency noise power (which we further refer as noise for simplicity) across a tunnel junction connecting two systems depend on the local density of states (DoS) of the systems and their occupation numbers. To characterize the charge and heat transport through a tunnel junction, it is convenient to introduce the energy-dependent transmission coefficient (more generally, imaginary part of the T-matrix) $\mathcal{T}(\varepsilon)=2\pi|\lambda|^2\rho_L(\varepsilon)\rho_S(\varepsilon)$ \cite{Andreev2002, Nguyen2020, Nguyen2024}, where $\lambda$ is the tunneling amplitude through the barrier and $\rho_i(\varepsilon) \, (i=L,S)$ are the DoS of the lead and the system. Since the probe is a metallic lead, its DoS can be approximated (close to the Fermi level) by a constant value $\rho_L=(2\pi v_F)^{-1}$, where $v_F$ is the Fermi velocity. In the weak tunneling limit, one can consider the T-matrix in the lowest order of the tunneling amplitudes. 
Employing Keldysh Green's functions \cite{Kamenev2011, Popoff2022}, one can express the currents and the noises in this regime as
\vspace{-0.5cm}
\begin{widetext}
\vspace{-0.7cm}
\begin{align} \label{currentDef}
&I_{c/h}=\int_{-\infty}^{\infty} d\varepsilon \left(\varepsilon-\Delta V\right)^{n}\mathcal{T}(\varepsilon)\left[n_L(\varepsilon-\Delta V, T+\Delta T)-n_S(\varepsilon, T)\right], &\\ \label{noiseDef}
&S_{c/m/h}=\int_{-\infty}^{\infty} d\varepsilon \left(\varepsilon-\Delta V\right)^{l}\mathcal{T}(\varepsilon)\left[n_{L}(\varepsilon-\Delta V, T+\Delta T)+n_S(\varepsilon, T)-2n_{L}(\varepsilon-\Delta V, T+\Delta T)n_S(\varepsilon, T)\right],&
\end{align} \vspace{-0.3cm}
\end{widetext}
where $n=0,1$ for charge $I_c$ and heat current $I_h$; $l=0,1,2$ for charge $S_c$, mixed $S_m$, and heat noise $S_h$, correspondingly \cite{Benenti2017, Pekola2021}. $\Delta V$ and $\Delta T$ are voltage and temperature drops, respectively, across the tunnel junction. $n(\varepsilon, T)=\left(e^{\frac{\varepsilon-\mu}{T}}+1\right)^{-1}$ is the Fermi-Dirac function ($\mu$ is the chemical potential).

The linear response theory of thermoelectric transport expresses currents via the transport coefficients that capture the system's response to small biases. These responses can be expressed in terms of the Onsager transport integrals \cite{Onsager1931}
\begin{align} \renewcommand\arraystretch{1.3}
\label{LCoeff}    \begin{pmatrix}
        I_c \\ I_h
    \end{pmatrix}=\begin{pmatrix}
        \mathcal{L}_0 & \frac{1}{T}\mathcal{L}_1\\
        \mathcal{L}_1 & \frac{1}{T}\mathcal{L}_2
    \end{pmatrix} \begin{pmatrix}
        \Delta V \\ \Delta T
    \end{pmatrix},
\end{align}
\vspace{-0.3cm}
\begin{align} \label{Lint}
    &\mathcal{L}_n=\frac{1}{4T}\int_{-\infty}^{\infty}d\varepsilon\mathcal{T}(\varepsilon)\frac{\varepsilon^n}{\cosh^2\left(\frac{\varepsilon}{2T}\right)}, \, n=0,1,2.&
\end{align}
This defines charge conductance $G=\mathcal{L}_0$, thermoelectric coefficient $G_T=\frac{1}{T}\mathcal{L}_1$,  heat conductance $G_H=\frac{1}{T}\mathcal{L}_2$, Peltier coefficient $\Pi=\frac{\mathcal{L}_1}{\mathcal{L}_0}$, Seebeck coefficient (also known as thermopower) $\mathcal{S}=\frac{\mathcal{L}_1}{T\mathcal{L}_0}$, and thermal conductance $K=G_H-T G_T\mathcal{S}$ \cite{Benenti2017}. With the same approach, we generalize this idea and introduce the \textit{noise coefficients}  \vspace{-0.4cm}
\begin{align} \label{Nint}
    \mathcal{N}_n=\frac{1}{4T}\int_{-\infty}^{\infty}d\varepsilon\mathcal{T}(\varepsilon)\frac{\varepsilon^n\tanh\left(\frac{\varepsilon}{2T}\right)}{\cosh^2\left(\frac{\varepsilon}{2T}\right)}, \, n=0,1,2,3. \vspace{-0.4cm}
\end{align}
Within the linear response, each type of noise can be decomposed into three components: equilibrium (also known as Johnson-Nyquist) noise $S_{c,m,h}^{JN}$, which is present at finite temperature even without any bias; shot noise $\delta S^{SN}_{c,m,h}\equiv\left.\frac{\partial S_{c,m,h}}{\partial\Delta V}\right\vert_{\Delta T=0}$ induced due to the voltage bias; and delta-T noise $\delta S^{\Delta T}_{c,m,h}\equiv\left.\frac{\partial S_{c,m,h}}{\partial\Delta T}\right\vert_{\Delta V= 0}$ induced due to the temperature bias \footnote{We assume that the voltage bias is applied to the lead, while the system is grounded. We also consider the lead's chemical potential shift $\mu_L=\Delta V=|\Delta V|$ by a positive voltage bias. This choice is arbitrary, for a negative bias, one can replace $\partial(\Delta V)\rightarrow \partial(-\Delta V)$ in the corresponding derivatives}. It brings us to nine different types of noise, but only four of them are independent. Using Eqs. (\ref{Lint}) and (\ref{Nint}), all types of noise in the weak tunneling regime can be brought into a form similar to Eq. (\ref{LCoeff}):
\begin{align} \label{SmatrN} \renewcommand\arraystretch{1.2}
   \begin{pmatrix}
        S_c\\ S_m\\ S_h
    \end{pmatrix}=
    \begin{pmatrix}
        2T\mathcal{L}_0 & \mathcal{N}_0 & \frac{1}{T}\mathcal{N}_1 \\
        2T\mathcal{L}_1 & \mathcal{N}_1-2 T\mathcal{L}_0& \frac{1}{T}\mathcal{N}_2\\
        2T\mathcal{L}_2 & \mathcal{N}_2-4T\mathcal{L}_1& \frac{1}{T}\mathcal{N}_3
    \end{pmatrix}
    \begin{pmatrix}
        1\\ \Delta V \\ \Delta T
    \end{pmatrix}.
\end{align} 
Note that the Johnson-Nyquist noise is fully determined by the transport coefficients $\mathcal{L}_n$ \cite{Crepieux2014, Eymeoud2016}, and none of the mixed noise $S_m$ components are independent; they are related to charge and heat noise as\\
\scalebox{1.}{\parbox{1.\linewidth}{
\begin{align}  \label{NoiseRec}
 \delta S_m^{SN}=T\, \delta S_c^{\Delta T}-S_c^{JN},\,   \delta S_h^{SN}=T\delta S_m^{\Delta T}-2S_m^{JN}. 
\end{align} }}\\
Detailed derivations of these results are given in \cite{Pavlov2025}. The expressions (\ref{NoiseRec}) extend the Onsager reciprocity relations \cite{Onsager1931b} to noise and have the same origin.

For our purpose, it will be more convenient to represent the transmission coefficient as a function of real time. For that, one switches to the time domain and makes an analytic continuation that accounts for finite temperature, as detailed in \cite{Matveev2002}:
\begin{align} \label{Tcoeff}
    \mathcal{T}(\varepsilon)=-\frac{1}{\pi}\cosh\frac{\varepsilon}{2T}\int_{-\infty}^{\infty}dt \mathcal{T}\left(\frac{1}{2T}+\textit{i}t\right)e^{\textit{i}\varepsilon t}.
\end{align}
This expression can be substituted to Eqs. (\ref{Lint}) and (\ref{Nint}) by integrating out the energy dependence and expressing all transport coefficients through integrals in the time domain, which are calculated explicitly in \cite{Pavlov2025}.

The transport integrals (\ref{Lint}) depend on the spectral symmetry of the system due to the structure of their kernels, which are either symmetric ($n=0,2$) or antisymmetric ($n=1$) functions of energy. For instance, the differential thermopower, which is proportional to $\mathcal{L}_1$, is zero for particle-hole symmetric systems \footnote{Because of nonlinear effects, the thermopower at the finite voltage and finite temperature bias is not necessarily zero at the particle-hole symmetric point \cite{Scheibner2005, Karki2017}, which is beyond the linear response theory.}. 
The same symmetry considerations apply to the noise coefficients (\ref{Nint}). Coefficients $\mathcal{L}_0$, $\mathcal{L}_2$, $\mathcal{N}_1$, and $\mathcal{N}_3$ depend only on the symmetric part of $\mathcal{T}(\varepsilon)$, while coefficients $\mathcal{L}_1$, $\mathcal{N}_0$, and $\mathcal{N}_2$ depend only on the antisymmetric part of $\mathcal{T}(\varepsilon)$. 

\paragraph*{Lorenz ratio ---}
There is a well-known relation between coefficients $\mathcal{L}_2$ and $\mathcal{L}_0$ that constitutes the WF law (conventionally, it is represented as a relation between $K$ and $G$ that holds for small values of the thermoelectric coefficient and is violated by the strong thermoelectric effects \cite{Benenti2017}). As we show in the following, as long as the ratio between these two transport coefficients is universal, there are universal ratios between all coefficients within the symmetric and antisymmetric categories. 

The WF law is known to hold (up to subleading finite temperature corrections) whenever charge and heat are transferred by the same carriers, which (in the zero-T limit) is equivalent to the condition of existing a well-defined Fermi surface \cite{Tanatar2007}, and scattering processes in the probed system are elastic (so there are no competition between the elastic and inelastic processes) \cite{Dong2013}. The generalized WF law $\frac{\mathcal{L}_2}{T^2 \mathcal{L}_0}=L_0 R_L$ goes beyond the Fermi liquid (FL) paradigm and constitutes the same relation between the heat conductance and the charge conductance up to a modified proportionality constant \cite{vanDalum2020}. Here, $L_0=\frac{\pi^2}{3}$ is the Lorenz number, and $R_L$ is the Lorenz ratio that accounts for the deviations from the FL relation ($R_L=1$ for the FL).

The Lorenz ratio can be written as
\begin{align} \label{RL}
    R_L=\frac{6\int_{-\infty}^{\infty}dt \cosh^{-3}\left(\pi Tt\right)\mathcal{T}\left(\frac{1}{2T}+\textit{i}t\right)}{\int_{-\infty}^{\infty}dt \cosh^{-1}\left(\pi Tt\right)\mathcal{T}\left(\frac{1}{2T}+\textit{i}t\right)}-3.
\end{align}
The generalized WF law holds at finite temperatures (meaning that $R_L$ is a constant universal number over a certain window of temperatures) as long as $\mathcal{T}$ in Eq. (\ref{RL}) scales as a function of a single parameter (the temperature). In general, an interplay of several energy scales for $\mathcal{T}$ and a competition between elastic and inelastic processes break this universality and violate the generalized WF law \cite{Kubala2008}. 
As an example, the WF law may be fulfilled at low and high temperatures but broken at the scale of mesoscopic fluctuations \cite{Vavilov2005}. We stress out that so far we did not put any constraints on the $\mathcal{T}$ structure. A quantum dot below the Kondo temperature (which is proportional to the charging energy) with negligible level spacing $\delta$ (so $T\gg \delta $ \cite{Aleiner1998}) is characterized by the transmission coefficient $\mathcal{T}\left(\frac{1}{2T}+\textit{i}t\right)\sim \frac{1}{\cosh^{\alpha}\left(\pi T t\right)}$. This structure of the transmission coefficient covers a broad variety of systems. For example, $\alpha=1$ for FL and $\alpha\in [1,3]$ for an $N$-channel Kondo quantum dot \cite{Kiselev2023} or a quantum Hall simulator \cite{Stabler2023} ($N\in [1,\infty)$, $\alpha=1+2/N$). The same structure with $\alpha=\frac{1}{2}$ is applicable to a quantum dot with SYK interactions in the conformal regime \cite{Pavlov2020}. In all these cases, the $\cosh$-like structure of the T-matrix stems from the conformal symmetry \footnote{The Lorenz ratio gives direct access to the scaling dimension of the leading irrelevant operator through the parameter $\alpha$, a related property of the delta-T noise for quantum Hall devices is discussed in \cite{Ebisu2022, Schiller2022, Zhang2022, Acciai2025}}.
The Lorenz ratio in this case reads
\begin{align} \label{RLcosh}
    R_L=\frac{3\alpha}{2+\alpha}.
\end{align}

\begin{figure}[t!]
\vspace{-0.6cm}
\center
\hspace*{-.5cm}
\includegraphics[width=1.1\linewidth]{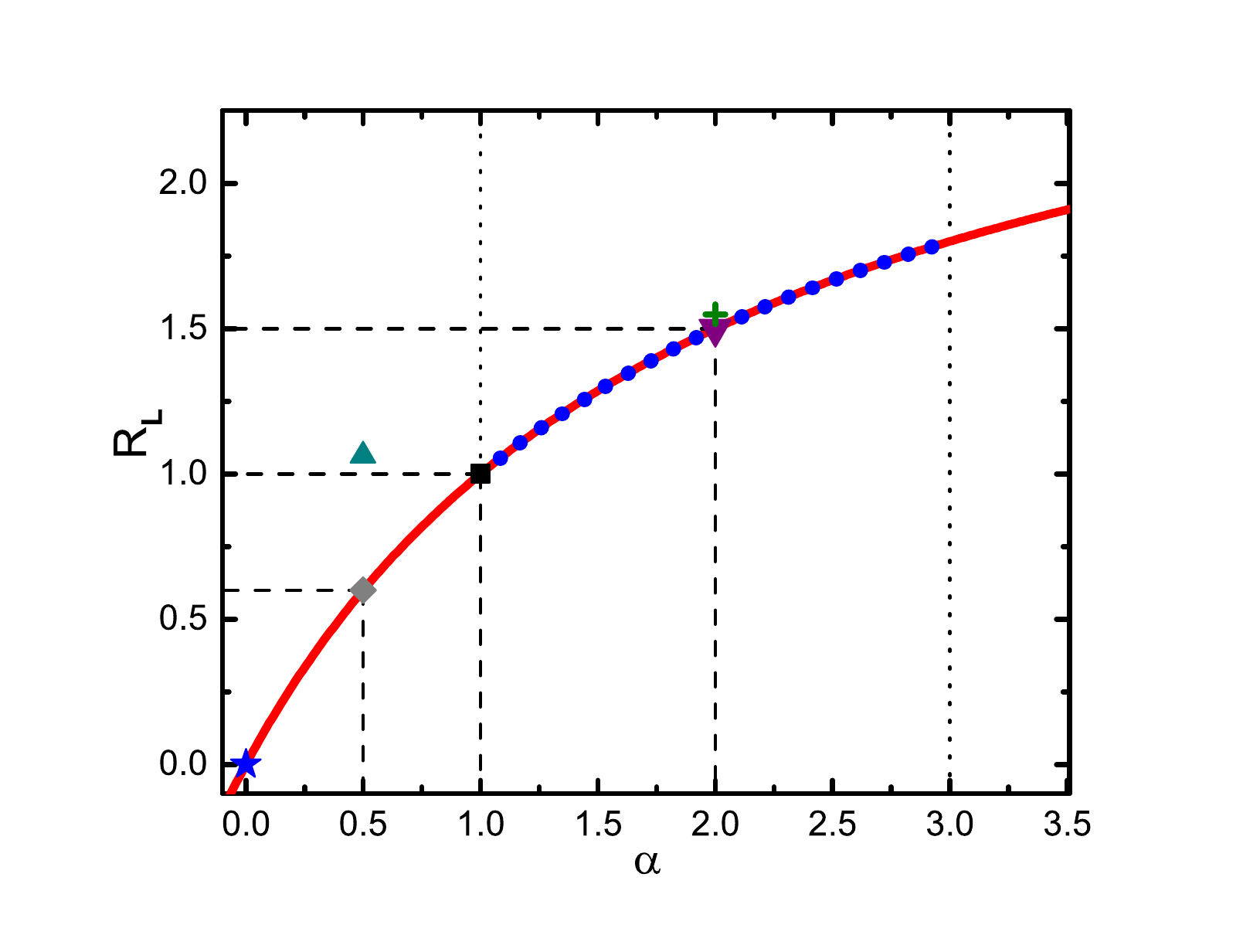} 
\vspace{-1cm}
\caption{Lorenz ratio given by Eq. (\ref{RL}). Red solid line, Lorenz ratio given by Eq. (\ref{RLcosh}). Blue dotted line, a range of Lorenz ratios for an N-channel Kondo or quantum Hall simulator; vertical dotted lines at $\alpha=1$ and $\alpha=3$ depict the range of $\alpha$ for the Kondo or quantum Hall simulator (they values reproduce the results obtained in \cite{Kiselev2023, Stabler2023, KiselevUn2} for these systems). The black square corresponds to the Fermi liquid regime. Gray diamond, SYK in the conformal regime. Downward violet triangle, inelastic tunneling regime for the conformal SYK. Blue star, large-$q$ conformal regime of the double-scaled SYK. Upward cyan triangle - SYK dot in the Schwarzian regime. Green cross, Schwarzian SYK regime with only inelastic tunneling. Dashed lines are used as eye guides.}
\label{fig:Lorenz_ratio} \vspace{-0.5cm}
\end{figure}

We illustrate these Lorenz ratios in Fig. \ref{fig:Lorenz_ratio}. The Lorenz ratio for Kondo and quantum Hall devices is always enhanced comparing to the FL due to the Anderson orthogonality catastrophe \cite{Anderson1967, Mahan1967, Furusaki1995, Kiselev2023}. In contrary, the reduced Lorenz ratio is realized for the SYK model, and an arbitrary small Lorenz ratio can be achieved within the double-scaled SYK (DSSYK) model \cite{Kitaev2018, Berkooz2025}, a $q$-body generalization of the SYK model (its saddle-point Green's function scales as $\cosh^{-\frac{2}{q}}\left(\pi T t\right)$, so $\alpha=\frac{2}{q}$), including $R_L=0$ at $q\rightarrow \infty$. In this limit, DSSYK belongs to the same universality class as the Random Matrix Theory \cite{Altland2024}. Furthermore, it can happen that the transmission coefficient is fully determined by inelastic processes, as is the case for an SYK dot deep within the Coulomb blockade \cite{Pavlov2020}. In this case, the universality of the generalized WF law can be restored whenever a single energy parameter governs the $T$-matrix (or at least its symmetric part) behavior over some range of temperatures (the detailed analysis is provided in \cite{Pavlov2025}). We illustrate it for the SYK model in conformal and Schwarzian regimes in Fig. \ref{fig:Lorenz_ratio} (for comparison, we put the corresponding points at the same $\alpha$ values that are realized for their conformal counterparts). Though the transmission coefficient of the system departs from the $\cosh$-like behavior within the Schwarzian regime, the generalized WF law is still obeyed. We analyze the WF law for the resonant impurity model and the two-stage Kondo model in \cite{Suppl}.

Within the elastic tunneling regime, the Lorenz ratio effectively probes the DoS in the vicinity of the zero energy. For FL, the DoS saturates in this area to some finite value and can be approximated by a constant (which results in $\alpha=1$). If DoS is suppressed due to strong electron-electron correlations, approaching zero at zero energy, the Lorenz ratio is boosted, as is the case for the multichannel Kondo simulators and quantum Hall devices (resulting in $\alpha>1$). The conformal regime of the (double-scaled) SYK model provides an opposite example - the DoS diverges in this case, leading to $\alpha<1$ and $R_L>1$. The SYK model provides a particular well-controlled example here, since its DoS switches between $\varepsilon^{-\frac{1}{2}}$ divergence within the conformal regime to $\varepsilon^{\frac{1}{2}}$ convergence within the Schwarzian regime, and the corresponding Lorenz ratio immediately changes from $R_L=\frac{3}{5}<1$ to $R_L\simeq 1.06>1$ (gray diamond and cyan upward triangle in Fig. \ref{fig:Lorenz_ratio}). The vanishing Lorenz ratio $R_L=0$ for the DSSYK at $q\rightarrow \infty$ arises from $\varepsilon^{-1}$ divergence of the DoS \footnote{The divergence of the DoS in the DSSYK model exists for the conformal saddle-point solution, and disappears in the infrared regularization of the saddle-point. The theory is well behaved for large $q$ and large $N$ as long as $q^2/N\ll 1$. The effective infrared theory in this regime coincides with the conventional SYK Schwarzian theory up to parametric prefactors and has the same energy scaling \cite{Berkooz2025}.}. Generally speaking, other scaling laws for divergent DoS may be realized. For instance, some conformal field theories exhibit the logarithmic DoS divergency \cite{Gurarie2013}. The Van Hove singularities \cite{vanHove1953} result into the logarithmic or power-law divergencies \cite{Efremov2019, Chandrasekaran2020, Yuan2020, Zervou2023} of the DoS (see also connections to multicritical Lifshitz points \cite{Shtyk2017}), that strongly affect their transport properties \cite{Piriou2011, Chen2011, Barber2018, Stangier2022}. The emergent noise relations for these theories may provide tools for distinguishing different divergencies in experiments, but such a consideration is beyond the scope of this work.

\paragraph*{Universal noise relations ---}
As long as the universality of Eq. (\ref{RL}) holds (i. e., the generalized WF law is obeyed), one can construct similar relations involving other transport coefficients. We detail this procedure in End Matter. Overall, there are five independent ratios (including the WF law) that relate all the coefficients between each other. For instance, one can define the ratio between the delta-T charge noise and the electric conductance. This is identical to the ratio of the following coefficients: $\delta S_c^{\Delta T}/G=\frac{\mathcal{N}_1}{T\mathcal{L}_0}=L_1R^{\Delta T}_C$. Here, we denoted $L_1=1$ as the first extended Lorenz number for FL, and $R^{\Delta T}_C$ is the deviation of this ratio from the FL value. Furthermore, one can define $\frac{\delta S_h^{\Delta T}}{T^2G}=\frac{\mathcal{N}_3}{T^3\mathcal{L}_0}=L_2R^{\Delta T}_H$ with $L_2=\pi^2$, $\frac{\delta S_c^{SN}}{G_T}=\frac{T\mathcal{N}_0}{\mathcal{L}_1}=L_3 R^V_C$ with $L_3=\frac{1}{3}$, and $\frac{\mathcal{N}_2}{T\mathcal{L}_1}=L_4 R^V_H$ with $L_4=\frac{12+\pi^2}{9}\simeq 2.43$. 

Having established these relations between the symmetric and antisymmetric coefficients, we can identify other characteristics of transport that acquire certain universality. Let us consider the Fano factor of charge transport, which constitutes a ratio between the shot noise and charge current due to the voltage bias, $F_c^{SN}=S_c/I_c$. Note that in the zero temperature limit this ratio becomes exactly unity, which is a universal property of a weak tunneling contact - this can be seen from Eqs. (\ref{currentDef}) and (\ref{noiseDef}) for an arbitrary $\mathcal{T}(\varepsilon)$, as the Fermi-Dirac distribution becomes an idempotent function $n_L(\varepsilon+\Delta V, T)\underset{T\rightarrow 0}{=}\theta(\varepsilon+\Delta V)$. For the linear response regime, which is valid at $\Delta T, \Delta V\ll T$, let us subtract the equilibrium component of the noise, defining $F_c^{SN}=\delta S_c^{SN}/G$. We introduce the delta-T Fano factor as the ratio between the delta-T charge noise and the charge current induced due to temperature imbalance, $F_c^{\Delta T}=\delta S_c^{\Delta T}/G_T$. We have
\begin{align} \label{FFc}
    F_c^{SN}F_c^{\Delta T}=\frac{\mathcal{N}_1}{\mathcal{L}_0}\frac{\mathcal{N}_0}{\mathcal{L}_1}=L_1L_3R_C^{\Delta T}R_C^V=const.
\end{align}
We can further use the Fano factors for the heat current in the same scheme ($F_h= S_h/TI_h$), obtaining
\begin{align} \label{FFh} \hspace{-.15cm}
    F_h^{SN}F_h^{\Delta T}=\frac{\mathcal{N}_2-4\mathcal{L}_1}{\mathcal{L}_1}\frac{\mathcal{N}_3}{\mathcal{L}_2}=\frac{L_2R_H^{\Delta T}}{L_0R_L}\left(L_4R_H^V-4\right).
\end{align}
Furthermore, we can write the thermopower $\mathcal{S}$ as
\begin{align} 
    \mathcal{S}=\frac{G_T}{G}=\frac{F_c^{SN}}{L_3 R^V_C}=\frac{1}{L_3 R^V_C}\frac{\mathcal{N}_0}{\mathcal{L}_0}.
\end{align}
Within the universality regime, the shot noise Fano factor provides us with the same information as the Seebeck coefficient. As is evident from Eqs. (\ref{FFc}) and (\ref{FFh}), a suppression of the Fano factor for the shot noise leads to an enhancement of the Fano factor for the delta-T noise (see also \cite{Nguyen2025}).\\
Moving beyond the linear response regime, one can construct multiple universal ratios between the currents and noises components quadratic with respect to biases, as demonstrated in \cite{Suppl}. In particular, the quadratic components of the currents can be expressed in a closed form through the linear-response components of noises, and a product of the generalized charge Fano factor (used to quantify nonlinear corrections to charge noise and current \cite{Mora2009, Mora2015}) and the linear-response charge Fano factor forms another universal relation.
\paragraph*{Conclusions ---}
The introduced \textit{noise coefficients} provide a complete characterization of the zero-frequency noise power in mesoscopic and nanoscopic transport (Eq. (\ref{SmatrN})) in a system weakly coupled to a metallic contact. There are seven independent differential observables that characterize current and noise. Indeed, the Johnson-Nyquist charge, mixed and heat noise exactly replicate the differential transport coefficients - conductance, thermoelectric coefficient, and heat conductance, correspondingly. Moreover, the mixed shot noise and the charge delta-T noise, and the heat shot noise and the mixed delta-T noise are connected though the reciprocity relations (\ref{NoiseRec}), similar to the Onsager reciprocity relation that connects the Seebeck and Peltier coefficients. Therefore, the mixed noise does not have independent components. It leaves only four potentially independent noise coefficients which can be treated on the same footing with the three independent transport coefficients characterizing currents. These coefficients are grouped into two categories: symmetric and antisymmetric ones with respect to their dependence on the spectral asymmetry. As we have shown, the generalized Wiedemann-Franz law, the relation that connects two symmetric transport coefficients, is just one of the relations that connect all the symmetric transport and noise coefficients with each other as long as the transmission coefficient of the system can be approximated as a single-parameter function. The same consideration holds for the antisymmetric coefficients, so in this universality regime all transport and noise features of the system are determined by just two independent observables. This redundancy of the noise features can be used to extract the same information about the system from different measurements. In particular, the thermoelectric measurements can be substituted by the shot noise measurements, which require only the voltage bias instead of tuning between the voltage and temperature biases. The established relations between the Fano factors signify that the delta-T noise can be a valuable experimental signature in cases when the shot noise is suppressed. These findings provide new tools for experimental studies of strongly correlated systems and understanding properties of non-Fermi-liquid materials, quantum information probes in nanotransport \cite{Koski2013, Koski2014, Kleeorin2019, Han2022}, studies of holographic systems \cite{Kruchkov2019}, design of quantum heat engines \cite{Pekola2015, Jaliel2019, Pekola2021}, insights into quantum Hall and fractional quantum Hall devices \cite{Sanchez2015, Rech2020, Zhang2022, Schiller2022, Ebisu2022, Rebora2022, Stabler2023, Iyer2023,  Acciai2025, Zhang2025, Zhang2025b}, and thermoelectric experiments in cold atoms \cite{Brantut2013, Hausler2021}. An interesting open question is how the relations between the noise and thermoelectrics behave beyond the weak tunneling limit, especially in the opposite limit of an open quantum point contact (QPC). A rigorous quantum field theoretical QPC investigation of the noise including large-scale numerical simulations \cite{Bauer2013} can be carried out in future studies.

\paragraph*{Acknowledgements} -- We thank Liliana Arrachea, Diana Duli\'c, Dmitry Efremov, Deepak Karki, Thierry Martin, Thanh Nguyen, Jerome Rech, Subir Sachdev, and Herre van der Zant for stimulating discussions.
A. I. P. appreciates the hospitality of ICTP Trieste, where part of this work was carried out. The work of M. N. K. is conducted within the
framework of the Trieste Institute for Theoretical Quantum Technologies (TQT).

\bibliography{refNoise}

\begin{thebibliography}{128}%
\makeatletter
\providecommand \@ifxundefined [1]{%
 \@ifx{#1\undefined}
}%
\providecommand \@ifnum [1]{%
 \ifnum #1\expandafter \@firstoftwo
 \else \expandafter \@secondoftwo
 \fi
}%
\providecommand \@ifx [1]{%
 \ifx #1\expandafter \@firstoftwo
 \else \expandafter \@secondoftwo
 \fi
}%
\providecommand \natexlab [1]{#1}%
\providecommand \enquote  [1]{``#1''}%
\providecommand \bibnamefont  [1]{#1}%
\providecommand \bibfnamefont [1]{#1}%
\providecommand \citenamefont [1]{#1}%
\providecommand \href@noop [0]{\@secondoftwo}%
\providecommand \href [0]{\begingroup \@sanitize@url \@href}%
\providecommand \@href[1]{\@@startlink{#1}\@@href}%
\providecommand \@@href[1]{\endgroup#1\@@endlink}%
\providecommand \@sanitize@url [0]{\catcode `\\12\catcode `\$12\catcode `\&12\catcode `\#12\catcode `\^12\catcode `\_12\catcode `\%12\relax}%
\providecommand \@@startlink[1]{}%
\providecommand \@@endlink[0]{}%
\providecommand \url  [0]{\begingroup\@sanitize@url \@url }%
\providecommand \@url [1]{\endgroup\@href {#1}{\urlprefix }}%
\providecommand \urlprefix  [0]{URL }%
\providecommand \Eprint [0]{\href }%
\providecommand \doibase [0]{https://doi.org/}%
\providecommand \selectlanguage [0]{\@gobble}%
\providecommand \bibinfo  [0]{\@secondoftwo}%
\providecommand \bibfield  [0]{\@secondoftwo}%
\providecommand \translation [1]{[#1]}%
\providecommand \BibitemOpen [0]{}%
\providecommand \bibitemStop [0]{}%
\providecommand \bibitemNoStop [0]{.\EOS\space}%
\providecommand \EOS [0]{\spacefactor3000\relax}%
\providecommand \BibitemShut  [1]{\csname bibitem#1\endcsname}%
\let\auto@bib@innerbib\@empty
\bibitem [{\citenamefont {Scheibner}\ \emph {et~al.}(2005)\citenamefont {Scheibner}, \citenamefont {Buhmann}, \citenamefont {Reuter}, \citenamefont {Kiselev},\ and\ \citenamefont {Molenkamp}}]{Scheibner2005}%
  \BibitemOpen
  \bibfield  {author} {\bibinfo {author} {\bibfnamefont {R.}~\bibnamefont {Scheibner}}, \bibinfo {author} {\bibfnamefont {H.}~\bibnamefont {Buhmann}}, \bibinfo {author} {\bibfnamefont {D.}~\bibnamefont {Reuter}}, \bibinfo {author} {\bibfnamefont {M.~N.}\ \bibnamefont {Kiselev}},\ and\ \bibinfo {author} {\bibfnamefont {L.~W.}\ \bibnamefont {Molenkamp}},\ }\bibfield  {title} {\bibinfo {title} {Thermopower of a {K}ondo spin-correlated quantum dot},\ }\href {https://doi.org/10.1103/PhysRevLett.95.176602} {\bibfield  {journal} {\bibinfo  {journal} {Phys. Rev. Lett.}\ }\textbf {\bibinfo {volume} {95}},\ \bibinfo {pages} {176602} (\bibinfo {year} {2005})}\BibitemShut {NoStop}%
\bibitem [{\citenamefont {Yamauchi}\ \emph {et~al.}(2011)\citenamefont {Yamauchi}, \citenamefont {Sekiguchi}, \citenamefont {Chida}, \citenamefont {Arakawa}, \citenamefont {Nakamura}, \citenamefont {Kobayashi}, \citenamefont {Ono}, \citenamefont {Fujii},\ and\ \citenamefont {Sakano}}]{Yamauchi2011}%
  \BibitemOpen
  \bibfield  {author} {\bibinfo {author} {\bibfnamefont {Y.}~\bibnamefont {Yamauchi}}, \bibinfo {author} {\bibfnamefont {K.}~\bibnamefont {Sekiguchi}}, \bibinfo {author} {\bibfnamefont {K.}~\bibnamefont {Chida}}, \bibinfo {author} {\bibfnamefont {T.}~\bibnamefont {Arakawa}}, \bibinfo {author} {\bibfnamefont {S.}~\bibnamefont {Nakamura}}, \bibinfo {author} {\bibfnamefont {K.}~\bibnamefont {Kobayashi}}, \bibinfo {author} {\bibfnamefont {T.}~\bibnamefont {Ono}}, \bibinfo {author} {\bibfnamefont {T.}~\bibnamefont {Fujii}},\ and\ \bibinfo {author} {\bibfnamefont {R.}~\bibnamefont {Sakano}},\ }\bibfield  {title} {\bibinfo {title} {Evolution of the {K}ondo effect in a quantum dot probed by shot noise},\ }\href {https://doi.org/10.1103/PhysRevLett.106.176601} {\bibfield  {journal} {\bibinfo  {journal} {Phys. Rev. Lett.}\ }\textbf {\bibinfo {volume} {106}},\ \bibinfo {pages} {176601} (\bibinfo {year} {2011})}\BibitemShut {NoStop}%
\bibitem [{\citenamefont {Borzenets}\ \emph {et~al.}(2020)\citenamefont {Borzenets}, \citenamefont {Shim}, \citenamefont {Chen}, \citenamefont {Ludwig}, \citenamefont {Wieck}, \citenamefont {Tarucha}, \citenamefont {Sim},\ and\ \citenamefont {Yamamoto}}]{Borzenets2020}%
  \BibitemOpen
  \bibfield  {author} {\bibinfo {author} {\bibfnamefont {I.~V.}\ \bibnamefont {Borzenets}}, \bibinfo {author} {\bibfnamefont {J.}~\bibnamefont {Shim}}, \bibinfo {author} {\bibfnamefont {J.~C.~H.}\ \bibnamefont {Chen}}, \bibinfo {author} {\bibfnamefont {A.}~\bibnamefont {Ludwig}}, \bibinfo {author} {\bibfnamefont {A.~D.}\ \bibnamefont {Wieck}}, \bibinfo {author} {\bibfnamefont {S.}~\bibnamefont {Tarucha}}, \bibinfo {author} {\bibfnamefont {H.-S.}\ \bibnamefont {Sim}},\ and\ \bibinfo {author} {\bibfnamefont {M.}~\bibnamefont {Yamamoto}},\ }\bibfield  {title} {\bibinfo {title} {Observation of the {K}ondo screening cloud},\ }\href {https://doi.org/10.1038/s41586-020-2058-6} {\bibfield  {journal} {\bibinfo  {journal} {Nature}\ }\textbf {\bibinfo {volume} {579}},\ \bibinfo {pages} {210} (\bibinfo {year} {2020})}\BibitemShut {NoStop}%
\bibitem [{\citenamefont {Godijn}\ \emph {et~al.}(1999)\citenamefont {Godijn}, \citenamefont {M\"oller}, \citenamefont {Buhmann}, \citenamefont {Molenkamp},\ and\ \citenamefont {van Langen}}]{Godijin1999}%
  \BibitemOpen
  \bibfield  {author} {\bibinfo {author} {\bibfnamefont {S.~F.}\ \bibnamefont {Godijn}}, \bibinfo {author} {\bibfnamefont {S.}~\bibnamefont {M\"oller}}, \bibinfo {author} {\bibfnamefont {H.}~\bibnamefont {Buhmann}}, \bibinfo {author} {\bibfnamefont {L.~W.}\ \bibnamefont {Molenkamp}},\ and\ \bibinfo {author} {\bibfnamefont {S.~A.}\ \bibnamefont {van Langen}},\ }\bibfield  {title} {\bibinfo {title} {Thermopower of a chaotic quantum dot},\ }\href {https://doi.org/10.1103/PhysRevLett.82.2927} {\bibfield  {journal} {\bibinfo  {journal} {Phys. Rev. Lett.}\ }\textbf {\bibinfo {volume} {82}},\ \bibinfo {pages} {2927} (\bibinfo {year} {1999})}\BibitemShut {NoStop}%
\bibitem [{\citenamefont {Barthold}\ \emph {et~al.}(2006)\citenamefont {Barthold}, \citenamefont {Hohls}, \citenamefont {Maire}, \citenamefont {Pierz},\ and\ \citenamefont {Haug}}]{Barthold2006}%
  \BibitemOpen
  \bibfield  {author} {\bibinfo {author} {\bibfnamefont {P.}~\bibnamefont {Barthold}}, \bibinfo {author} {\bibfnamefont {F.}~\bibnamefont {Hohls}}, \bibinfo {author} {\bibfnamefont {N.}~\bibnamefont {Maire}}, \bibinfo {author} {\bibfnamefont {K.}~\bibnamefont {Pierz}},\ and\ \bibinfo {author} {\bibfnamefont {R.~J.}\ \bibnamefont {Haug}},\ }\bibfield  {title} {\bibinfo {title} {Enhanced shot noise in tunneling through a stack of coupled quantum dots},\ }\href {https://doi.org/10.1103/PhysRevLett.96.246804} {\bibfield  {journal} {\bibinfo  {journal} {Phys. Rev. Lett.}\ }\textbf {\bibinfo {volume} {96}},\ \bibinfo {pages} {246804} (\bibinfo {year} {2006})}\BibitemShut {NoStop}%
\bibitem [{\citenamefont {Hartman}\ \emph {et~al.}(2018)\citenamefont {Hartman}, \citenamefont {Olsen}, \citenamefont {L\"uscher}, \citenamefont {Samani}, \citenamefont {Fallahi}, \citenamefont {Gardner}, \citenamefont {Manfra},\ and\ \citenamefont {Folk}}]{Hartman2018}%
  \BibitemOpen
  \bibfield  {author} {\bibinfo {author} {\bibfnamefont {N.}~\bibnamefont {Hartman}}, \bibinfo {author} {\bibfnamefont {C.}~\bibnamefont {Olsen}}, \bibinfo {author} {\bibfnamefont {S.}~\bibnamefont {L\"uscher}}, \bibinfo {author} {\bibfnamefont {M.}~\bibnamefont {Samani}}, \bibinfo {author} {\bibfnamefont {S.}~\bibnamefont {Fallahi}}, \bibinfo {author} {\bibfnamefont {G.~C.}\ \bibnamefont {Gardner}}, \bibinfo {author} {\bibfnamefont {M.}~\bibnamefont {Manfra}},\ and\ \bibinfo {author} {\bibfnamefont {J.}~\bibnamefont {Folk}},\ }\bibfield  {title} {\bibinfo {title} {Direct entropy measurement in a mesoscopic quantum system},\ }\href {https://doi.org/10.1038/s41567-018-0250-5} {\bibfield  {journal} {\bibinfo  {journal} {Nature Phys.}\ }\textbf {\bibinfo {volume} {14}},\ \bibinfo {pages} {1083} (\bibinfo {year} {2018})}\BibitemShut {NoStop}%
\bibitem [{\citenamefont {Dong}\ \emph {et~al.}(2013)\citenamefont {Dong}, \citenamefont {Tokiwa}, \citenamefont {Bud'ko}, \citenamefont {Canfield},\ and\ \citenamefont {Gegenwart}}]{Dong2013}%
  \BibitemOpen
  \bibfield  {author} {\bibinfo {author} {\bibfnamefont {J.~K.}\ \bibnamefont {Dong}}, \bibinfo {author} {\bibfnamefont {Y.}~\bibnamefont {Tokiwa}}, \bibinfo {author} {\bibfnamefont {S.~L.}\ \bibnamefont {Bud'ko}}, \bibinfo {author} {\bibfnamefont {P.~C.}\ \bibnamefont {Canfield}},\ and\ \bibinfo {author} {\bibfnamefont {P.}~\bibnamefont {Gegenwart}},\ }\bibfield  {title} {\bibinfo {title} {Anomalous reduction of the {L}orenz ratio at the quantum critical point in {YbAgGe}},\ }\href {https://doi.org/10.1103/PhysRevLett.110.176402} {\bibfield  {journal} {\bibinfo  {journal} {Phys. Rev. Lett.}\ }\textbf {\bibinfo {volume} {110}},\ \bibinfo {pages} {176402} (\bibinfo {year} {2013})}\BibitemShut {NoStop}%
\bibitem [{\citenamefont {Chen}\ \emph {et~al.}(2023)\citenamefont {Chen}, \citenamefont {Lowder}, \citenamefont {Bakali}, \citenamefont {Andrews}, \citenamefont {Schrenk}, \citenamefont {Waas}, \citenamefont {Svagera}, \citenamefont {Eguchi}, \citenamefont {Prochaska}, \citenamefont {Wang}, \citenamefont {Setty}, \citenamefont {Sur}, \citenamefont {Si}, \citenamefont {Paschen},\ and\ \citenamefont {Natelson}}]{Chen2023}%
  \BibitemOpen
  \bibfield  {author} {\bibinfo {author} {\bibfnamefont {L.}~\bibnamefont {Chen}}, \bibinfo {author} {\bibfnamefont {D.~T.}\ \bibnamefont {Lowder}}, \bibinfo {author} {\bibfnamefont {E.}~\bibnamefont {Bakali}}, \bibinfo {author} {\bibfnamefont {A.~M.}\ \bibnamefont {Andrews}}, \bibinfo {author} {\bibfnamefont {W.}~\bibnamefont {Schrenk}}, \bibinfo {author} {\bibfnamefont {M.}~\bibnamefont {Waas}}, \bibinfo {author} {\bibfnamefont {R.}~\bibnamefont {Svagera}}, \bibinfo {author} {\bibfnamefont {G.}~\bibnamefont {Eguchi}}, \bibinfo {author} {\bibfnamefont {L.}~\bibnamefont {Prochaska}}, \bibinfo {author} {\bibfnamefont {Y.}~\bibnamefont {Wang}}, \bibinfo {author} {\bibfnamefont {C.}~\bibnamefont {Setty}}, \bibinfo {author} {\bibfnamefont {S.}~\bibnamefont {Sur}}, \bibinfo {author} {\bibfnamefont {Q.}~\bibnamefont {Si}}, \bibinfo {author} {\bibfnamefont {S.}~\bibnamefont {Paschen}},\ and\ \bibinfo {author} {\bibfnamefont {D.}~\bibnamefont {Natelson}},\ }\bibfield  {title} {\bibinfo {title} {Shot noise in a
  strange metal},\ }\href {https://doi.org/10.1126/science.abq6100} {\bibfield  {journal} {\bibinfo  {journal} {Science}\ }\textbf {\bibinfo {volume} {382}},\ \bibinfo {pages} {907} (\bibinfo {year} {2023})}\BibitemShut {NoStop}%
\bibitem [{\citenamefont {Wang}\ \emph {et~al.}(2024)\citenamefont {Wang}, \citenamefont {Setty}, \citenamefont {Sur}, \citenamefont {Chen}, \citenamefont {Paschen}, \citenamefont {Natelson},\ and\ \citenamefont {Si}}]{Wang2024}%
  \BibitemOpen
  \bibfield  {author} {\bibinfo {author} {\bibfnamefont {Y.}~\bibnamefont {Wang}}, \bibinfo {author} {\bibfnamefont {C.}~\bibnamefont {Setty}}, \bibinfo {author} {\bibfnamefont {S.}~\bibnamefont {Sur}}, \bibinfo {author} {\bibfnamefont {L.}~\bibnamefont {Chen}}, \bibinfo {author} {\bibfnamefont {S.}~\bibnamefont {Paschen}}, \bibinfo {author} {\bibfnamefont {D.}~\bibnamefont {Natelson}},\ and\ \bibinfo {author} {\bibfnamefont {Q.}~\bibnamefont {Si}},\ }\bibfield  {title} {\bibinfo {title} {Shot noise and universal {F}ano factor as a characterization of strongly correlated metals},\ }\href {https://doi.org/10.1103/PhysRevResearch.6.L042045} {\bibfield  {journal} {\bibinfo  {journal} {Phys. Rev. Res.}\ }\textbf {\bibinfo {volume} {6}},\ \bibinfo {pages} {L042045} (\bibinfo {year} {2024})}\BibitemShut {NoStop}%
\bibitem [{\citenamefont {Gleis}\ \emph {et~al.}(2025)\citenamefont {Gleis}, \citenamefont {Lee}, \citenamefont {Kotliar},\ and\ \citenamefont {von Delft}}]{Gleis2025}%
  \BibitemOpen
  \bibfield  {author} {\bibinfo {author} {\bibfnamefont {A.}~\bibnamefont {Gleis}}, \bibinfo {author} {\bibfnamefont {S.-S.~B.}\ \bibnamefont {Lee}}, \bibinfo {author} {\bibfnamefont {G.}~\bibnamefont {Kotliar}},\ and\ \bibinfo {author} {\bibfnamefont {J.}~\bibnamefont {von Delft}},\ }\bibfield  {title} {\bibinfo {title} {Dynamical scaling and {P}lanckian dissipation due to heavy-fermion quantum criticality},\ }\href {https://doi.org/10.1103/PhysRevLett.134.106501} {\bibfield  {journal} {\bibinfo  {journal} {Phys. Rev. Lett.}\ }\textbf {\bibinfo {volume} {134}},\ \bibinfo {pages} {106501} (\bibinfo {year} {2025})}\BibitemShut {NoStop}%
\bibitem [{\citenamefont {Mravlje}\ and\ \citenamefont {Georges}(2016)}]{Mravlje2016}%
  \BibitemOpen
  \bibfield  {author} {\bibinfo {author} {\bibfnamefont {J.}~\bibnamefont {Mravlje}}\ and\ \bibinfo {author} {\bibfnamefont {A.}~\bibnamefont {Georges}},\ }\bibfield  {title} {\bibinfo {title} {Thermopower and entropy: Lessons from {${\mathrm{Sr}}_{2}{\mathrm{RuO}}_{4}$}},\ }\href {https://doi.org/10.1103/PhysRevLett.117.036401} {\bibfield  {journal} {\bibinfo  {journal} {Phys. Rev. Lett.}\ }\textbf {\bibinfo {volume} {117}},\ \bibinfo {pages} {036401} (\bibinfo {year} {2016})}\BibitemShut {NoStop}%
\bibitem [{\citenamefont {Sachdev}(2015)}]{Sachdev2015}%
  \BibitemOpen
  \bibfield  {author} {\bibinfo {author} {\bibfnamefont {S.}~\bibnamefont {Sachdev}},\ }\bibfield  {title} {\bibinfo {title} {Bekenstein-{H}awking entropy and strange metals},\ }\href {https://doi.org/10.1103/PhysRevX.5.041025} {\bibfield  {journal} {\bibinfo  {journal} {Phys. Rev. X}\ }\textbf {\bibinfo {volume} {5}},\ \bibinfo {pages} {041025} (\bibinfo {year} {2015})}\BibitemShut {NoStop}%
\bibitem [{\citenamefont {Davison}\ \emph {et~al.}(2017)\citenamefont {Davison}, \citenamefont {Fu}, \citenamefont {Georges}, \citenamefont {Gu}, \citenamefont {Jensen},\ and\ \citenamefont {Sachdev}}]{Davison2017}%
  \BibitemOpen
  \bibfield  {author} {\bibinfo {author} {\bibfnamefont {R.~A.}\ \bibnamefont {Davison}}, \bibinfo {author} {\bibfnamefont {W.}~\bibnamefont {Fu}}, \bibinfo {author} {\bibfnamefont {A.}~\bibnamefont {Georges}}, \bibinfo {author} {\bibfnamefont {Y.}~\bibnamefont {Gu}}, \bibinfo {author} {\bibfnamefont {K.}~\bibnamefont {Jensen}},\ and\ \bibinfo {author} {\bibfnamefont {S.}~\bibnamefont {Sachdev}},\ }\bibfield  {title} {\bibinfo {title} {Thermoelectric transport in disordered metals without quasiparticles: The {S}achdev-{Y}e-{K}itaev models and holography},\ }\href {https://doi.org/10.1103/PhysRevB.95.155131} {\bibfield  {journal} {\bibinfo  {journal} {Phys. Rev. B}\ }\textbf {\bibinfo {volume} {95}},\ \bibinfo {pages} {155131} (\bibinfo {year} {2017})}\BibitemShut {NoStop}%
\bibitem [{\citenamefont {Inkof}\ \emph {et~al.}(2020)\citenamefont {Inkof}, \citenamefont {K\"uppers}, \citenamefont {Link}, \citenamefont {Gout\'eraux},\ and\ \citenamefont {Schmalian}}]{Inkof2020}%
  \BibitemOpen
  \bibfield  {author} {\bibinfo {author} {\bibfnamefont {G.~A.}\ \bibnamefont {Inkof}}, \bibinfo {author} {\bibfnamefont {J.~M.~C.}\ \bibnamefont {K\"uppers}}, \bibinfo {author} {\bibfnamefont {J.~M.}\ \bibnamefont {Link}}, \bibinfo {author} {\bibfnamefont {B.}~\bibnamefont {Gout\'eraux}},\ and\ \bibinfo {author} {\bibfnamefont {J.}~\bibnamefont {Schmalian}},\ }\bibfield  {title} {\bibinfo {title} {Quantum critical scaling and holographic bound for transport coefficients near {L}ifshitz points},\ }\href {https://doi.org/10.1007/jhep11(2020)088} {\bibfield  {journal} {\bibinfo  {journal} {J. High Energ. Phys.}\ }\textbf {\bibinfo {volume} {2020}}\bibinfo  {number} { (11)},\ \bibinfo {pages} {88}}\BibitemShut {NoStop}%
\bibitem [{\citenamefont {Martin}(2005)}]{Martin2005}%
  \BibitemOpen
\bibfield  {number} {  }\bibfield  {author} {\bibinfo {author} {\bibfnamefont {T.}~\bibnamefont {Martin}},\ }\bibinfo {title} {Noise in mesoscopic physics},\ in\ \href {https://doi.org/https://doi.org/10.1016/S0924-8099(05)80047-2} {\emph {\bibinfo {booktitle} {Nanophysics: Coherence and Transport}}},\ \bibinfo {series and number} {Les Houches, Session LXXXI},\ \bibinfo {editor} {edited by\ \bibinfo {editor} {\bibfnamefont {H.}~\bibnamefont {Bouchiat}}, \bibinfo {editor} {\bibfnamefont {Y.}~\bibnamefont {Gefen}}, \bibinfo {editor} {\bibfnamefont {S.}~\bibnamefont {Gu{\'e}ron}}, \bibinfo {editor} {\bibfnamefont {G.}~\bibnamefont {Montambaux}},\ and\ \bibinfo {editor} {\bibfnamefont {J.}~\bibnamefont {Dalibard}}}\ (\bibinfo  {publisher} {Elsevier},\ \bibinfo {year} {2005})\ p.\ \bibinfo {pages} {283}\BibitemShut {NoStop}%
\bibitem [{\citenamefont {Lumbroso}\ \emph {et~al.}(2018)\citenamefont {Lumbroso}, \citenamefont {Simine}, \citenamefont {Nitzan}, \citenamefont {Segal},\ and\ \citenamefont {Tal}}]{Lumbroso2018}%
  \BibitemOpen
  \bibfield  {author} {\bibinfo {author} {\bibfnamefont {O.~S.}\ \bibnamefont {Lumbroso}}, \bibinfo {author} {\bibfnamefont {L.}~\bibnamefont {Simine}}, \bibinfo {author} {\bibfnamefont {A.}~\bibnamefont {Nitzan}}, \bibinfo {author} {\bibfnamefont {D.}~\bibnamefont {Segal}},\ and\ \bibinfo {author} {\bibfnamefont {O.}~\bibnamefont {Tal}},\ }\bibfield  {title} {\bibinfo {title} {Electronic noise due to temperature differences in atomic-scale junctions},\ }\href {https://doi.org/10.1038/s41586-018-0592-2} {\bibfield  {journal} {\bibinfo  {journal} {Nature}\ }\textbf {\bibinfo {volume} {562}},\ \bibinfo {pages} {240} (\bibinfo {year} {2018})}\BibitemShut {NoStop}%
\bibitem [{\citenamefont {Sivre}\ \emph {et~al.}(2019)\citenamefont {Sivre}, \citenamefont {Duprez}, \citenamefont {Anthore}, \citenamefont {Aassime}, \citenamefont {Parmentier}, \citenamefont {Cavanna}, \citenamefont {Ouerghi}, \citenamefont {Gennser},\ and\ \citenamefont {Pierre}}]{Sivre2019}%
  \BibitemOpen
  \bibfield  {author} {\bibinfo {author} {\bibfnamefont {E.}~\bibnamefont {Sivre}}, \bibinfo {author} {\bibfnamefont {H.}~\bibnamefont {Duprez}}, \bibinfo {author} {\bibfnamefont {A.}~\bibnamefont {Anthore}}, \bibinfo {author} {\bibfnamefont {A.}~\bibnamefont {Aassime}}, \bibinfo {author} {\bibfnamefont {F.~D.}\ \bibnamefont {Parmentier}}, \bibinfo {author} {\bibfnamefont {A.}~\bibnamefont {Cavanna}}, \bibinfo {author} {\bibfnamefont {A.}~\bibnamefont {Ouerghi}}, \bibinfo {author} {\bibfnamefont {U.}~\bibnamefont {Gennser}},\ and\ \bibinfo {author} {\bibfnamefont {F.}~\bibnamefont {Pierre}},\ }\bibfield  {title} {\bibinfo {title} {Electronic heat flow and thermal shot noise in quantum circuits},\ }\href {https://doi.org/10.1038/s41467-019-13566-8} {\bibfield  {journal} {\bibinfo  {journal} {Nat. Commun.}\ }\textbf {\bibinfo {volume} {10}},\ \bibinfo {pages} {5638} (\bibinfo {year} {2019})}\BibitemShut {NoStop}%
\bibitem [{\citenamefont {Larocque}\ \emph {et~al.}(2020)\citenamefont {Larocque}, \citenamefont {Pinsolle}, \citenamefont {Lupien},\ and\ \citenamefont {Reulet}}]{Larocque2020}%
  \BibitemOpen
  \bibfield  {author} {\bibinfo {author} {\bibfnamefont {S.}~\bibnamefont {Larocque}}, \bibinfo {author} {\bibfnamefont {E.}~\bibnamefont {Pinsolle}}, \bibinfo {author} {\bibfnamefont {C.}~\bibnamefont {Lupien}},\ and\ \bibinfo {author} {\bibfnamefont {B.}~\bibnamefont {Reulet}},\ }\bibfield  {title} {\bibinfo {title} {Shot noise of a temperature-biased tunnel junction},\ }\href {https://doi.org/10.1103/PhysRevLett.125.106801} {\bibfield  {journal} {\bibinfo  {journal} {Phys. Rev. Lett.}\ }\textbf {\bibinfo {volume} {125}},\ \bibinfo {pages} {106801} (\bibinfo {year} {2020})}\BibitemShut {NoStop}%
\bibitem [{\citenamefont {Vavilov}\ and\ \citenamefont {Stone}(2005)}]{Vavilov2005}%
  \BibitemOpen
  \bibfield  {author} {\bibinfo {author} {\bibfnamefont {M.~G.}\ \bibnamefont {Vavilov}}\ and\ \bibinfo {author} {\bibfnamefont {A.~D.}\ \bibnamefont {Stone}},\ }\bibfield  {title} {\bibinfo {title} {Failure of the {W}iedemann-{F}ranz law in mesoscopic conductors},\ }\href {https://doi.org/10.1103/PhysRevB.72.205107} {\bibfield  {journal} {\bibinfo  {journal} {Phys. Rev. B}\ }\textbf {\bibinfo {volume} {72}},\ \bibinfo {pages} {205107} (\bibinfo {year} {2005})}\BibitemShut {NoStop}%
\bibitem [{\citenamefont {Tanatar}\ \emph {et~al.}(2007)\citenamefont {Tanatar}, \citenamefont {Paglione}, \citenamefont {Petrovic},\ and\ \citenamefont {Taillefer}}]{Tanatar2007}%
  \BibitemOpen
  \bibfield  {author} {\bibinfo {author} {\bibfnamefont {M.~A.}\ \bibnamefont {Tanatar}}, \bibinfo {author} {\bibfnamefont {J.}~\bibnamefont {Paglione}}, \bibinfo {author} {\bibfnamefont {C.}~\bibnamefont {Petrovic}},\ and\ \bibinfo {author} {\bibfnamefont {L.}~\bibnamefont {Taillefer}},\ }\bibfield  {title} {\bibinfo {title} {Anisotropic violation of the {W}iedemann-{F}ranz law at a quantum critical point},\ }\href {https://doi.org/10.1126/science.1140762} {\bibfield  {journal} {\bibinfo  {journal} {Science}\ }\textbf {\bibinfo {volume} {316}},\ \bibinfo {pages} {1320} (\bibinfo {year} {2007})}\BibitemShut {NoStop}%
\bibitem [{\citenamefont {Kubala}\ \emph {et~al.}(2008)\citenamefont {Kubala}, \citenamefont {K\"onig},\ and\ \citenamefont {Pekola}}]{Kubala2008}%
  \BibitemOpen
  \bibfield  {author} {\bibinfo {author} {\bibfnamefont {B.}~\bibnamefont {Kubala}}, \bibinfo {author} {\bibfnamefont {J.}~\bibnamefont {K\"onig}},\ and\ \bibinfo {author} {\bibfnamefont {J.}~\bibnamefont {Pekola}},\ }\bibfield  {title} {\bibinfo {title} {Violation of the {W}iedemann-{F}ranz law in a single-electron transistor},\ }\href {https://doi.org/10.1103/PhysRevLett.100.066801} {\bibfield  {journal} {\bibinfo  {journal} {Phys. Rev. Lett.}\ }\textbf {\bibinfo {volume} {100}},\ \bibinfo {pages} {066801} (\bibinfo {year} {2008})}\BibitemShut {NoStop}%
\bibitem [{\citenamefont {Tikhanovskaya}\ \emph {et~al.}(2021)\citenamefont {Tikhanovskaya}, \citenamefont {Guo}, \citenamefont {Sachdev},\ and\ \citenamefont {Tarnopolsky}}]{Tikhanovskaya2021}%
  \BibitemOpen
  \bibfield  {author} {\bibinfo {author} {\bibfnamefont {M.}~\bibnamefont {Tikhanovskaya}}, \bibinfo {author} {\bibfnamefont {H.}~\bibnamefont {Guo}}, \bibinfo {author} {\bibfnamefont {S.}~\bibnamefont {Sachdev}},\ and\ \bibinfo {author} {\bibfnamefont {G.}~\bibnamefont {Tarnopolsky}},\ }\bibfield  {title} {\bibinfo {title} {Excitation spectra of quantum matter without quasiparticles. {I}. {S}achdev-{Y}e-{K}itaev models},\ }\href {https://doi.org/10.1103/PhysRevB.103.075141} {\bibfield  {journal} {\bibinfo  {journal} {Phys. Rev. B}\ }\textbf {\bibinfo {volume} {103}},\ \bibinfo {pages} {075141} (\bibinfo {year} {2021})}\BibitemShut {NoStop}%
\bibitem [{\citenamefont {Chowdhury}\ \emph {et~al.}(2022)\citenamefont {Chowdhury}, \citenamefont {Georges}, \citenamefont {Parcollet},\ and\ \citenamefont {Sachdev}}]{Chowdhury2022}%
  \BibitemOpen
  \bibfield  {author} {\bibinfo {author} {\bibfnamefont {D.}~\bibnamefont {Chowdhury}}, \bibinfo {author} {\bibfnamefont {A.}~\bibnamefont {Georges}}, \bibinfo {author} {\bibfnamefont {O.}~\bibnamefont {Parcollet}},\ and\ \bibinfo {author} {\bibfnamefont {S.}~\bibnamefont {Sachdev}},\ }\bibfield  {title} {\bibinfo {title} {Sachdev-{Y}e-{K}itaev models and beyond: Window into non-{F}ermi liquids},\ }\href {https://doi.org/10.1103/RevModPhys.94.035004} {\bibfield  {journal} {\bibinfo  {journal} {Rev. Mod. Phys.}\ }\textbf {\bibinfo {volume} {94}},\ \bibinfo {pages} {035004} (\bibinfo {year} {2022})}\BibitemShut {NoStop}%
\bibitem [{\citenamefont {van~der Vaart}\ \emph {et~al.}(1995)\citenamefont {van~der Vaart}, \citenamefont {Godijn}, \citenamefont {Nazarov}, \citenamefont {Harmans}, \citenamefont {Mooij}, \citenamefont {Molenkamp},\ and\ \citenamefont {Foxon}}]{vanderVaart1995}%
  \BibitemOpen
  \bibfield  {author} {\bibinfo {author} {\bibfnamefont {N.~C.}\ \bibnamefont {van~der Vaart}}, \bibinfo {author} {\bibfnamefont {S.~F.}\ \bibnamefont {Godijn}}, \bibinfo {author} {\bibfnamefont {Y.~V.}\ \bibnamefont {Nazarov}}, \bibinfo {author} {\bibfnamefont {C.~J. P.~M.}\ \bibnamefont {Harmans}}, \bibinfo {author} {\bibfnamefont {J.~E.}\ \bibnamefont {Mooij}}, \bibinfo {author} {\bibfnamefont {L.~W.}\ \bibnamefont {Molenkamp}},\ and\ \bibinfo {author} {\bibfnamefont {C.~T.}\ \bibnamefont {Foxon}},\ }\bibfield  {title} {\bibinfo {title} {Resonant tunneling through two discrete energy states},\ }\href {https://doi.org/10.1103/PhysRevLett.74.4702} {\bibfield  {journal} {\bibinfo  {journal} {Phys. Rev. Lett.}\ }\textbf {\bibinfo {volume} {74}},\ \bibinfo {pages} {4702} (\bibinfo {year} {1995})}\BibitemShut {NoStop}%
\bibitem [{\citenamefont {Goldhaber-Gordon}\ \emph {et~al.}(1998{\natexlab{a}})\citenamefont {Goldhaber-Gordon}, \citenamefont {Shtrikman}, \citenamefont {Mahalu}, \citenamefont {Abusch-Magder}, \citenamefont {Meirav},\ and\ \citenamefont {Kastner}}]{GoldhaberGordon1998}%
  \BibitemOpen
  \bibfield  {author} {\bibinfo {author} {\bibfnamefont {D.}~\bibnamefont {Goldhaber-Gordon}}, \bibinfo {author} {\bibfnamefont {H.}~\bibnamefont {Shtrikman}}, \bibinfo {author} {\bibfnamefont {D.}~\bibnamefont {Mahalu}}, \bibinfo {author} {\bibfnamefont {D.}~\bibnamefont {Abusch-Magder}}, \bibinfo {author} {\bibfnamefont {U.}~\bibnamefont {Meirav}},\ and\ \bibinfo {author} {\bibfnamefont {M.~A.}\ \bibnamefont {Kastner}},\ }\bibfield  {title} {\bibinfo {title} {Kondo effect in a single-electron transistor},\ }\href {https://doi.org/10.1038/34373} {\bibfield  {journal} {\bibinfo  {journal} {Nature}\ }\textbf {\bibinfo {volume} {391}},\ \bibinfo {pages} {156} (\bibinfo {year} {1998}{\natexlab{a}})}\BibitemShut {NoStop}%
\bibitem [{\citenamefont {Goldhaber-Gordon}\ \emph {et~al.}(1998{\natexlab{b}})\citenamefont {Goldhaber-Gordon}, \citenamefont {G\"ores}, \citenamefont {Kastner}, \citenamefont {Shtrikman}, \citenamefont {Mahalu},\ and\ \citenamefont {Meirav}}]{GoldhaberGordon1998b}%
  \BibitemOpen
  \bibfield  {author} {\bibinfo {author} {\bibfnamefont {D.}~\bibnamefont {Goldhaber-Gordon}}, \bibinfo {author} {\bibfnamefont {J.}~\bibnamefont {G\"ores}}, \bibinfo {author} {\bibfnamefont {M.~A.}\ \bibnamefont {Kastner}}, \bibinfo {author} {\bibfnamefont {H.}~\bibnamefont {Shtrikman}}, \bibinfo {author} {\bibfnamefont {D.}~\bibnamefont {Mahalu}},\ and\ \bibinfo {author} {\bibfnamefont {U.}~\bibnamefont {Meirav}},\ }\bibfield  {title} {\bibinfo {title} {From the {K}ondo regime to the mixed-valence regime in a single-electron transistor},\ }\href {https://doi.org/10.1103/PhysRevLett.81.5225} {\bibfield  {journal} {\bibinfo  {journal} {Phys. Rev. Lett.}\ }\textbf {\bibinfo {volume} {81}},\ \bibinfo {pages} {5225} (\bibinfo {year} {1998}{\natexlab{b}})}\BibitemShut {NoStop}%
\bibitem [{\citenamefont {Nauen}\ \emph {et~al.}(2002)\citenamefont {Nauen}, \citenamefont {Hapke-Wurst}, \citenamefont {Hohls}, \citenamefont {Zeitler}, \citenamefont {Haug},\ and\ \citenamefont {Pierz}}]{Nauen2002}%
  \BibitemOpen
  \bibfield  {author} {\bibinfo {author} {\bibfnamefont {A.}~\bibnamefont {Nauen}}, \bibinfo {author} {\bibfnamefont {I.}~\bibnamefont {Hapke-Wurst}}, \bibinfo {author} {\bibfnamefont {F.}~\bibnamefont {Hohls}}, \bibinfo {author} {\bibfnamefont {U.}~\bibnamefont {Zeitler}}, \bibinfo {author} {\bibfnamefont {R.~J.}\ \bibnamefont {Haug}},\ and\ \bibinfo {author} {\bibfnamefont {K.}~\bibnamefont {Pierz}},\ }\bibfield  {title} {\bibinfo {title} {Shot noise in self-assembled {InAs} quantum dots},\ }\href {https://doi.org/10.1103/PhysRevB.66.161303} {\bibfield  {journal} {\bibinfo  {journal} {Phys. Rev. B}\ }\textbf {\bibinfo {volume} {66}},\ \bibinfo {pages} {161303} (\bibinfo {year} {2002})}\BibitemShut {NoStop}%
\bibitem [{\citenamefont {Nauen}\ \emph {et~al.}(2004)\citenamefont {Nauen}, \citenamefont {Hohls}, \citenamefont {Maire}, \citenamefont {Pierz},\ and\ \citenamefont {Haug}}]{Nauen2004}%
  \BibitemOpen
  \bibfield  {author} {\bibinfo {author} {\bibfnamefont {A.}~\bibnamefont {Nauen}}, \bibinfo {author} {\bibfnamefont {F.}~\bibnamefont {Hohls}}, \bibinfo {author} {\bibfnamefont {N.}~\bibnamefont {Maire}}, \bibinfo {author} {\bibfnamefont {K.}~\bibnamefont {Pierz}},\ and\ \bibinfo {author} {\bibfnamefont {R.~J.}\ \bibnamefont {Haug}},\ }\bibfield  {title} {\bibinfo {title} {Shot noise in tunneling through a single quantum dot},\ }\href {https://doi.org/10.1103/PhysRevB.70.033305} {\bibfield  {journal} {\bibinfo  {journal} {Phys. Rev. B}\ }\textbf {\bibinfo {volume} {70}},\ \bibinfo {pages} {033305} (\bibinfo {year} {2004})}\BibitemShut {NoStop}%
\bibitem [{\citenamefont {Slot}\ \emph {et~al.}(2004)\citenamefont {Slot}, \citenamefont {Holst}, \citenamefont {van~der Zant},\ and\ \citenamefont {Zaitsev-Zotov}}]{Slot2004}%
  \BibitemOpen
  \bibfield  {author} {\bibinfo {author} {\bibfnamefont {E.}~\bibnamefont {Slot}}, \bibinfo {author} {\bibfnamefont {M.~A.}\ \bibnamefont {Holst}}, \bibinfo {author} {\bibfnamefont {H.~S.~J.}\ \bibnamefont {van~der Zant}},\ and\ \bibinfo {author} {\bibfnamefont {S.~V.}\ \bibnamefont {Zaitsev-Zotov}},\ }\bibfield  {title} {\bibinfo {title} {One-dimensional conduction in charge-density-wave nanowires},\ }\href {https://doi.org/10.1103/PhysRevLett.93.176602} {\bibfield  {journal} {\bibinfo  {journal} {Phys. Rev. Lett.}\ }\textbf {\bibinfo {volume} {93}},\ \bibinfo {pages} {176602} (\bibinfo {year} {2004})}\BibitemShut {NoStop}%
\bibitem [{\citenamefont {Heersche}\ \emph {et~al.}(2006)\citenamefont {Heersche}, \citenamefont {de~Groot}, \citenamefont {Folk}, \citenamefont {van~der Zant}, \citenamefont {Romeike}, \citenamefont {Wegewijs}, \citenamefont {Zobbi}, \citenamefont {Barreca}, \citenamefont {Tondello},\ and\ \citenamefont {Cornia}}]{Heersche2006}%
  \BibitemOpen
  \bibfield  {author} {\bibinfo {author} {\bibfnamefont {H.~B.}\ \bibnamefont {Heersche}}, \bibinfo {author} {\bibfnamefont {Z.}~\bibnamefont {de~Groot}}, \bibinfo {author} {\bibfnamefont {J.~A.}\ \bibnamefont {Folk}}, \bibinfo {author} {\bibfnamefont {H.~S.~J.}\ \bibnamefont {van~der Zant}}, \bibinfo {author} {\bibfnamefont {C.}~\bibnamefont {Romeike}}, \bibinfo {author} {\bibfnamefont {M.~R.}\ \bibnamefont {Wegewijs}}, \bibinfo {author} {\bibfnamefont {L.}~\bibnamefont {Zobbi}}, \bibinfo {author} {\bibfnamefont {D.}~\bibnamefont {Barreca}}, \bibinfo {author} {\bibfnamefont {E.}~\bibnamefont {Tondello}},\ and\ \bibinfo {author} {\bibfnamefont {A.}~\bibnamefont {Cornia}},\ }\bibfield  {title} {\bibinfo {title} {Electron transport through single {${\mathrm{Mn}}_{12}$} molecular magnets},\ }\href {https://doi.org/10.1103/PhysRevLett.96.206801} {\bibfield  {journal} {\bibinfo  {journal} {Phys. Rev. Lett.}\ }\textbf {\bibinfo {volume} {96}},\ \bibinfo {pages} {206801} (\bibinfo {year} {2006})}\BibitemShut
  {NoStop}%
\bibitem [{\citenamefont {Huard}\ \emph {et~al.}(2007)\citenamefont {Huard}, \citenamefont {Sulpizio}, \citenamefont {Stander}, \citenamefont {Todd}, \citenamefont {Yang},\ and\ \citenamefont {Goldhaber-Gordon}}]{Huard2007}%
  \BibitemOpen
  \bibfield  {author} {\bibinfo {author} {\bibfnamefont {B.}~\bibnamefont {Huard}}, \bibinfo {author} {\bibfnamefont {J.~A.}\ \bibnamefont {Sulpizio}}, \bibinfo {author} {\bibfnamefont {N.}~\bibnamefont {Stander}}, \bibinfo {author} {\bibfnamefont {K.}~\bibnamefont {Todd}}, \bibinfo {author} {\bibfnamefont {B.}~\bibnamefont {Yang}},\ and\ \bibinfo {author} {\bibfnamefont {D.}~\bibnamefont {Goldhaber-Gordon}},\ }\bibfield  {title} {\bibinfo {title} {Transport measurements across a tunable potential barrier in graphene},\ }\href {https://doi.org/10.1103/PhysRevLett.98.236803} {\bibfield  {journal} {\bibinfo  {journal} {Phys. Rev. Lett.}\ }\textbf {\bibinfo {volume} {98}},\ \bibinfo {pages} {236803} (\bibinfo {year} {2007})}\BibitemShut {NoStop}%
\bibitem [{\citenamefont {Osorio}\ \emph {et~al.}(2008)\citenamefont {Osorio}, \citenamefont {Bj{\o}rnholm}, \citenamefont {Lehn}, \citenamefont {Ruben},\ and\ \citenamefont {van~der Zant}}]{Osorio2008}%
  \BibitemOpen
  \bibfield  {author} {\bibinfo {author} {\bibfnamefont {E.~A.}\ \bibnamefont {Osorio}}, \bibinfo {author} {\bibfnamefont {T.}~\bibnamefont {Bj{\o}rnholm}}, \bibinfo {author} {\bibfnamefont {J.-M.}\ \bibnamefont {Lehn}}, \bibinfo {author} {\bibfnamefont {M.}~\bibnamefont {Ruben}},\ and\ \bibinfo {author} {\bibfnamefont {H.~S.~J.}\ \bibnamefont {van~der Zant}},\ }\bibfield  {title} {\bibinfo {title} {Single-molecule transport in three-terminal devices},\ }\href {https://doi.org/10.1088/0953-8984/20/37/374121} {\bibfield  {journal} {\bibinfo  {journal} {J. Phys.: Condens. Matter}\ }\textbf {\bibinfo {volume} {20}},\ \bibinfo {pages} {374121} (\bibinfo {year} {2008})}\BibitemShut {NoStop}%
\bibitem [{\citenamefont {Pr\"{u}ser}\ \emph {et~al.}(2011)\citenamefont {Pr\"{u}ser}, \citenamefont {Wenderoth}, \citenamefont {Dargel}, \citenamefont {Weismann}, \citenamefont {Peters}, \citenamefont {Pruschke},\ and\ \citenamefont {Ulbrich}}]{Pruser2011}%
  \BibitemOpen
  \bibfield  {author} {\bibinfo {author} {\bibfnamefont {H.}~\bibnamefont {Pr\"{u}ser}}, \bibinfo {author} {\bibfnamefont {M.}~\bibnamefont {Wenderoth}}, \bibinfo {author} {\bibfnamefont {P.~E.}\ \bibnamefont {Dargel}}, \bibinfo {author} {\bibfnamefont {A.}~\bibnamefont {Weismann}}, \bibinfo {author} {\bibfnamefont {R.}~\bibnamefont {Peters}}, \bibinfo {author} {\bibfnamefont {T.}~\bibnamefont {Pruschke}},\ and\ \bibinfo {author} {\bibfnamefont {R.~G.}\ \bibnamefont {Ulbrich}},\ }\bibfield  {title} {\bibinfo {title} {Long-range {K}ondo signature of a single magnetic impurity},\ }\href {https://doi.org/10.1038/nphys1876} {\bibfield  {journal} {\bibinfo  {journal} {Nature Phys.}\ }\textbf {\bibinfo {volume} {7}},\ \bibinfo {pages} {203} (\bibinfo {year} {2011})}\BibitemShut {NoStop}%
\bibitem [{\citenamefont {Perrin}\ \emph {et~al.}(2013)\citenamefont {Perrin}, \citenamefont {Verzijl}, \citenamefont {Martin}, \citenamefont {Shaikh}, \citenamefont {Eelkema}, \citenamefont {van Esch}, \citenamefont {van Ruitenbeek}, \citenamefont {Thijssen}, \citenamefont {van~der Zant},\ and\ \citenamefont {Duli\'c}}]{Perrin2013}%
  \BibitemOpen
  \bibfield  {author} {\bibinfo {author} {\bibfnamefont {M.~L.}\ \bibnamefont {Perrin}}, \bibinfo {author} {\bibfnamefont {C.~J.~O.}\ \bibnamefont {Verzijl}}, \bibinfo {author} {\bibfnamefont {C.~A.}\ \bibnamefont {Martin}}, \bibinfo {author} {\bibfnamefont {A.~J.}\ \bibnamefont {Shaikh}}, \bibinfo {author} {\bibfnamefont {R.}~\bibnamefont {Eelkema}}, \bibinfo {author} {\bibfnamefont {J.~H.}\ \bibnamefont {van Esch}}, \bibinfo {author} {\bibfnamefont {J.~M.}\ \bibnamefont {van Ruitenbeek}}, \bibinfo {author} {\bibfnamefont {J.~M.}\ \bibnamefont {Thijssen}}, \bibinfo {author} {\bibfnamefont {H.~S.~J.}\ \bibnamefont {van~der Zant}},\ and\ \bibinfo {author} {\bibfnamefont {D.}~\bibnamefont {Duli\'c}},\ }\bibfield  {title} {\bibinfo {title} {Large tunable image-charge effects in single-molecule junctions},\ }\href {https://doi.org/10.1038/nnano.2013.26} {\bibfield  {journal} {\bibinfo  {journal} {Nature Nanotech.}\ }\textbf {\bibinfo {volume} {8}},\ \bibinfo {pages} {282} (\bibinfo {year} {2013})}\BibitemShut
  {NoStop}%
\bibitem [{\citenamefont {Burzur\'i}\ \emph {et~al.}(2014)\citenamefont {Burzur\'i}, \citenamefont {Yamamoto}, \citenamefont {Warnock}, \citenamefont {Zhong}, \citenamefont {Park}, \citenamefont {Cornia},\ and\ \citenamefont {van~der Zant}}]{Burzuri2014}%
  \BibitemOpen
  \bibfield  {author} {\bibinfo {author} {\bibfnamefont {E.}~\bibnamefont {Burzur\'i}}, \bibinfo {author} {\bibfnamefont {Y.}~\bibnamefont {Yamamoto}}, \bibinfo {author} {\bibfnamefont {M.}~\bibnamefont {Warnock}}, \bibinfo {author} {\bibfnamefont {X.}~\bibnamefont {Zhong}}, \bibinfo {author} {\bibfnamefont {K.}~\bibnamefont {Park}}, \bibinfo {author} {\bibfnamefont {A.}~\bibnamefont {Cornia}},\ and\ \bibinfo {author} {\bibfnamefont {H.~S.~J.}\ \bibnamefont {van~der Zant}},\ }\bibfield  {title} {\bibinfo {title} {Franck-{C}ondon blockade in a single-molecule transistor},\ }\href {https://doi.org/10.1021/nl500524w} {\bibfield  {journal} {\bibinfo  {journal} {Nano Lett.}\ }\textbf {\bibinfo {volume} {14}},\ \bibinfo {pages} {3191} (\bibinfo {year} {2014})}\BibitemShut {NoStop}%
\bibitem [{\citenamefont {Iftikhar}\ \emph {et~al.}(2015)\citenamefont {Iftikhar}, \citenamefont {Jezouin}, \citenamefont {Anthore}, \citenamefont {Gennser}, \citenamefont {Parmentier}, \citenamefont {Cavanna},\ and\ \citenamefont {Pierre}}]{Iftikhar2015}%
  \BibitemOpen
  \bibfield  {author} {\bibinfo {author} {\bibfnamefont {Z.}~\bibnamefont {Iftikhar}}, \bibinfo {author} {\bibfnamefont {S.}~\bibnamefont {Jezouin}}, \bibinfo {author} {\bibfnamefont {A.}~\bibnamefont {Anthore}}, \bibinfo {author} {\bibfnamefont {U.}~\bibnamefont {Gennser}}, \bibinfo {author} {\bibfnamefont {F.~D.}\ \bibnamefont {Parmentier}}, \bibinfo {author} {\bibfnamefont {A.}~\bibnamefont {Cavanna}},\ and\ \bibinfo {author} {\bibfnamefont {F.}~\bibnamefont {Pierre}},\ }\bibfield  {title} {\bibinfo {title} {Two-channel {K}ondo effect and renormalization flow with macroscopic quantum charge states},\ }\href {https://doi.org/10.1038/nature15384} {\bibfield  {journal} {\bibinfo  {journal} {Nature}\ }\textbf {\bibinfo {volume} {526}},\ \bibinfo {pages} {233} (\bibinfo {year} {2015})}\BibitemShut {NoStop}%
\bibitem [{\citenamefont {Iftikhar}\ \emph {et~al.}(2018)\citenamefont {Iftikhar}, \citenamefont {Anthore}, \citenamefont {Mitchell}, \citenamefont {Parmentier}, \citenamefont {Gennser}, \citenamefont {Ouerghi}, \citenamefont {Cavanna}, \citenamefont {Mora}, \citenamefont {Simon},\ and\ \citenamefont {Pierre}}]{Iftikhar2018}%
  \BibitemOpen
  \bibfield  {author} {\bibinfo {author} {\bibfnamefont {Z.}~\bibnamefont {Iftikhar}}, \bibinfo {author} {\bibfnamefont {A.}~\bibnamefont {Anthore}}, \bibinfo {author} {\bibfnamefont {A.~K.}\ \bibnamefont {Mitchell}}, \bibinfo {author} {\bibfnamefont {F.~D.}\ \bibnamefont {Parmentier}}, \bibinfo {author} {\bibfnamefont {U.}~\bibnamefont {Gennser}}, \bibinfo {author} {\bibfnamefont {A.}~\bibnamefont {Ouerghi}}, \bibinfo {author} {\bibfnamefont {A.}~\bibnamefont {Cavanna}}, \bibinfo {author} {\bibfnamefont {C.}~\bibnamefont {Mora}}, \bibinfo {author} {\bibfnamefont {P.}~\bibnamefont {Simon}},\ and\ \bibinfo {author} {\bibfnamefont {F.}~\bibnamefont {Pierre}},\ }\bibfield  {title} {\bibinfo {title} {Tunable quantum criticality and super-ballistic transport in a ``charge'' {K}ondo circuit},\ }\href {https://doi.org/10.1126/science.aan5592} {\bibfield  {journal} {\bibinfo  {journal} {Science}\ }\textbf {\bibinfo {volume} {360}},\ \bibinfo {pages} {1315} (\bibinfo {year} {2018})}\BibitemShut {NoStop}%
\bibitem [{\citenamefont {Gehring}\ \emph {et~al.}(2019)\citenamefont {Gehring}, \citenamefont {Thijssen},\ and\ \citenamefont {van~der Zant}}]{Gehring2019}%
  \BibitemOpen
  \bibfield  {author} {\bibinfo {author} {\bibfnamefont {P.}~\bibnamefont {Gehring}}, \bibinfo {author} {\bibfnamefont {J.~M.}\ \bibnamefont {Thijssen}},\ and\ \bibinfo {author} {\bibfnamefont {H.~S.~J.}\ \bibnamefont {van~der Zant}},\ }\bibfield  {title} {\bibinfo {title} {Single-molecule quantum-transport phenomena in break junctions},\ }\href {https://doi.org/10.1038/s42254-019-0055-1} {\bibfield  {journal} {\bibinfo  {journal} {Nat. Rev. Phys.}\ }\textbf {\bibinfo {volume} {1}},\ \bibinfo {pages} {381} (\bibinfo {year} {2019})}\BibitemShut {NoStop}%
\bibitem [{\citenamefont {Guo}\ \emph {et~al.}(2021)\citenamefont {Guo}, \citenamefont {Zhu}, \citenamefont {Zhou}, \citenamefont {Yu}, \citenamefont {Lu},\ and\ \citenamefont {Liang}}]{Guo2021}%
  \BibitemOpen
  \bibfield  {author} {\bibinfo {author} {\bibfnamefont {X.}~\bibnamefont {Guo}}, \bibinfo {author} {\bibfnamefont {Q.}~\bibnamefont {Zhu}}, \bibinfo {author} {\bibfnamefont {L.}~\bibnamefont {Zhou}}, \bibinfo {author} {\bibfnamefont {W.}~\bibnamefont {Yu}}, \bibinfo {author} {\bibfnamefont {W.}~\bibnamefont {Lu}},\ and\ \bibinfo {author} {\bibfnamefont {W.}~\bibnamefont {Liang}},\ }\bibfield  {title} {\bibinfo {title} {Evolution and universality of two-stage {K}ondo effect in single manganese phthalocyanine molecule transistors},\ }\href {https://doi.org/10.1038/s41467-021-21492-x} {\bibfield  {journal} {\bibinfo  {journal} {Nat. Commun.}\ }\textbf {\bibinfo {volume} {12}},\ \bibinfo {pages} {1566} (\bibinfo {year} {2021})}\BibitemShut {NoStop}%
\bibitem [{\citenamefont {Dani}\ \emph {et~al.}(2022)\citenamefont {Dani}, \citenamefont {Hussein}, \citenamefont {Bayer}, \citenamefont {Kohler},\ and\ \citenamefont {Haug}}]{Dani2022}%
  \BibitemOpen
  \bibfield  {author} {\bibinfo {author} {\bibfnamefont {O.}~\bibnamefont {Dani}}, \bibinfo {author} {\bibfnamefont {R.}~\bibnamefont {Hussein}}, \bibinfo {author} {\bibfnamefont {J.~C.}\ \bibnamefont {Bayer}}, \bibinfo {author} {\bibfnamefont {S.}~\bibnamefont {Kohler}},\ and\ \bibinfo {author} {\bibfnamefont {R.~J.}\ \bibnamefont {Haug}},\ }\bibfield  {title} {\bibinfo {title} {Temperature-dependent broadening of coherent current peaks in {InAs} double quantum dots},\ }\href {https://doi.org/10.1038/s42005-022-01074-z} {\bibfield  {journal} {\bibinfo  {journal} {Commun. Phys.}\ }\textbf {\bibinfo {volume} {5}},\ \bibinfo {pages} {5638} (\bibinfo {year} {2022})}\BibitemShut {NoStop}%
\bibitem [{\citenamefont {Pouse}\ \emph {et~al.}(2023)\citenamefont {Pouse}, \citenamefont {Peeters}, \citenamefont {Hsueh}, \citenamefont {Gennser}, \citenamefont {Cavanna}, \citenamefont {Kastner}, \citenamefont {Mitchell},\ and\ \citenamefont {Goldhaber-Gordon}}]{Pouse2023}%
  \BibitemOpen
  \bibfield  {author} {\bibinfo {author} {\bibfnamefont {W.}~\bibnamefont {Pouse}}, \bibinfo {author} {\bibfnamefont {L.}~\bibnamefont {Peeters}}, \bibinfo {author} {\bibfnamefont {C.~L.}\ \bibnamefont {Hsueh}}, \bibinfo {author} {\bibfnamefont {U.}~\bibnamefont {Gennser}}, \bibinfo {author} {\bibfnamefont {A.}~\bibnamefont {Cavanna}}, \bibinfo {author} {\bibfnamefont {M.~A.}\ \bibnamefont {Kastner}}, \bibinfo {author} {\bibfnamefont {A.~K.}\ \bibnamefont {Mitchell}},\ and\ \bibinfo {author} {\bibfnamefont {D.}~\bibnamefont {Goldhaber-Gordon}},\ }\bibfield  {title} {\bibinfo {title} {Quantum simulation of an exotic quantum critical point in a two-site charge {K}ondo circuit},\ }\href {https://doi.org/10.1038/s41567-022-01905-4} {\bibfield  {journal} {\bibinfo  {journal} {Nature Phys.}\ }\textbf {\bibinfo {volume} {19}},\ \bibinfo {pages} {492} (\bibinfo {year} {2023})}\BibitemShut {NoStop}%
\bibitem [{\citenamefont {Piquard}\ \emph {et~al.}(2023)\citenamefont {Piquard}, \citenamefont {Glidic}, \citenamefont {Han}, \citenamefont {Aassime}, \citenamefont {Cavanna}, \citenamefont {Gennser}, \citenamefont {Meir}, \citenamefont {Sela}, \citenamefont {Anthore},\ and\ \citenamefont {Pierre}}]{Piquard2023}%
  \BibitemOpen
  \bibfield  {author} {\bibinfo {author} {\bibfnamefont {C.}~\bibnamefont {Piquard}}, \bibinfo {author} {\bibfnamefont {P.}~\bibnamefont {Glidic}}, \bibinfo {author} {\bibfnamefont {C.}~\bibnamefont {Han}}, \bibinfo {author} {\bibfnamefont {A.}~\bibnamefont {Aassime}}, \bibinfo {author} {\bibfnamefont {A.}~\bibnamefont {Cavanna}}, \bibinfo {author} {\bibfnamefont {U.}~\bibnamefont {Gennser}}, \bibinfo {author} {\bibfnamefont {Y.}~\bibnamefont {Meir}}, \bibinfo {author} {\bibfnamefont {E.}~\bibnamefont {Sela}}, \bibinfo {author} {\bibfnamefont {A.}~\bibnamefont {Anthore}},\ and\ \bibinfo {author} {\bibfnamefont {F.}~\bibnamefont {Pierre}},\ }\bibfield  {title} {\bibinfo {title} {Observing the universal screening of a {K}ondo impurity},\ }\href {https://doi.org/10.1038/s41467-023-42857-4} {\bibfield  {journal} {\bibinfo  {journal} {Nat. Commun.}\ }\textbf {\bibinfo {volume} {14}},\ \bibinfo {pages} {7263} (\bibinfo {year} {2023})}\BibitemShut {NoStop}%
\bibitem [{\citenamefont {Anderson}\ \emph {et~al.}(2024)\citenamefont {Anderson}, \citenamefont {Laitinen}, \citenamefont {Zimmerman}, \citenamefont {Werkmeister}, \citenamefont {Shackleton}, \citenamefont {Kruchkov}, \citenamefont {Taniguchi}, \citenamefont {Watanabe}, \citenamefont {Sachdev},\ and\ \citenamefont {Kim}}]{Anderson2024}%
  \BibitemOpen
  \bibfield  {author} {\bibinfo {author} {\bibfnamefont {L.~E.}\ \bibnamefont {Anderson}}, \bibinfo {author} {\bibfnamefont {A.}~\bibnamefont {Laitinen}}, \bibinfo {author} {\bibfnamefont {A.}~\bibnamefont {Zimmerman}}, \bibinfo {author} {\bibfnamefont {T.}~\bibnamefont {Werkmeister}}, \bibinfo {author} {\bibfnamefont {L.}~\bibnamefont {Shackleton}}, \bibinfo {author} {\bibfnamefont {A.}~\bibnamefont {Kruchkov}}, \bibinfo {author} {\bibfnamefont {T.}~\bibnamefont {Taniguchi}}, \bibinfo {author} {\bibfnamefont {K.}~\bibnamefont {Watanabe}}, \bibinfo {author} {\bibfnamefont {S.}~\bibnamefont {Sachdev}},\ and\ \bibinfo {author} {\bibfnamefont {P.}~\bibnamefont {Kim}},\ }\bibfield  {title} {\bibinfo {title} {Magneto-thermoelectric transport in graphene quantum dot with strong correlations},\ }\href {https://doi.org/10.1103/PhysRevLett.132.246502} {\bibfield  {journal} {\bibinfo  {journal} {Phys. Rev. Lett.}\ }\textbf {\bibinfo {volume} {132}},\ \bibinfo {pages} {246502} (\bibinfo {year} {2024})}\BibitemShut
  {NoStop}%
\bibitem [{Sup()}]{Suppl}%
  \BibitemOpen
  \href@noop {} {}\bibinfo {note} {See Supplemental Material for discussion on connections to experimental setups, applications of the developed theory to various models, and generalization beyond the linear response regime, which includes Refs. \cite{Spietz2003,Fevrier2018, Tamir2022, Flensberg1993, Matveev1995, Farahi2023, Hu2025, Gnezdilov2018, Manaparambil2025, Aristov2007, Nguyen2010, Pustilnik2001, Wojcik2016, Brown2009, Karki2020, Oguri2018}}\BibitemShut {NoStop}%
\bibitem [{\citenamefont {S\'anchez}\ \emph {et~al.}(2013)\citenamefont {S\'anchez}, \citenamefont {Sothmann}, \citenamefont {Jordan},\ and\ \citenamefont {B\"uttiker}}]{Sanchez2013}%
  \BibitemOpen
  \bibfield  {author} {\bibinfo {author} {\bibfnamefont {R.}~\bibnamefont {S\'anchez}}, \bibinfo {author} {\bibfnamefont {B.}~\bibnamefont {Sothmann}}, \bibinfo {author} {\bibfnamefont {A.~N.}\ \bibnamefont {Jordan}},\ and\ \bibinfo {author} {\bibfnamefont {M.}~\bibnamefont {B\"uttiker}},\ }\bibfield  {title} {\bibinfo {title} {Correlations of heat and charge currents in quantum-dot thermoelectric engines},\ }\href {https://doi.org/10.1088/1367-2630/15/12/125001} {\bibfield  {journal} {\bibinfo  {journal} {New J. Phys.}\ }\textbf {\bibinfo {volume} {15}},\ \bibinfo {pages} {125001} (\bibinfo {year} {2013})}\BibitemShut {NoStop}%
\bibitem [{\citenamefont {Battista}\ \emph {et~al.}(2014)\citenamefont {Battista}, \citenamefont {Haupt},\ and\ \citenamefont {Splettstoesser}}]{Battista2014}%
  \BibitemOpen
  \bibfield  {author} {\bibinfo {author} {\bibfnamefont {F.}~\bibnamefont {Battista}}, \bibinfo {author} {\bibfnamefont {F.}~\bibnamefont {Haupt}},\ and\ \bibinfo {author} {\bibfnamefont {J.}~\bibnamefont {Splettstoesser}},\ }\bibfield  {title} {\bibinfo {title} {Correlations between charge and energy current in ac-driven coherent conductors},\ }\href {https://doi.org/10.1088/1742-6596/568/5/052008} {\bibfield  {journal} {\bibinfo  {journal} {J. Phys.: Conf. Ser.}\ }\textbf {\bibinfo {volume} {568}},\ \bibinfo {pages} {052008} (\bibinfo {year} {2014})}\BibitemShut {NoStop}%
\bibitem [{\citenamefont {Cr\'epieux}\ and\ \citenamefont {Michelini}(2014)}]{Crepieux2014}%
  \BibitemOpen
  \bibfield  {author} {\bibinfo {author} {\bibfnamefont {A.}~\bibnamefont {Cr\'epieux}}\ and\ \bibinfo {author} {\bibfnamefont {F.}~\bibnamefont {Michelini}},\ }\bibfield  {title} {\bibinfo {title} {Mixed, charge and heat noises in thermoelectric nanosystems},\ }\href {https://doi.org/10.1088/0953-8984/27/1/015302} {\bibfield  {journal} {\bibinfo  {journal} {J. Phys.: Condens. Matter}\ }\textbf {\bibinfo {volume} {27}},\ \bibinfo {pages} {015302} (\bibinfo {year} {2014})}\BibitemShut {NoStop}%
\bibitem [{\citenamefont {Jacquod}\ \emph {et~al.}(2012)\citenamefont {Jacquod}, \citenamefont {Whitney}, \citenamefont {Meair},\ and\ \citenamefont {B\"uttiker}}]{Jacquod2012}%
  \BibitemOpen
  \bibfield  {author} {\bibinfo {author} {\bibfnamefont {P.}~\bibnamefont {Jacquod}}, \bibinfo {author} {\bibfnamefont {R.~S.}\ \bibnamefont {Whitney}}, \bibinfo {author} {\bibfnamefont {J.}~\bibnamefont {Meair}},\ and\ \bibinfo {author} {\bibfnamefont {M.}~\bibnamefont {B\"uttiker}},\ }\bibfield  {title} {\bibinfo {title} {Onsager relations in coupled electric, thermoelectric, and spin transport: The tenfold way},\ }\href {https://doi.org/10.1103/PhysRevB.86.155118} {\bibfield  {journal} {\bibinfo  {journal} {Phys. Rev. B}\ }\textbf {\bibinfo {volume} {86}},\ \bibinfo {pages} {155118} (\bibinfo {year} {2012})}\BibitemShut {NoStop}%
\bibitem [{\citenamefont {Mazza}\ \emph {et~al.}(2014)\citenamefont {Mazza}, \citenamefont {Bosisio}, \citenamefont {Benenti}, \citenamefont {Giovannetti}, \citenamefont {Fazio},\ and\ \citenamefont {Taddei}}]{Mazza2014}%
  \BibitemOpen
  \bibfield  {author} {\bibinfo {author} {\bibfnamefont {F.}~\bibnamefont {Mazza}}, \bibinfo {author} {\bibfnamefont {R.}~\bibnamefont {Bosisio}}, \bibinfo {author} {\bibfnamefont {G.}~\bibnamefont {Benenti}}, \bibinfo {author} {\bibfnamefont {V.}~\bibnamefont {Giovannetti}}, \bibinfo {author} {\bibfnamefont {R.}~\bibnamefont {Fazio}},\ and\ \bibinfo {author} {\bibfnamefont {F.}~\bibnamefont {Taddei}},\ }\bibfield  {title} {\bibinfo {title} {Thermoelectric efficiency of three-terminal quantum thermal machines},\ }\href {https://doi.org/10.1088/1367-2630/16/8/085001} {\bibfield  {journal} {\bibinfo  {journal} {New J. Phys.}\ }\textbf {\bibinfo {volume} {16}},\ \bibinfo {pages} {085001} (\bibinfo {year} {2014})}\BibitemShut {NoStop}%
\bibitem [{\citenamefont {Benenti}\ \emph {et~al.}(2017)\citenamefont {Benenti}, \citenamefont {Casati}, \citenamefont {Saito},\ and\ \citenamefont {Whitney}}]{Benenti2017}%
  \BibitemOpen
  \bibfield  {author} {\bibinfo {author} {\bibfnamefont {G.}~\bibnamefont {Benenti}}, \bibinfo {author} {\bibfnamefont {G.}~\bibnamefont {Casati}}, \bibinfo {author} {\bibfnamefont {K.}~\bibnamefont {Saito}},\ and\ \bibinfo {author} {\bibfnamefont {R.~S.}\ \bibnamefont {Whitney}},\ }\bibfield  {title} {\bibinfo {title} {Fundamental aspects of steady-state conversion of heat to work at the nanoscale},\ }\href {https://doi.org/10.1016/j.physrep.2017.05.008} {\bibfield  {journal} {\bibinfo  {journal} {Phys. Rep.}\ }\textbf {\bibinfo {volume} {694}},\ \bibinfo {pages} {1} (\bibinfo {year} {2017})}\BibitemShut {NoStop}%
\bibitem [{\citenamefont {Andreev}\ and\ \citenamefont {Matveev}(2001)}]{Andreev2002}%
  \BibitemOpen
  \bibfield  {author} {\bibinfo {author} {\bibfnamefont {A.~V.}\ \bibnamefont {Andreev}}\ and\ \bibinfo {author} {\bibfnamefont {K.~A.}\ \bibnamefont {Matveev}},\ }\bibfield  {title} {\bibinfo {title} {Coulomb blockade oscillations in the thermopower of open quantum dots},\ }\href {https://doi.org/10.1103/PhysRevLett.86.280} {\bibfield  {journal} {\bibinfo  {journal} {Phys. Rev. Lett.}\ }\textbf {\bibinfo {volume} {86}},\ \bibinfo {pages} {280} (\bibinfo {year} {2001})}\BibitemShut {NoStop}%
\bibitem [{\citenamefont {Nguyen}\ and\ \citenamefont {Kiselev}(2020)}]{Nguyen2020}%
  \BibitemOpen
  \bibfield  {author} {\bibinfo {author} {\bibfnamefont {T.~K.~T.}\ \bibnamefont {Nguyen}}\ and\ \bibinfo {author} {\bibfnamefont {M.~N.}\ \bibnamefont {Kiselev}},\ }\bibfield  {title} {\bibinfo {title} {Thermoelectric transport in a three-channel charge {K}ondo circuit},\ }\href {https://doi.org/10.1103/PhysRevLett.125.026801} {\bibfield  {journal} {\bibinfo  {journal} {Phys. Rev. Lett.}\ }\textbf {\bibinfo {volume} {125}},\ \bibinfo {pages} {026801} (\bibinfo {year} {2020})}\BibitemShut {NoStop}%
\bibitem [{\citenamefont {Nguyen}\ \emph {et~al.}(2024)\citenamefont {Nguyen}, \citenamefont {Nguyen},\ and\ \citenamefont {Kiselev}}]{Nguyen2024}%
  \BibitemOpen
  \bibfield  {author} {\bibinfo {author} {\bibfnamefont {T.~K.~T.}\ \bibnamefont {Nguyen}}, \bibinfo {author} {\bibfnamefont {H.~Q.}\ \bibnamefont {Nguyen}},\ and\ \bibinfo {author} {\bibfnamefont {M.~N.}\ \bibnamefont {Kiselev}},\ }\bibfield  {title} {\bibinfo {title} {Thermoelectric transport across a tunnel contact between two charge {K}ondo circuits: Beyond perturbation theory},\ }\href {https://doi.org/10.1103/PhysRevB.109.115139} {\bibfield  {journal} {\bibinfo  {journal} {Phys. Rev. B}\ }\textbf {\bibinfo {volume} {109}},\ \bibinfo {pages} {115139} (\bibinfo {year} {2024})}\BibitemShut {NoStop}%
\bibitem [{\citenamefont {Kamenev}(2011)}]{Kamenev2011}%
  \BibitemOpen
  \bibfield  {author} {\bibinfo {author} {\bibfnamefont {A.}~\bibnamefont {Kamenev}},\ }\href {https://doi.org/10.1017/cbo9781139003667} {\emph {\bibinfo {title} {Field Theory of Non-Equilibrium Systems}}}\ (\bibinfo  {publisher} {Cambridge University Press, New York},\ \bibinfo {year} {2011})\BibitemShut {NoStop}%
\bibitem [{\citenamefont {Popoff}\ \emph {et~al.}(2022)\citenamefont {Popoff}, \citenamefont {Rech}, \citenamefont {Jonckheere}, \citenamefont {Raymond}, \citenamefont {Gr\'emaud}, \citenamefont {Malherbe},\ and\ \citenamefont {Martin}}]{Popoff2022}%
  \BibitemOpen
  \bibfield  {author} {\bibinfo {author} {\bibfnamefont {A.}~\bibnamefont {Popoff}}, \bibinfo {author} {\bibfnamefont {J.}~\bibnamefont {Rech}}, \bibinfo {author} {\bibfnamefont {T.}~\bibnamefont {Jonckheere}}, \bibinfo {author} {\bibfnamefont {L.}~\bibnamefont {Raymond}}, \bibinfo {author} {\bibfnamefont {B.}~\bibnamefont {Gr\'emaud}}, \bibinfo {author} {\bibfnamefont {S.}~\bibnamefont {Malherbe}},\ and\ \bibinfo {author} {\bibfnamefont {T.}~\bibnamefont {Martin}},\ }\bibfield  {title} {\bibinfo {title} {Scattering theory of non-equilibrium noise and delta {T} current fluctuations through a quantum dot},\ }\href {https://doi.org/10.1088/1361-648x/ac5200} {\bibfield  {journal} {\bibinfo  {journal} {J. Phys.: Condens. Matter}\ }\textbf {\bibinfo {volume} {34}},\ \bibinfo {pages} {185301} (\bibinfo {year} {2022})}\BibitemShut {NoStop}%
\bibitem [{\citenamefont {Pekola}\ and\ \citenamefont {Karimi}(2021)}]{Pekola2021}%
  \BibitemOpen
  \bibfield  {author} {\bibinfo {author} {\bibfnamefont {J.~P.}\ \bibnamefont {Pekola}}\ and\ \bibinfo {author} {\bibfnamefont {B.}~\bibnamefont {Karimi}},\ }\bibfield  {title} {\bibinfo {title} {Colloquium: Quantum heat transport in condensed matter systems},\ }\href {https://doi.org/10.1103/RevModPhys.93.041001} {\bibfield  {journal} {\bibinfo  {journal} {Rev. Mod. Phys.}\ }\textbf {\bibinfo {volume} {93}},\ \bibinfo {pages} {041001} (\bibinfo {year} {2021})}\BibitemShut {NoStop}%
\bibitem [{\citenamefont {Onsager}(1931{\natexlab{a}})}]{Onsager1931}%
  \BibitemOpen
  \bibfield  {author} {\bibinfo {author} {\bibfnamefont {L.}~\bibnamefont {Onsager}},\ }\bibfield  {title} {\bibinfo {title} {Reciprocal relations in irreversible processes. {I.}},\ }\href {https://doi.org/10.1103/PhysRev.37.405} {\bibfield  {journal} {\bibinfo  {journal} {Phys. Rev.}\ }\textbf {\bibinfo {volume} {37}},\ \bibinfo {pages} {405} (\bibinfo {year} {1931}{\natexlab{a}})}\BibitemShut {NoStop}%
\bibitem [{Note1()}]{Note1}%
  \BibitemOpen
  \bibinfo {note} {We assume that the voltage bias is applied to the lead, while the system is grounded. We also consider the lead's chemical potential shift $\mu _L=\Delta V=|\Delta V|$ by a positive voltage bias. This choice is arbitrary, for a negative bias, one can replace $\partial (\Delta V)\rightarrow \partial (-\Delta V)$ in the corresponding derivatives}\BibitemShut {NoStop}%
\bibitem [{\citenamefont {Eymeoud}\ and\ \citenamefont {Crepieux}(2016)}]{Eymeoud2016}%
  \BibitemOpen
  \bibfield  {author} {\bibinfo {author} {\bibfnamefont {P.}~\bibnamefont {Eymeoud}}\ and\ \bibinfo {author} {\bibfnamefont {A.}~\bibnamefont {Crepieux}},\ }\bibfield  {title} {\bibinfo {title} {Mixed electrical-heat noise spectrum in a quantum dot},\ }\href {https://doi.org/10.1103/PhysRevB.94.205416} {\bibfield  {journal} {\bibinfo  {journal} {Phys. Rev. B}\ }\textbf {\bibinfo {volume} {94}},\ \bibinfo {pages} {205416} (\bibinfo {year} {2016})}\BibitemShut {NoStop}%
\bibitem [{\citenamefont {Pavlov}\ and\ \citenamefont {Kiselev}(2025)}]{Pavlov2025}%
  \BibitemOpen
  \bibfield  {author} {\bibinfo {author} {\bibfnamefont {A.~I.}\ \bibnamefont {Pavlov}}\ and\ \bibinfo {author} {\bibfnamefont {M.~N.}\ \bibnamefont {Kiselev}},\ }\href@noop {} {\bibinfo {title} {Noise signatures of a charged {S}achdev-{Y}e-{K}itaev dot in mesoscopic transport}} (\bibinfo {year} {2025}),\ \Eprint {https://arxiv.org/abs/2508.13098} {arXiv:2508.13098 [cond-mat.mes-hall]} \BibitemShut {NoStop}%
\bibitem [{\citenamefont {Onsager}(1931{\natexlab{b}})}]{Onsager1931b}%
  \BibitemOpen
  \bibfield  {author} {\bibinfo {author} {\bibfnamefont {L.}~\bibnamefont {Onsager}},\ }\bibfield  {title} {\bibinfo {title} {Reciprocal relations in irreversible processes. {II}.},\ }\href {https://doi.org/10.1103/PhysRev.38.2265} {\bibfield  {journal} {\bibinfo  {journal} {Phys. Rev.}\ }\textbf {\bibinfo {volume} {38}},\ \bibinfo {pages} {2265} (\bibinfo {year} {1931}{\natexlab{b}})}\BibitemShut {NoStop}%
\bibitem [{\citenamefont {Matveev}\ and\ \citenamefont {Andreev}(2002)}]{Matveev2002}%
  \BibitemOpen
  \bibfield  {author} {\bibinfo {author} {\bibfnamefont {K.~A.}\ \bibnamefont {Matveev}}\ and\ \bibinfo {author} {\bibfnamefont {A.~V.}\ \bibnamefont {Andreev}},\ }\bibfield  {title} {\bibinfo {title} {Thermopower of a single-electron transistor in the regime of strong inelastic cotunneling},\ }\href {https://doi.org/10.1103/PhysRevB.66.045301} {\bibfield  {journal} {\bibinfo  {journal} {Phys. Rev. B}\ }\textbf {\bibinfo {volume} {66}},\ \bibinfo {pages} {045301} (\bibinfo {year} {2002})}\BibitemShut {NoStop}%
\bibitem [{Note2()}]{Note2}%
  \BibitemOpen
  \bibinfo {note} {Because of nonlinear effects, the thermopower at the finite voltage and finite temperature bias is not necessarily zero at the particle-hole symmetric point \cite {Scheibner2005, Karki2017}, which is beyond the linear response theory.}\BibitemShut {Stop}%
\bibitem [{\citenamefont {van Dalum}\ \emph {et~al.}(2020)\citenamefont {van Dalum}, \citenamefont {Mitchell},\ and\ \citenamefont {Fritz}}]{vanDalum2020}%
  \BibitemOpen
  \bibfield  {author} {\bibinfo {author} {\bibfnamefont {G.~A.~R.}\ \bibnamefont {van Dalum}}, \bibinfo {author} {\bibfnamefont {A.~K.}\ \bibnamefont {Mitchell}},\ and\ \bibinfo {author} {\bibfnamefont {L.}~\bibnamefont {Fritz}},\ }\bibfield  {title} {\bibinfo {title} {Wiedemann-{F}ranz law in a non-{F}ermi liquid and {M}ajorana central charge: Thermoelectric transport in a two-channel {K}ondo system},\ }\href {https://doi.org/10.1103/PhysRevB.102.041111} {\bibfield  {journal} {\bibinfo  {journal} {Phys. Rev. B}\ }\textbf {\bibinfo {volume} {102}},\ \bibinfo {pages} {041111} (\bibinfo {year} {2020})}\BibitemShut {NoStop}%
\bibitem [{\citenamefont {Aleiner}\ and\ \citenamefont {Glazman}(1998)}]{Aleiner1998}%
  \BibitemOpen
  \bibfield  {author} {\bibinfo {author} {\bibfnamefont {I.~L.}\ \bibnamefont {Aleiner}}\ and\ \bibinfo {author} {\bibfnamefont {L.~I.}\ \bibnamefont {Glazman}},\ }\bibfield  {title} {\bibinfo {title} {Mesoscopic charge quantization},\ }\href {https://doi.org/10.1103/PhysRevB.57.9608} {\bibfield  {journal} {\bibinfo  {journal} {Phys. Rev. B}\ }\textbf {\bibinfo {volume} {57}},\ \bibinfo {pages} {9608} (\bibinfo {year} {1998})}\BibitemShut {NoStop}%
\bibitem [{\citenamefont {Kiselev}(2023)}]{Kiselev2023}%
  \BibitemOpen
  \bibfield  {author} {\bibinfo {author} {\bibfnamefont {M.~N.}\ \bibnamefont {Kiselev}},\ }\bibfield  {title} {\bibinfo {title} {Generalized {W}iedemann-{F}ranz law in a two-site charge {K}ondo circuit: {L}orenz ratio as a manifestation of the orthogonality catastrophe},\ }\href {https://doi.org/10.1103/PhysRevB.108.L081108} {\bibfield  {journal} {\bibinfo  {journal} {Phys. Rev. B}\ }\textbf {\bibinfo {volume} {108}},\ \bibinfo {pages} {L081108} (\bibinfo {year} {2023})}\BibitemShut {NoStop}%
\bibitem [{\citenamefont {St\"abler}\ and\ \citenamefont {Sukhorukov}(2023)}]{Stabler2023}%
  \BibitemOpen
  \bibfield  {author} {\bibinfo {author} {\bibfnamefont {F.}~\bibnamefont {St\"abler}}\ and\ \bibinfo {author} {\bibfnamefont {E.}~\bibnamefont {Sukhorukov}},\ }\bibfield  {title} {\bibinfo {title} {Mesoscopic heat multiplier and fractionalizer},\ }\href {https://doi.org/10.1103/PhysRevB.108.235405} {\bibfield  {journal} {\bibinfo  {journal} {Phys. Rev. B}\ }\textbf {\bibinfo {volume} {108}},\ \bibinfo {pages} {235405} (\bibinfo {year} {2023})}\BibitemShut {NoStop}%
\bibitem [{\citenamefont {Pavlov}\ and\ \citenamefont {Kiselev}(2021)}]{Pavlov2020}%
  \BibitemOpen
  \bibfield  {author} {\bibinfo {author} {\bibfnamefont {A.~I.}\ \bibnamefont {Pavlov}}\ and\ \bibinfo {author} {\bibfnamefont {M.~N.}\ \bibnamefont {Kiselev}},\ }\bibfield  {title} {\bibinfo {title} {Quantum thermal transport in the charged {S}achdev-{Y}e-{K}itaev model: Thermoelectric {C}oulomb blockade},\ }\href {https://doi.org/10.1103/PhysRevB.103.L201107} {\bibfield  {journal} {\bibinfo  {journal} {Phys. Rev. B}\ }\textbf {\bibinfo {volume} {103}},\ \bibinfo {pages} {L201107} (\bibinfo {year} {2021})}\BibitemShut {NoStop}%
\bibitem [{Note3()}]{Note3}%
  \BibitemOpen
  \bibinfo {note} {The Lorenz ratio gives direct access to the scaling dimension of the leading irrelevant operator through the parameter $\alpha $, a related property of the delta-T noise for quantum Hall devices is discussed in \cite {Ebisu2022, Schiller2022, Zhang2022, Acciai2025}}\BibitemShut {NoStop}%
\bibitem [{\citenamefont {Kiselev}()}]{KiselevUn2}%
  \BibitemOpen
  \bibfield  {author} {\bibinfo {author} {\bibfnamefont {M.~N.}\ \bibnamefont {Kiselev}},\ }\bibfield  {title} {\bibinfo {title} {Universal scaling functions for a quantum transport through single-site and double-site charge {K}ondo circuits, {L}ecture notes in {P}hysics},\ }\bibinfo {note} {(to be published)}\BibitemShut {NoStop}%
\bibitem [{\citenamefont {Anderson}(1967)}]{Anderson1967}%
  \BibitemOpen
  \bibfield  {author} {\bibinfo {author} {\bibfnamefont {P.~W.}\ \bibnamefont {Anderson}},\ }\bibfield  {title} {\bibinfo {title} {Infrared catastrophe in {F}ermi gases with local scattering potentials},\ }\href {https://doi.org/10.1103/PhysRevLett.18.1049} {\bibfield  {journal} {\bibinfo  {journal} {Phys. Rev. Lett.}\ }\textbf {\bibinfo {volume} {18}},\ \bibinfo {pages} {1049} (\bibinfo {year} {1967})}\BibitemShut {NoStop}%
\bibitem [{\citenamefont {Mahan}(1967)}]{Mahan1967}%
  \BibitemOpen
  \bibfield  {author} {\bibinfo {author} {\bibfnamefont {G.~D.}\ \bibnamefont {Mahan}},\ }\bibfield  {title} {\bibinfo {title} {Excitons in metals: Infinite hole mass},\ }\href {https://doi.org/10.1103/PhysRev.163.612} {\bibfield  {journal} {\bibinfo  {journal} {Phys. Rev.}\ }\textbf {\bibinfo {volume} {163}},\ \bibinfo {pages} {612} (\bibinfo {year} {1967})}\BibitemShut {NoStop}%
\bibitem [{\citenamefont {Furusaki}\ and\ \citenamefont {Matveev}(1995)}]{Furusaki1995}%
  \BibitemOpen
  \bibfield  {author} {\bibinfo {author} {\bibfnamefont {A.}~\bibnamefont {Furusaki}}\ and\ \bibinfo {author} {\bibfnamefont {K.~A.}\ \bibnamefont {Matveev}},\ }\bibfield  {title} {\bibinfo {title} {Theory of strong inelastic cotunneling},\ }\href {https://doi.org/10.1103/PhysRevB.52.16676} {\bibfield  {journal} {\bibinfo  {journal} {Phys. Rev. B}\ }\textbf {\bibinfo {volume} {52}},\ \bibinfo {pages} {16676} (\bibinfo {year} {1995})}\BibitemShut {NoStop}%
\bibitem [{\citenamefont {Kitaev}\ and\ \citenamefont {Suh}(2018)}]{Kitaev2018}%
  \BibitemOpen
  \bibfield  {author} {\bibinfo {author} {\bibfnamefont {A.}~\bibnamefont {Kitaev}}\ and\ \bibinfo {author} {\bibfnamefont {S.~J.}\ \bibnamefont {Suh}},\ }\bibfield  {title} {\bibinfo {title} {The soft mode in the {S}achdev-{Y}e-{K}itaev model and its gravity dual},\ }\href {https://doi.org/10.1007/jhep05(2018)183} {\bibfield  {journal} {\bibinfo  {journal} {J. High Energ. Phys.}\ }\textbf {\bibinfo {volume} {2018}}\bibinfo  {number} { (5)},\ \bibinfo {pages} {183}}\BibitemShut {NoStop}%
\bibitem [{\citenamefont {Berkooz}\ and\ \citenamefont {Mamroud}(2025)}]{Berkooz2025}%
  \BibitemOpen
\bibfield  {number} {  }\bibfield  {author} {\bibinfo {author} {\bibfnamefont {M.}~\bibnamefont {Berkooz}}\ and\ \bibinfo {author} {\bibfnamefont {O.}~\bibnamefont {Mamroud}},\ }\bibfield  {title} {\bibinfo {title} {A cordial introduction to double scaled {SYK}},\ }\href {https://doi.org/10.1088/1361-6633/ada889} {\bibfield  {journal} {\bibinfo  {journal} {Rep. Prog. Phys.}\ }\textbf {\bibinfo {volume} {88}},\ \bibinfo {pages} {036001} (\bibinfo {year} {2025})}\BibitemShut {NoStop}%
\bibitem [{\citenamefont {Altland}\ \emph {et~al.}(2024)\citenamefont {Altland}, \citenamefont {Kim}, \citenamefont {Micklitz}, \citenamefont {Rezaei}, \citenamefont {Sonner},\ and\ \citenamefont {Verbaarschot}}]{Altland2024}%
  \BibitemOpen
  \bibfield  {author} {\bibinfo {author} {\bibfnamefont {A.}~\bibnamefont {Altland}}, \bibinfo {author} {\bibfnamefont {K.~W.}\ \bibnamefont {Kim}}, \bibinfo {author} {\bibfnamefont {T.}~\bibnamefont {Micklitz}}, \bibinfo {author} {\bibfnamefont {M.}~\bibnamefont {Rezaei}}, \bibinfo {author} {\bibfnamefont {J.}~\bibnamefont {Sonner}},\ and\ \bibinfo {author} {\bibfnamefont {J.~J.~M.}\ \bibnamefont {Verbaarschot}},\ }\bibfield  {title} {\bibinfo {title} {Quantum chaos on edge},\ }\href {https://doi.org/10.1103/PhysRevResearch.6.033286} {\bibfield  {journal} {\bibinfo  {journal} {Phys. Rev. Res.}\ }\textbf {\bibinfo {volume} {6}},\ \bibinfo {pages} {033286} (\bibinfo {year} {2024})}\BibitemShut {NoStop}%
\bibitem [{Note4()}]{Note4}%
  \BibitemOpen
  \bibinfo {note} {The divergence of the DoS in the DSSYK model exists for the conformal saddle-point solution, and disappears in the infrared regularization of the saddle-point. The theory is well behaved for large $q$ and large $N$ as long as $q^2/N\ll 1$. The effective infrared theory in this regime coincides with the conventional SYK Schwarzian theory up to parametric prefactors and has the same energy scaling \cite {Berkooz2025}.}\BibitemShut {Stop}%
\bibitem [{\citenamefont {Gurarie}(2013)}]{Gurarie2013}%
  \BibitemOpen
  \bibfield  {author} {\bibinfo {author} {\bibfnamefont {V.}~\bibnamefont {Gurarie}},\ }\bibfield  {title} {\bibinfo {title} {Logarithmic operators and logarithmic conformal field theories},\ }\href {https://doi.org/10.1088/1751-8113/46/49/494003} {\bibfield  {journal} {\bibinfo  {journal} {J. Phys. A: Math. Theor.}\ }\textbf {\bibinfo {volume} {46}},\ \bibinfo {pages} {494003} (\bibinfo {year} {2013})}\BibitemShut {NoStop}%
\bibitem [{\citenamefont {Van~Hove}(1953)}]{vanHove1953}%
  \BibitemOpen
  \bibfield  {author} {\bibinfo {author} {\bibfnamefont {L.}~\bibnamefont {Van~Hove}},\ }\bibfield  {title} {\bibinfo {title} {The occurrence of singularities in the elastic frequency distribution of a crystal},\ }\href {https://doi.org/10.1103/PhysRev.89.1189} {\bibfield  {journal} {\bibinfo  {journal} {Phys. Rev.}\ }\textbf {\bibinfo {volume} {89}},\ \bibinfo {pages} {1189} (\bibinfo {year} {1953})}\BibitemShut {NoStop}%
\bibitem [{\citenamefont {Efremov}\ \emph {et~al.}(2019)\citenamefont {Efremov}, \citenamefont {Shtyk}, \citenamefont {Rost}, \citenamefont {Chamon}, \citenamefont {Mackenzie},\ and\ \citenamefont {Betouras}}]{Efremov2019}%
  \BibitemOpen
  \bibfield  {author} {\bibinfo {author} {\bibfnamefont {D.~V.}\ \bibnamefont {Efremov}}, \bibinfo {author} {\bibfnamefont {A.}~\bibnamefont {Shtyk}}, \bibinfo {author} {\bibfnamefont {A.~W.}\ \bibnamefont {Rost}}, \bibinfo {author} {\bibfnamefont {C.}~\bibnamefont {Chamon}}, \bibinfo {author} {\bibfnamefont {A.~P.}\ \bibnamefont {Mackenzie}},\ and\ \bibinfo {author} {\bibfnamefont {J.~J.}\ \bibnamefont {Betouras}},\ }\bibfield  {title} {\bibinfo {title} {Multicritical {F}ermi surface topological transitions},\ }\href {https://doi.org/10.1103/PhysRevLett.123.207202} {\bibfield  {journal} {\bibinfo  {journal} {Phys. Rev. Lett.}\ }\textbf {\bibinfo {volume} {123}},\ \bibinfo {pages} {207202} (\bibinfo {year} {2019})}\BibitemShut {NoStop}%
\bibitem [{\citenamefont {Chandrasekaran}\ \emph {et~al.}(2020)\citenamefont {Chandrasekaran}, \citenamefont {Shtyk}, \citenamefont {Betouras},\ and\ \citenamefont {Chamon}}]{Chandrasekaran2020}%
  \BibitemOpen
  \bibfield  {author} {\bibinfo {author} {\bibfnamefont {A.}~\bibnamefont {Chandrasekaran}}, \bibinfo {author} {\bibfnamefont {A.}~\bibnamefont {Shtyk}}, \bibinfo {author} {\bibfnamefont {J.~J.}\ \bibnamefont {Betouras}},\ and\ \bibinfo {author} {\bibfnamefont {C.}~\bibnamefont {Chamon}},\ }\bibfield  {title} {\bibinfo {title} {Catastrophe theory classification of {F}ermi surface topological transitions in two dimensions},\ }\href {https://doi.org/10.1103/PhysRevResearch.2.013355} {\bibfield  {journal} {\bibinfo  {journal} {Phys. Rev. Res.}\ }\textbf {\bibinfo {volume} {2}},\ \bibinfo {pages} {013355} (\bibinfo {year} {2020})}\BibitemShut {NoStop}%
\bibitem [{\citenamefont {Yuan}\ and\ \citenamefont {Fu}(2020)}]{Yuan2020}%
  \BibitemOpen
  \bibfield  {author} {\bibinfo {author} {\bibfnamefont {N.~F.~Q.}\ \bibnamefont {Yuan}}\ and\ \bibinfo {author} {\bibfnamefont {L.}~\bibnamefont {Fu}},\ }\bibfield  {title} {\bibinfo {title} {Classification of critical points in energy bands based on topology, scaling, and symmetry},\ }\href {https://doi.org/10.1103/PhysRevB.101.125120} {\bibfield  {journal} {\bibinfo  {journal} {Phys. Rev. B}\ }\textbf {\bibinfo {volume} {101}},\ \bibinfo {pages} {125120} (\bibinfo {year} {2020})}\BibitemShut {NoStop}%
\bibitem [{\citenamefont {Zervou}\ \emph {et~al.}(2023)\citenamefont {Zervou}, \citenamefont {Efremov},\ and\ \citenamefont {Betouras}}]{Zervou2023}%
  \BibitemOpen
  \bibfield  {author} {\bibinfo {author} {\bibfnamefont {A.}~\bibnamefont {Zervou}}, \bibinfo {author} {\bibfnamefont {D.~V.}\ \bibnamefont {Efremov}},\ and\ \bibinfo {author} {\bibfnamefont {J.~J.}\ \bibnamefont {Betouras}},\ }\bibfield  {title} {\bibinfo {title} {Fate of density waves in the presence of a higher-order van {H}ove singularity},\ }\href {https://doi.org/10.1103/PhysRevResearch.5.L042006} {\bibfield  {journal} {\bibinfo  {journal} {Phys. Rev. Res.}\ }\textbf {\bibinfo {volume} {5}},\ \bibinfo {pages} {L042006} (\bibinfo {year} {2023})}\BibitemShut {NoStop}%
\bibitem [{\citenamefont {Shtyk}\ \emph {et~al.}(2017)\citenamefont {Shtyk}, \citenamefont {Goldstein},\ and\ \citenamefont {Chamon}}]{Shtyk2017}%
  \BibitemOpen
  \bibfield  {author} {\bibinfo {author} {\bibfnamefont {A.}~\bibnamefont {Shtyk}}, \bibinfo {author} {\bibfnamefont {G.}~\bibnamefont {Goldstein}},\ and\ \bibinfo {author} {\bibfnamefont {C.}~\bibnamefont {Chamon}},\ }\bibfield  {title} {\bibinfo {title} {Electrons at the monkey saddle: A multicritical {L}ifshitz point},\ }\href {https://doi.org/10.1103/PhysRevB.95.035137} {\bibfield  {journal} {\bibinfo  {journal} {Phys. Rev. B}\ }\textbf {\bibinfo {volume} {95}},\ \bibinfo {pages} {035137} (\bibinfo {year} {2017})}\BibitemShut {NoStop}%
\bibitem [{\citenamefont {Piriou}\ \emph {et~al.}(2011)\citenamefont {Piriou}, \citenamefont {Jenkins}, \citenamefont {Berthod}, \citenamefont {Maggio-Aprile},\ and\ \citenamefont {Fischer}}]{Piriou2011}%
  \BibitemOpen
  \bibfield  {author} {\bibinfo {author} {\bibfnamefont {A.}~\bibnamefont {Piriou}}, \bibinfo {author} {\bibfnamefont {N.}~\bibnamefont {Jenkins}}, \bibinfo {author} {\bibfnamefont {C.}~\bibnamefont {Berthod}}, \bibinfo {author} {\bibfnamefont {I.}~\bibnamefont {Maggio-Aprile}},\ and\ \bibinfo {author} {\bibfnamefont {{\O}.}~\bibnamefont {Fischer}},\ }\bibfield  {title} {\bibinfo {title} {First direct observation of the {V}an {H}ove singularity in the tunnelling spectra of cuprates},\ }\href {https://doi.org/10.1038/ncomms1229} {\bibfield  {journal} {\bibinfo  {journal} {Nat. Commun.}\ }\textbf {\bibinfo {volume} {2}},\ \bibinfo {pages} {221} (\bibinfo {year} {2011})}\BibitemShut {NoStop}%
\bibitem [{\citenamefont {Chen}\ \emph {et~al.}(2011)\citenamefont {Chen}, \citenamefont {Pathak}, \citenamefont {Yang}, \citenamefont {Su}, \citenamefont {Galanakis}, \citenamefont {Mikelsons}, \citenamefont {Jarrell},\ and\ \citenamefont {Moreno}}]{Chen2011}%
  \BibitemOpen
  \bibfield  {author} {\bibinfo {author} {\bibfnamefont {K.-S.}\ \bibnamefont {Chen}}, \bibinfo {author} {\bibfnamefont {S.}~\bibnamefont {Pathak}}, \bibinfo {author} {\bibfnamefont {S.-X.}\ \bibnamefont {Yang}}, \bibinfo {author} {\bibfnamefont {S.-Q.}\ \bibnamefont {Su}}, \bibinfo {author} {\bibfnamefont {D.}~\bibnamefont {Galanakis}}, \bibinfo {author} {\bibfnamefont {K.}~\bibnamefont {Mikelsons}}, \bibinfo {author} {\bibfnamefont {M.}~\bibnamefont {Jarrell}},\ and\ \bibinfo {author} {\bibfnamefont {J.}~\bibnamefont {Moreno}},\ }\bibfield  {title} {\bibinfo {title} {Role of the van {H}ove singularity in the quantum criticality of the {H}ubbard model},\ }\href {https://doi.org/10.1103/PhysRevB.84.245107} {\bibfield  {journal} {\bibinfo  {journal} {Phys. Rev. B}\ }\textbf {\bibinfo {volume} {84}},\ \bibinfo {pages} {245107} (\bibinfo {year} {2011})}\BibitemShut {NoStop}%
\bibitem [{\citenamefont {Barber}\ \emph {et~al.}(2018)\citenamefont {Barber}, \citenamefont {Gibbs}, \citenamefont {Maeno}, \citenamefont {Mackenzie},\ and\ \citenamefont {Hicks}}]{Barber2018}%
  \BibitemOpen
  \bibfield  {author} {\bibinfo {author} {\bibfnamefont {M.~E.}\ \bibnamefont {Barber}}, \bibinfo {author} {\bibfnamefont {A.~S.}\ \bibnamefont {Gibbs}}, \bibinfo {author} {\bibfnamefont {Y.}~\bibnamefont {Maeno}}, \bibinfo {author} {\bibfnamefont {A.~P.}\ \bibnamefont {Mackenzie}},\ and\ \bibinfo {author} {\bibfnamefont {C.~W.}\ \bibnamefont {Hicks}},\ }\bibfield  {title} {\bibinfo {title} {Resistivity in the vicinity of a van {H}ove singularity: {${\mathrm{Sr}}_{2}{\mathrm{RuO}}_{4}$} under uniaxial pressure},\ }\href {https://doi.org/10.1103/PhysRevLett.120.076602} {\bibfield  {journal} {\bibinfo  {journal} {Phys. Rev. Lett.}\ }\textbf {\bibinfo {volume} {120}},\ \bibinfo {pages} {076602} (\bibinfo {year} {2018})}\BibitemShut {NoStop}%
\bibitem [{\citenamefont {Stangier}\ \emph {et~al.}(2022)\citenamefont {Stangier}, \citenamefont {Berg},\ and\ \citenamefont {Schmalian}}]{Stangier2022}%
  \BibitemOpen
  \bibfield  {author} {\bibinfo {author} {\bibfnamefont {V.~C.}\ \bibnamefont {Stangier}}, \bibinfo {author} {\bibfnamefont {E.}~\bibnamefont {Berg}},\ and\ \bibinfo {author} {\bibfnamefont {J.}~\bibnamefont {Schmalian}},\ }\bibfield  {title} {\bibinfo {title} {Breakdown of the {W}iedemann-{F}ranz law at the {L}ifshitz point of strained {${\mathrm{Sr}}_{2}{\mathrm{RuO}}_{4}$}},\ }\href {https://doi.org/10.1103/PhysRevB.105.115113} {\bibfield  {journal} {\bibinfo  {journal} {Phys. Rev. B}\ }\textbf {\bibinfo {volume} {105}},\ \bibinfo {pages} {115113} (\bibinfo {year} {2022})}\BibitemShut {NoStop}%
\bibitem [{\citenamefont {Nguyen}\ \emph {et~al.}(2025)\citenamefont {Nguyen}, \citenamefont {Rech}, \citenamefont {Martin},\ and\ \citenamefont {Kiselev}}]{Nguyen2025}%
  \BibitemOpen
  \bibfield  {author} {\bibinfo {author} {\bibfnamefont {T.~K.~T.}\ \bibnamefont {Nguyen}}, \bibinfo {author} {\bibfnamefont {J.}~\bibnamefont {Rech}}, \bibinfo {author} {\bibfnamefont {T.}~\bibnamefont {Martin}},\ and\ \bibinfo {author} {\bibfnamefont {M.~N.}\ \bibnamefont {Kiselev}},\ }\href@noop {} {\bibinfo {title} {Noises in a two-channel charge {K}ondo model}} (\bibinfo {year} {2025}),\ \Eprint {https://arxiv.org/abs/2511.02590} {arXiv:2511.02590 [cond-mat.mes-hall]} \BibitemShut {NoStop}%
\bibitem [{\citenamefont {Mora}\ \emph {et~al.}(2009)\citenamefont {Mora}, \citenamefont {Vitushinsky}, \citenamefont {Leyronas}, \citenamefont {Clerk},\ and\ \citenamefont {Le~Hur}}]{Mora2009}%
  \BibitemOpen
  \bibfield  {author} {\bibinfo {author} {\bibfnamefont {C.}~\bibnamefont {Mora}}, \bibinfo {author} {\bibfnamefont {P.}~\bibnamefont {Vitushinsky}}, \bibinfo {author} {\bibfnamefont {X.}~\bibnamefont {Leyronas}}, \bibinfo {author} {\bibfnamefont {A.~A.}\ \bibnamefont {Clerk}},\ and\ \bibinfo {author} {\bibfnamefont {K.}~\bibnamefont {Le~Hur}},\ }\bibfield  {title} {\bibinfo {title} {Theory of nonequilibrium transport in the $\text{SU}({N})$ {K}ondo regime},\ }\href {https://doi.org/10.1103/PhysRevB.80.155322} {\bibfield  {journal} {\bibinfo  {journal} {Phys. Rev. B}\ }\textbf {\bibinfo {volume} {80}},\ \bibinfo {pages} {155322} (\bibinfo {year} {2009})}\BibitemShut {NoStop}%
\bibitem [{\citenamefont {Mora}\ \emph {et~al.}(2015)\citenamefont {Mora}, \citenamefont {Moca}, \citenamefont {von Delft},\ and\ \citenamefont {Zar\'and}}]{Mora2015}%
  \BibitemOpen
  \bibfield  {author} {\bibinfo {author} {\bibfnamefont {C.}~\bibnamefont {Mora}}, \bibinfo {author} {\bibfnamefont {C.~P.}\ \bibnamefont {Moca}}, \bibinfo {author} {\bibfnamefont {J.}~\bibnamefont {von Delft}},\ and\ \bibinfo {author} {\bibfnamefont {G.}~\bibnamefont {Zar\'and}},\ }\bibfield  {title} {\bibinfo {title} {Fermi-liquid theory for the single-impurity {A}nderson model},\ }\href {https://doi.org/10.1103/PhysRevB.92.075120} {\bibfield  {journal} {\bibinfo  {journal} {Phys. Rev. B}\ }\textbf {\bibinfo {volume} {92}},\ \bibinfo {pages} {075120} (\bibinfo {year} {2015})}\BibitemShut {NoStop}%
\bibitem [{\citenamefont {Koski}\ \emph {et~al.}(2013)\citenamefont {Koski}, \citenamefont {Sagawa}, \citenamefont {Saira}, \citenamefont {Yoon}, \citenamefont {Kutvonen}, \citenamefont {Solinas}, \citenamefont {M\"{o}tt\"{o}nen}, \citenamefont {Ala-Nissila},\ and\ \citenamefont {Pekola}}]{Koski2013}%
  \BibitemOpen
  \bibfield  {author} {\bibinfo {author} {\bibfnamefont {J.~V.}\ \bibnamefont {Koski}}, \bibinfo {author} {\bibfnamefont {T.}~\bibnamefont {Sagawa}}, \bibinfo {author} {\bibfnamefont {O.-P.}\ \bibnamefont {Saira}}, \bibinfo {author} {\bibfnamefont {Y.}~\bibnamefont {Yoon}}, \bibinfo {author} {\bibfnamefont {A.}~\bibnamefont {Kutvonen}}, \bibinfo {author} {\bibfnamefont {P.}~\bibnamefont {Solinas}}, \bibinfo {author} {\bibfnamefont {M.}~\bibnamefont {M\"{o}tt\"{o}nen}}, \bibinfo {author} {\bibfnamefont {T.}~\bibnamefont {Ala-Nissila}},\ and\ \bibinfo {author} {\bibfnamefont {J.~P.}\ \bibnamefont {Pekola}},\ }\bibfield  {title} {\bibinfo {title} {Distribution of entropy production in a single-electron box},\ }\href {https://doi.org/10.1038/nphys2711} {\bibfield  {journal} {\bibinfo  {journal} {Nature Phys.}\ }\textbf {\bibinfo {volume} {9}},\ \bibinfo {pages} {644} (\bibinfo {year} {2013})}\BibitemShut {NoStop}%
\bibitem [{\citenamefont {Koski}\ \emph {et~al.}(2014)\citenamefont {Koski}, \citenamefont {Maisi}, \citenamefont {Sagawa},\ and\ \citenamefont {Pekola}}]{Koski2014}%
  \BibitemOpen
  \bibfield  {author} {\bibinfo {author} {\bibfnamefont {J.~V.}\ \bibnamefont {Koski}}, \bibinfo {author} {\bibfnamefont {V.~F.}\ \bibnamefont {Maisi}}, \bibinfo {author} {\bibfnamefont {T.}~\bibnamefont {Sagawa}},\ and\ \bibinfo {author} {\bibfnamefont {J.~P.}\ \bibnamefont {Pekola}},\ }\bibfield  {title} {\bibinfo {title} {Experimental observation of the role of mutual information in the nonequilibrium dynamics of a {M}axwell demon},\ }\href {https://doi.org/10.1103/PhysRevLett.113.030601} {\bibfield  {journal} {\bibinfo  {journal} {Phys. Rev. Lett.}\ }\textbf {\bibinfo {volume} {113}},\ \bibinfo {pages} {030601} (\bibinfo {year} {2014})}\BibitemShut {NoStop}%
\bibitem [{\citenamefont {Kleeorin}\ \emph {et~al.}(2019)\citenamefont {Kleeorin}, \citenamefont {Thierschmann}, \citenamefont {Buhmann}, \citenamefont {Georges}, \citenamefont {Molenkamp},\ and\ \citenamefont {Meir}}]{Kleeorin2019}%
  \BibitemOpen
  \bibfield  {author} {\bibinfo {author} {\bibfnamefont {Y.}~\bibnamefont {Kleeorin}}, \bibinfo {author} {\bibfnamefont {H.}~\bibnamefont {Thierschmann}}, \bibinfo {author} {\bibfnamefont {H.}~\bibnamefont {Buhmann}}, \bibinfo {author} {\bibfnamefont {A.}~\bibnamefont {Georges}}, \bibinfo {author} {\bibfnamefont {L.~W.}\ \bibnamefont {Molenkamp}},\ and\ \bibinfo {author} {\bibfnamefont {Y.}~\bibnamefont {Meir}},\ }\bibfield  {title} {\bibinfo {title} {How to measure the entropy of a mesoscopic system via thermoelectric transport},\ }\href {https://doi.org/10.1038/s41467-019-13630-3} {\bibfield  {journal} {\bibinfo  {journal} {Nat. Commun.}\ }\textbf {\bibinfo {volume} {10}},\ \bibinfo {pages} {5801} (\bibinfo {year} {2019})}\BibitemShut {NoStop}%
\bibitem [{\citenamefont {Han}\ \emph {et~al.}(2022)\citenamefont {Han}, \citenamefont {Iftikhar}, \citenamefont {Kleeorin}, \citenamefont {Anthore}, \citenamefont {Pierre}, \citenamefont {Meir}, \citenamefont {Mitchell},\ and\ \citenamefont {Sela}}]{Han2022}%
  \BibitemOpen
  \bibfield  {author} {\bibinfo {author} {\bibfnamefont {C.}~\bibnamefont {Han}}, \bibinfo {author} {\bibfnamefont {Z.}~\bibnamefont {Iftikhar}}, \bibinfo {author} {\bibfnamefont {Y.}~\bibnamefont {Kleeorin}}, \bibinfo {author} {\bibfnamefont {A.}~\bibnamefont {Anthore}}, \bibinfo {author} {\bibfnamefont {F.}~\bibnamefont {Pierre}}, \bibinfo {author} {\bibfnamefont {Y.}~\bibnamefont {Meir}}, \bibinfo {author} {\bibfnamefont {A.~K.}\ \bibnamefont {Mitchell}},\ and\ \bibinfo {author} {\bibfnamefont {E.}~\bibnamefont {Sela}},\ }\bibfield  {title} {\bibinfo {title} {Fractional entropy of multichannel {K}ondo systems from conductance-charge relations},\ }\href {https://doi.org/10.1103/PhysRevLett.128.146803} {\bibfield  {journal} {\bibinfo  {journal} {Phys. Rev. Lett.}\ }\textbf {\bibinfo {volume} {128}},\ \bibinfo {pages} {146803} (\bibinfo {year} {2022})}\BibitemShut {NoStop}%
\bibitem [{\citenamefont {Kruchkov}\ \emph {et~al.}(2020)\citenamefont {Kruchkov}, \citenamefont {Patel}, \citenamefont {Kim},\ and\ \citenamefont {Sachdev}}]{Kruchkov2019}%
  \BibitemOpen
  \bibfield  {author} {\bibinfo {author} {\bibfnamefont {A.}~\bibnamefont {Kruchkov}}, \bibinfo {author} {\bibfnamefont {A.~A.}\ \bibnamefont {Patel}}, \bibinfo {author} {\bibfnamefont {P.}~\bibnamefont {Kim}},\ and\ \bibinfo {author} {\bibfnamefont {S.}~\bibnamefont {Sachdev}},\ }\bibfield  {title} {\bibinfo {title} {Thermoelectric power of {S}achdev-{Y}e-{K}itaev islands: Probing {B}ekenstein-{H}awking entropy in quantum matter experiments},\ }\href {https://doi.org/10.1103/PhysRevB.101.205148} {\bibfield  {journal} {\bibinfo  {journal} {Phys. Rev. B}\ }\textbf {\bibinfo {volume} {101}},\ \bibinfo {pages} {205148} (\bibinfo {year} {2020})}\BibitemShut {NoStop}%
\bibitem [{\citenamefont {Pekola}(2015)}]{Pekola2015}%
  \BibitemOpen
  \bibfield  {author} {\bibinfo {author} {\bibfnamefont {J.~P.}\ \bibnamefont {Pekola}},\ }\bibfield  {title} {\bibinfo {title} {Towards quantum thermodynamics in electronic circuits},\ }\href {https://doi.org/10.1038/nphys3169} {\bibfield  {journal} {\bibinfo  {journal} {Nature Phys.}\ }\textbf {\bibinfo {volume} {11}},\ \bibinfo {pages} {118} (\bibinfo {year} {2015})}\BibitemShut {NoStop}%
\bibitem [{\citenamefont {Jaliel}\ \emph {et~al.}(2019)\citenamefont {Jaliel}, \citenamefont {Puddy}, \citenamefont {S\'anchez}, \citenamefont {Jordan}, \citenamefont {Sothmann}, \citenamefont {Farrer}, \citenamefont {Griffiths}, \citenamefont {Ritchie},\ and\ \citenamefont {Smith}}]{Jaliel2019}%
  \BibitemOpen
  \bibfield  {author} {\bibinfo {author} {\bibfnamefont {G.}~\bibnamefont {Jaliel}}, \bibinfo {author} {\bibfnamefont {R.~K.}\ \bibnamefont {Puddy}}, \bibinfo {author} {\bibfnamefont {R.}~\bibnamefont {S\'anchez}}, \bibinfo {author} {\bibfnamefont {A.~N.}\ \bibnamefont {Jordan}}, \bibinfo {author} {\bibfnamefont {B.}~\bibnamefont {Sothmann}}, \bibinfo {author} {\bibfnamefont {I.}~\bibnamefont {Farrer}}, \bibinfo {author} {\bibfnamefont {J.~P.}\ \bibnamefont {Griffiths}}, \bibinfo {author} {\bibfnamefont {D.~A.}\ \bibnamefont {Ritchie}},\ and\ \bibinfo {author} {\bibfnamefont {C.~G.}\ \bibnamefont {Smith}},\ }\bibfield  {title} {\bibinfo {title} {Experimental realization of a quantum dot energy harvester},\ }\href {https://doi.org/10.1103/PhysRevLett.123.117701} {\bibfield  {journal} {\bibinfo  {journal} {Phys. Rev. Lett.}\ }\textbf {\bibinfo {volume} {123}},\ \bibinfo {pages} {117701} (\bibinfo {year} {2019})}\BibitemShut {NoStop}%
\bibitem [{\citenamefont {S\'anchez}\ \emph {et~al.}(2015)\citenamefont {S\'anchez}, \citenamefont {Sothmann},\ and\ \citenamefont {Jordan}}]{Sanchez2015}%
  \BibitemOpen
  \bibfield  {author} {\bibinfo {author} {\bibfnamefont {R.}~\bibnamefont {S\'anchez}}, \bibinfo {author} {\bibfnamefont {B.}~\bibnamefont {Sothmann}},\ and\ \bibinfo {author} {\bibfnamefont {A.~N.}\ \bibnamefont {Jordan}},\ }\bibfield  {title} {\bibinfo {title} {Chiral thermoelectrics with quantum {H}all edge states},\ }\href {https://doi.org/10.1103/PhysRevLett.114.146801} {\bibfield  {journal} {\bibinfo  {journal} {Phys. Rev. Lett.}\ }\textbf {\bibinfo {volume} {114}},\ \bibinfo {pages} {146801} (\bibinfo {year} {2015})}\BibitemShut {NoStop}%
\bibitem [{\citenamefont {Rech}\ \emph {et~al.}(2020)\citenamefont {Rech}, \citenamefont {Jonckheere}, \citenamefont {Gr\'emaud},\ and\ \citenamefont {Martin}}]{Rech2020}%
  \BibitemOpen
  \bibfield  {author} {\bibinfo {author} {\bibfnamefont {J.}~\bibnamefont {Rech}}, \bibinfo {author} {\bibfnamefont {T.}~\bibnamefont {Jonckheere}}, \bibinfo {author} {\bibfnamefont {B.}~\bibnamefont {Gr\'emaud}},\ and\ \bibinfo {author} {\bibfnamefont {T.}~\bibnamefont {Martin}},\ }\bibfield  {title} {\bibinfo {title} {Negative delta-{T} noise in the fractional quantum {H}all effect},\ }\href {https://doi.org/10.1103/PhysRevLett.125.086801} {\bibfield  {journal} {\bibinfo  {journal} {Phys. Rev. Lett.}\ }\textbf {\bibinfo {volume} {125}},\ \bibinfo {pages} {086801} (\bibinfo {year} {2020})}\BibitemShut {NoStop}%
\bibitem [{\citenamefont {Zhang}\ \emph {et~al.}(2022)\citenamefont {Zhang}, \citenamefont {Gornyi},\ and\ \citenamefont {Sp\aa{}nsl\"att}}]{Zhang2022}%
  \BibitemOpen
  \bibfield  {author} {\bibinfo {author} {\bibfnamefont {G.}~\bibnamefont {Zhang}}, \bibinfo {author} {\bibfnamefont {I.~V.}\ \bibnamefont {Gornyi}},\ and\ \bibinfo {author} {\bibfnamefont {C.}~\bibnamefont {Sp\aa{}nsl\"att}},\ }\bibfield  {title} {\bibinfo {title} {Delta-{$T$} noise for weak tunneling in one-dimensional systems: Interactions versus quantum statistics},\ }\href {https://doi.org/10.1103/PhysRevB.105.195423} {\bibfield  {journal} {\bibinfo  {journal} {Phys. Rev. B}\ }\textbf {\bibinfo {volume} {105}},\ \bibinfo {pages} {195423} (\bibinfo {year} {2022})}\BibitemShut {NoStop}%
\bibitem [{\citenamefont {Schiller}\ \emph {et~al.}(2022)\citenamefont {Schiller}, \citenamefont {Oreg},\ and\ \citenamefont {Snizhko}}]{Schiller2022}%
  \BibitemOpen
  \bibfield  {author} {\bibinfo {author} {\bibfnamefont {N.}~\bibnamefont {Schiller}}, \bibinfo {author} {\bibfnamefont {Y.}~\bibnamefont {Oreg}},\ and\ \bibinfo {author} {\bibfnamefont {K.}~\bibnamefont {Snizhko}},\ }\bibfield  {title} {\bibinfo {title} {Extracting the scaling dimension of quantum {H}all quasiparticles from current correlations},\ }\href {https://doi.org/10.1103/PhysRevB.105.165150} {\bibfield  {journal} {\bibinfo  {journal} {Phys. Rev. B}\ }\textbf {\bibinfo {volume} {105}},\ \bibinfo {pages} {165150} (\bibinfo {year} {2022})}\BibitemShut {NoStop}%
\bibitem [{\citenamefont {Ebisu}\ \emph {et~al.}(2022)\citenamefont {Ebisu}, \citenamefont {Schiller},\ and\ \citenamefont {Oreg}}]{Ebisu2022}%
  \BibitemOpen
  \bibfield  {author} {\bibinfo {author} {\bibfnamefont {H.}~\bibnamefont {Ebisu}}, \bibinfo {author} {\bibfnamefont {N.}~\bibnamefont {Schiller}},\ and\ \bibinfo {author} {\bibfnamefont {Y.}~\bibnamefont {Oreg}},\ }\bibfield  {title} {\bibinfo {title} {Fluctuations in heat current and scaling dimension},\ }\href {https://doi.org/10.1103/PhysRevLett.128.215901} {\bibfield  {journal} {\bibinfo  {journal} {Phys. Rev. Lett.}\ }\textbf {\bibinfo {volume} {128}},\ \bibinfo {pages} {215901} (\bibinfo {year} {2022})}\BibitemShut {NoStop}%
\bibitem [{\citenamefont {Rebora}\ \emph {et~al.}(2022)\citenamefont {Rebora}, \citenamefont {Rech}, \citenamefont {Ferraro}, \citenamefont {Jonckheere}, \citenamefont {Martin},\ and\ \citenamefont {Sassetti}}]{Rebora2022}%
  \BibitemOpen
  \bibfield  {author} {\bibinfo {author} {\bibfnamefont {G.}~\bibnamefont {Rebora}}, \bibinfo {author} {\bibfnamefont {J.}~\bibnamefont {Rech}}, \bibinfo {author} {\bibfnamefont {D.}~\bibnamefont {Ferraro}}, \bibinfo {author} {\bibfnamefont {T.}~\bibnamefont {Jonckheere}}, \bibinfo {author} {\bibfnamefont {T.}~\bibnamefont {Martin}},\ and\ \bibinfo {author} {\bibfnamefont {M.}~\bibnamefont {Sassetti}},\ }\bibfield  {title} {\bibinfo {title} {Delta-{$T$} noise for fractional quantum {H}all states at different filling factor},\ }\href {https://doi.org/10.1103/PhysRevResearch.4.043191} {\bibfield  {journal} {\bibinfo  {journal} {Phys. Rev. Res.}\ }\textbf {\bibinfo {volume} {4}},\ \bibinfo {pages} {043191} (\bibinfo {year} {2022})}\BibitemShut {NoStop}%
\bibitem [{\citenamefont {Iyer}\ \emph {et~al.}(2023)\citenamefont {Iyer}, \citenamefont {Rech}, \citenamefont {Jonckheere}, \citenamefont {Raymond}, \citenamefont {Gr\'emaud},\ and\ \citenamefont {Martin}}]{Iyer2023}%
  \BibitemOpen
  \bibfield  {author} {\bibinfo {author} {\bibfnamefont {K.}~\bibnamefont {Iyer}}, \bibinfo {author} {\bibfnamefont {J.}~\bibnamefont {Rech}}, \bibinfo {author} {\bibfnamefont {T.}~\bibnamefont {Jonckheere}}, \bibinfo {author} {\bibfnamefont {L.}~\bibnamefont {Raymond}}, \bibinfo {author} {\bibfnamefont {B.}~\bibnamefont {Gr\'emaud}},\ and\ \bibinfo {author} {\bibfnamefont {T.}~\bibnamefont {Martin}},\ }\bibfield  {title} {\bibinfo {title} {Colored delta-{$T$} noise in fractional quantum {H}all liquids},\ }\href {https://doi.org/10.1103/PhysRevB.108.245427} {\bibfield  {journal} {\bibinfo  {journal} {Phys. Rev. B}\ }\textbf {\bibinfo {volume} {108}},\ \bibinfo {pages} {245427} (\bibinfo {year} {2023})}\BibitemShut {NoStop}%
\bibitem [{\citenamefont {Acciai}\ \emph {et~al.}(2025)\citenamefont {Acciai}, \citenamefont {Zhang},\ and\ \citenamefont {Sp\aa{}nsl\"att}}]{Acciai2025}%
  \BibitemOpen
  \bibfield  {author} {\bibinfo {author} {\bibfnamefont {M.}~\bibnamefont {Acciai}}, \bibinfo {author} {\bibfnamefont {G.}~\bibnamefont {Zhang}},\ and\ \bibinfo {author} {\bibfnamefont {C.}~\bibnamefont {Sp\aa{}nsl\"att}},\ }\bibfield  {title} {\bibinfo {title} {{Role of scaling dimensions in generalized noises in fractional quantum {H}all tunneling due to a temperature bias}},\ }\href {https://doi.org/10.21468/SciPostPhys.18.2.058} {\bibfield  {journal} {\bibinfo  {journal} {SciPost Phys.}\ }\textbf {\bibinfo {volume} {18}},\ \bibinfo {pages} {058} (\bibinfo {year} {2025})}\BibitemShut {NoStop}%
\bibitem [{\citenamefont {Zhang}\ \emph {et~al.}(2025{\natexlab{a}})\citenamefont {Zhang}, \citenamefont {Gornyi},\ and\ \citenamefont {Gefen}}]{Zhang2025}%
  \BibitemOpen
  \bibfield  {author} {\bibinfo {author} {\bibfnamefont {G.}~\bibnamefont {Zhang}}, \bibinfo {author} {\bibfnamefont {I.}~\bibnamefont {Gornyi}},\ and\ \bibinfo {author} {\bibfnamefont {Y.}~\bibnamefont {Gefen}},\ }\bibfield  {title} {\bibinfo {title} {Landscapes of an out-of-equilibrium anyonic sea},\ }\href {https://doi.org/10.1103/PhysRevLett.134.096303} {\bibfield  {journal} {\bibinfo  {journal} {Phys. Rev. Lett.}\ }\textbf {\bibinfo {volume} {134}},\ \bibinfo {pages} {096303} (\bibinfo {year} {2025}{\natexlab{a}})}\BibitemShut {NoStop}%
\bibitem [{\citenamefont {Zhang}\ \emph {et~al.}(2025{\natexlab{b}})\citenamefont {Zhang}, \citenamefont {Gornyi},\ and\ \citenamefont {Gefen}}]{Zhang2025b}%
  \BibitemOpen
  \bibfield  {author} {\bibinfo {author} {\bibfnamefont {G.}~\bibnamefont {Zhang}}, \bibinfo {author} {\bibfnamefont {I.}~\bibnamefont {Gornyi}},\ and\ \bibinfo {author} {\bibfnamefont {Y.}~\bibnamefont {Gefen}},\ }\href@noop {} {\bibinfo {title} {Effective linear response in non-equilibrium anyonic systems}} (\bibinfo {year} {2025}{\natexlab{b}}),\ \Eprint {https://arxiv.org/abs/2510.03985} {arXiv:2510.03985 [cond-mat.mes-hall]} \BibitemShut {NoStop}%
\bibitem [{\citenamefont {Brantut}\ \emph {et~al.}(2013)\citenamefont {Brantut}, \citenamefont {Grenier}, \citenamefont {Meineke}, \citenamefont {Stadler}, \citenamefont {Krinner}, \citenamefont {Kollath}, \citenamefont {Esslinger},\ and\ \citenamefont {Georges}}]{Brantut2013}%
  \BibitemOpen
  \bibfield  {author} {\bibinfo {author} {\bibfnamefont {J.-P.}\ \bibnamefont {Brantut}}, \bibinfo {author} {\bibfnamefont {C.}~\bibnamefont {Grenier}}, \bibinfo {author} {\bibfnamefont {J.}~\bibnamefont {Meineke}}, \bibinfo {author} {\bibfnamefont {D.}~\bibnamefont {Stadler}}, \bibinfo {author} {\bibfnamefont {S.}~\bibnamefont {Krinner}}, \bibinfo {author} {\bibfnamefont {C.}~\bibnamefont {Kollath}}, \bibinfo {author} {\bibfnamefont {T.}~\bibnamefont {Esslinger}},\ and\ \bibinfo {author} {\bibfnamefont {A.}~\bibnamefont {Georges}},\ }\bibfield  {title} {\bibinfo {title} {A thermoelectric heat engine with ultracold atoms},\ }\href {https://doi.org/10.1126/science.1242308} {\bibfield  {journal} {\bibinfo  {journal} {Science}\ }\textbf {\bibinfo {volume} {342}},\ \bibinfo {pages} {713} (\bibinfo {year} {2013})}\BibitemShut {NoStop}%
\bibitem [{\citenamefont {H\"ausler}\ \emph {et~al.}(2021)\citenamefont {H\"ausler}, \citenamefont {Fabritius}, \citenamefont {Mohan}, \citenamefont {Lebrat}, \citenamefont {Corman},\ and\ \citenamefont {Esslinger}}]{Hausler2021}%
  \BibitemOpen
  \bibfield  {author} {\bibinfo {author} {\bibfnamefont {S.}~\bibnamefont {H\"ausler}}, \bibinfo {author} {\bibfnamefont {P.}~\bibnamefont {Fabritius}}, \bibinfo {author} {\bibfnamefont {J.}~\bibnamefont {Mohan}}, \bibinfo {author} {\bibfnamefont {M.}~\bibnamefont {Lebrat}}, \bibinfo {author} {\bibfnamefont {L.}~\bibnamefont {Corman}},\ and\ \bibinfo {author} {\bibfnamefont {T.}~\bibnamefont {Esslinger}},\ }\bibfield  {title} {\bibinfo {title} {Interaction-assisted reversal of thermopower with ultracold atoms},\ }\href {https://doi.org/10.1103/PhysRevX.11.021034} {\bibfield  {journal} {\bibinfo  {journal} {Phys. Rev. X}\ }\textbf {\bibinfo {volume} {11}},\ \bibinfo {pages} {021034} (\bibinfo {year} {2021})}\BibitemShut {NoStop}%
\bibitem [{\citenamefont {Bauer}\ \emph {et~al.}(2013)\citenamefont {Bauer}, \citenamefont {Heyder}, \citenamefont {Schubert}, \citenamefont {Borowsky}, \citenamefont {Taubert}, \citenamefont {Bruognolo}, \citenamefont {Schuh}, \citenamefont {Wegscheider}, \citenamefont {von Delft},\ and\ \citenamefont {Ludwig}}]{Bauer2013}%
  \BibitemOpen
  \bibfield  {author} {\bibinfo {author} {\bibfnamefont {F.}~\bibnamefont {Bauer}}, \bibinfo {author} {\bibfnamefont {J.}~\bibnamefont {Heyder}}, \bibinfo {author} {\bibfnamefont {E.}~\bibnamefont {Schubert}}, \bibinfo {author} {\bibfnamefont {D.}~\bibnamefont {Borowsky}}, \bibinfo {author} {\bibfnamefont {D.}~\bibnamefont {Taubert}}, \bibinfo {author} {\bibfnamefont {B.}~\bibnamefont {Bruognolo}}, \bibinfo {author} {\bibfnamefont {D.}~\bibnamefont {Schuh}}, \bibinfo {author} {\bibfnamefont {W.}~\bibnamefont {Wegscheider}}, \bibinfo {author} {\bibfnamefont {J.}~\bibnamefont {von Delft}},\ and\ \bibinfo {author} {\bibfnamefont {S.}~\bibnamefont {Ludwig}},\ }\bibfield  {title} {\bibinfo {title} {Microscopic origin of the `0.7-anomaly' in quantum point contacts},\ }\href {https://doi.org/10.1038/nature12421} {\bibfield  {journal} {\bibinfo  {journal} {Nature}\ }\textbf {\bibinfo {volume} {501}},\ \bibinfo {pages} {73} (\bibinfo {year} {2013})}\BibitemShut {NoStop}%
\bibitem [{\citenamefont {Spietz}\ \emph {et~al.}(2003)\citenamefont {Spietz}, \citenamefont {Lehnert}, \citenamefont {Siddiqi},\ and\ \citenamefont {Schoelkopf}}]{Spietz2003}%
  \BibitemOpen
  \bibfield  {author} {\bibinfo {author} {\bibfnamefont {L.}~\bibnamefont {Spietz}}, \bibinfo {author} {\bibfnamefont {K.~W.}\ \bibnamefont {Lehnert}}, \bibinfo {author} {\bibfnamefont {I.}~\bibnamefont {Siddiqi}},\ and\ \bibinfo {author} {\bibfnamefont {R.~J.}\ \bibnamefont {Schoelkopf}},\ }\bibfield  {title} {\bibinfo {title} {Primary electronic thermometry using the shot noise of a tunnel junction},\ }\href {https://doi.org/10.1126/science.1084647} {\bibfield  {journal} {\bibinfo  {journal} {Science}\ }\textbf {\bibinfo {volume} {300}},\ \bibinfo {pages} {1929} (\bibinfo {year} {2003})}\BibitemShut {NoStop}%
\bibitem [{\citenamefont {F\'evrier}\ and\ \citenamefont {Gabelli}(2018)}]{Fevrier2018}%
  \BibitemOpen
  \bibfield  {author} {\bibinfo {author} {\bibfnamefont {P.}~\bibnamefont {F\'evrier}}\ and\ \bibinfo {author} {\bibfnamefont {J.}~\bibnamefont {Gabelli}},\ }\bibfield  {title} {\bibinfo {title} {Tunneling time probed by quantum shot noise},\ }\href {https://doi.org/10.1038/s41467-018-07369-6} {\bibfield  {journal} {\bibinfo  {journal} {Nat. Commun.}\ }\textbf {\bibinfo {volume} {9}},\ \bibinfo {pages} {4940} (\bibinfo {year} {2018})}\BibitemShut {NoStop}%
\bibitem [{\citenamefont {Tamir}\ \emph {et~al.}(2022)\citenamefont {Tamir}, \citenamefont {Caspari}, \citenamefont {Rolf}, \citenamefont {Lotze},\ and\ \citenamefont {Franke}}]{Tamir2022}%
  \BibitemOpen
  \bibfield  {author} {\bibinfo {author} {\bibfnamefont {I.}~\bibnamefont {Tamir}}, \bibinfo {author} {\bibfnamefont {V.}~\bibnamefont {Caspari}}, \bibinfo {author} {\bibfnamefont {D.}~\bibnamefont {Rolf}}, \bibinfo {author} {\bibfnamefont {C.}~\bibnamefont {Lotze}},\ and\ \bibinfo {author} {\bibfnamefont {K.~J.}\ \bibnamefont {Franke}},\ }\bibfield  {title} {\bibinfo {title} {Shot-noise measurements of single-atom junctions using a scanning tunneling microscope},\ }\href {https://doi.org/10.1063/5.0078917} {\bibfield  {journal} {\bibinfo  {journal} {Rev. Sci. Instrum.}\ }\textbf {\bibinfo {volume} {93}},\ \bibinfo {pages} {023702} (\bibinfo {year} {2022})}\BibitemShut {NoStop}%
\bibitem [{\citenamefont {Flensberg}(1993)}]{Flensberg1993}%
  \BibitemOpen
  \bibfield  {author} {\bibinfo {author} {\bibfnamefont {K.}~\bibnamefont {Flensberg}},\ }\bibfield  {title} {\bibinfo {title} {Capacitance and conductance of mesoscopic systems connected by quantum point contacts},\ }\href {https://doi.org/10.1103/PhysRevB.48.11156} {\bibfield  {journal} {\bibinfo  {journal} {Phys. Rev. B}\ }\textbf {\bibinfo {volume} {48}},\ \bibinfo {pages} {11156} (\bibinfo {year} {1993})}\BibitemShut {NoStop}%
\bibitem [{\citenamefont {Matveev}(1995)}]{Matveev1995}%
  \BibitemOpen
  \bibfield  {author} {\bibinfo {author} {\bibfnamefont {K.~A.}\ \bibnamefont {Matveev}},\ }\bibfield  {title} {\bibinfo {title} {Coulomb blockade at almost perfect transmission},\ }\href {https://doi.org/10.1103/PhysRevB.51.1743} {\bibfield  {journal} {\bibinfo  {journal} {Phys. Rev. B}\ }\textbf {\bibinfo {volume} {51}},\ \bibinfo {pages} {1743} (\bibinfo {year} {1995})}\BibitemShut {NoStop}%
\bibitem [{\citenamefont {Farahi}\ \emph {et~al.}(2023)\citenamefont {Farahi}, \citenamefont {Chiu}, \citenamefont {Liu}, \citenamefont {Papic}, \citenamefont {Watanabe}, \citenamefont {Taniguchi}, \citenamefont {Zaletel},\ and\ \citenamefont {Yazdani}}]{Farahi2023}%
  \BibitemOpen
  \bibfield  {author} {\bibinfo {author} {\bibfnamefont {G.}~\bibnamefont {Farahi}}, \bibinfo {author} {\bibfnamefont {C.-L.}\ \bibnamefont {Chiu}}, \bibinfo {author} {\bibfnamefont {X.}~\bibnamefont {Liu}}, \bibinfo {author} {\bibfnamefont {Z.}~\bibnamefont {Papic}}, \bibinfo {author} {\bibfnamefont {K.}~\bibnamefont {Watanabe}}, \bibinfo {author} {\bibfnamefont {T.}~\bibnamefont {Taniguchi}}, \bibinfo {author} {\bibfnamefont {M.~P.}\ \bibnamefont {Zaletel}},\ and\ \bibinfo {author} {\bibfnamefont {A.}~\bibnamefont {Yazdani}},\ }\bibfield  {title} {\bibinfo {title} {Broken symmetries and excitation spectra of interacting electrons in partially filled {L}andau levels},\ }\href {https://doi.org/10.1038/s41567-023-02126-z} {\bibfield  {journal} {\bibinfo  {journal} {Nature Physics}\ }\textbf {\bibinfo {volume} {19}},\ \bibinfo {pages} {1482} (\bibinfo {year} {2023})}\BibitemShut {NoStop}%
\bibitem [{\citenamefont {Hu}\ \emph {et~al.}(2025)\citenamefont {Hu}, \citenamefont {Tsui}, \citenamefont {He}, \citenamefont {Kamber}, \citenamefont {Wang}, \citenamefont {Mohammadi}, \citenamefont {Watanabe}, \citenamefont {Taniguchi}, \citenamefont {Papi\'c}, \citenamefont {Zaletel},\ and\ \citenamefont {Yazdani}}]{Hu2025}%
  \BibitemOpen
  \bibfield  {author} {\bibinfo {author} {\bibfnamefont {Y.}~\bibnamefont {Hu}}, \bibinfo {author} {\bibfnamefont {Y.-C.}\ \bibnamefont {Tsui}}, \bibinfo {author} {\bibfnamefont {M.}~\bibnamefont {He}}, \bibinfo {author} {\bibfnamefont {U.}~\bibnamefont {Kamber}}, \bibinfo {author} {\bibfnamefont {T.}~\bibnamefont {Wang}}, \bibinfo {author} {\bibfnamefont {A.~S.}\ \bibnamefont {Mohammadi}}, \bibinfo {author} {\bibfnamefont {K.}~\bibnamefont {Watanabe}}, \bibinfo {author} {\bibfnamefont {T.}~\bibnamefont {Taniguchi}}, \bibinfo {author} {\bibfnamefont {Z.}~\bibnamefont {Papi\'c}}, \bibinfo {author} {\bibfnamefont {M.~P.}\ \bibnamefont {Zaletel}},\ and\ \bibinfo {author} {\bibfnamefont {A.}~\bibnamefont {Yazdani}},\ }\bibfield  {title} {\bibinfo {title} {High-resolution tunnelling spectroscopy of fractional quantum {H}all states},\ }\href {https://doi.org/10.1038/s41567-025-02830-y} {\bibfield  {journal} {\bibinfo  {journal} {Nature Physics}\ }\textbf {\bibinfo {volume} {21}},\ \bibinfo {pages} {716} (\bibinfo
  {year} {2025})}\BibitemShut {NoStop}%
\bibitem [{\citenamefont {Gnezdilov}\ \emph {et~al.}(2018)\citenamefont {Gnezdilov}, \citenamefont {Hutasoit},\ and\ \citenamefont {Beenakker}}]{Gnezdilov2018}%
  \BibitemOpen
  \bibfield  {author} {\bibinfo {author} {\bibfnamefont {N.~V.}\ \bibnamefont {Gnezdilov}}, \bibinfo {author} {\bibfnamefont {J.~A.}\ \bibnamefont {Hutasoit}},\ and\ \bibinfo {author} {\bibfnamefont {C.~W.~J.}\ \bibnamefont {Beenakker}},\ }\bibfield  {title} {\bibinfo {title} {Low-high voltage duality in tunneling spectroscopy of the {S}achdev-{Y}e-{K}itaev model},\ }\href {https://doi.org/10.1103/PhysRevB.98.081413} {\bibfield  {journal} {\bibinfo  {journal} {Phys. Rev. B}\ }\textbf {\bibinfo {volume} {98}},\ \bibinfo {pages} {081413} (\bibinfo {year} {2018})}\BibitemShut {NoStop}%
\bibitem [{\citenamefont {Manaparambil}\ \emph {et~al.}(2025)\citenamefont {Manaparambil}, \citenamefont {Weichselbaum}, \citenamefont {von Delft},\ and\ \citenamefont {Weymann}}]{Manaparambil2025}%
  \BibitemOpen
  \bibfield  {author} {\bibinfo {author} {\bibfnamefont {A.}~\bibnamefont {Manaparambil}}, \bibinfo {author} {\bibfnamefont {A.}~\bibnamefont {Weichselbaum}}, \bibinfo {author} {\bibfnamefont {J.}~\bibnamefont {von Delft}},\ and\ \bibinfo {author} {\bibfnamefont {I.}~\bibnamefont {Weymann}},\ }\bibfield  {title} {\bibinfo {title} {Nonequilibrium steady-state thermoelectrics of {K}ondo-correlated quantum dots},\ }\href {https://doi.org/10.1103/PhysRevB.111.035445} {\bibfield  {journal} {\bibinfo  {journal} {Phys. Rev. B}\ }\textbf {\bibinfo {volume} {111}},\ \bibinfo {pages} {035445} (\bibinfo {year} {2025})}\BibitemShut {NoStop}%
\bibitem [{\citenamefont {Aristov}(2007)}]{Aristov2007}%
  \BibitemOpen
  \bibfield  {author} {\bibinfo {author} {\bibfnamefont {D.~N.}\ \bibnamefont {Aristov}},\ }\bibfield  {title} {\bibinfo {title} {Luttinger liquids with curvature: {D}ensity correlations and {C}oulomb drag effect},\ }\href {https://doi.org/10.1103/PhysRevB.76.085327} {\bibfield  {journal} {\bibinfo  {journal} {Phys. Rev. B}\ }\textbf {\bibinfo {volume} {76}},\ \bibinfo {pages} {085327} (\bibinfo {year} {2007})}\BibitemShut {NoStop}%
\bibitem [{\citenamefont {Nguyen}\ \emph {et~al.}(2010)\citenamefont {Nguyen}, \citenamefont {Kiselev},\ and\ \citenamefont {Kravtsov}}]{Nguyen2010}%
  \BibitemOpen
  \bibfield  {author} {\bibinfo {author} {\bibfnamefont {T.~K.~T.}\ \bibnamefont {Nguyen}}, \bibinfo {author} {\bibfnamefont {M.~N.}\ \bibnamefont {Kiselev}},\ and\ \bibinfo {author} {\bibfnamefont {V.~E.}\ \bibnamefont {Kravtsov}},\ }\bibfield  {title} {\bibinfo {title} {Thermoelectric transport through a quantum dot: {E}ffects of asymmetry in {K}ondo channels},\ }\href {https://doi.org/10.1103/PhysRevB.82.113306} {\bibfield  {journal} {\bibinfo  {journal} {Phys. Rev. B}\ }\textbf {\bibinfo {volume} {82}},\ \bibinfo {pages} {113306} (\bibinfo {year} {2010})}\BibitemShut {NoStop}%
\bibitem [{\citenamefont {Pustilnik}\ and\ \citenamefont {Glazman}(2001)}]{Pustilnik2001}%
  \BibitemOpen
  \bibfield  {author} {\bibinfo {author} {\bibfnamefont {M.}~\bibnamefont {Pustilnik}}\ and\ \bibinfo {author} {\bibfnamefont {L.~I.}\ \bibnamefont {Glazman}},\ }\bibfield  {title} {\bibinfo {title} {Kondo effect in real quantum dots},\ }\href {https://doi.org/10.1103/PhysRevLett.87.216601} {\bibfield  {journal} {\bibinfo  {journal} {Phys. Rev. Lett.}\ }\textbf {\bibinfo {volume} {87}},\ \bibinfo {pages} {216601} (\bibinfo {year} {2001})}\BibitemShut {NoStop}%
\bibitem [{\citenamefont {W\'ojcik}\ and\ \citenamefont {Weymann}(2016)}]{Wojcik2016}%
  \BibitemOpen
  \bibfield  {author} {\bibinfo {author} {\bibfnamefont {K.~P.}\ \bibnamefont {W\'ojcik}}\ and\ \bibinfo {author} {\bibfnamefont {I.}~\bibnamefont {Weymann}},\ }\bibfield  {title} {\bibinfo {title} {Thermopower of strongly correlated {T}-shaped double quantum dots},\ }\href {https://doi.org/10.1103/PhysRevB.93.085428} {\bibfield  {journal} {\bibinfo  {journal} {Phys. Rev. B}\ }\textbf {\bibinfo {volume} {93}},\ \bibinfo {pages} {085428} (\bibinfo {year} {2016})}\BibitemShut {NoStop}%
\bibitem [{\citenamefont {Brown}\ \emph {et~al.}(2009)\citenamefont {Brown}, \citenamefont {Crisan},\ and\ \citenamefont {\c{T}ifrea}}]{Brown2009}%
  \BibitemOpen
  \bibfield  {author} {\bibinfo {author} {\bibfnamefont {K.}~\bibnamefont {Brown}}, \bibinfo {author} {\bibfnamefont {M.}~\bibnamefont {Crisan}},\ and\ \bibinfo {author} {\bibfnamefont {I.}~\bibnamefont {\c{T}ifrea}},\ }\bibfield  {title} {\bibinfo {title} {Transport and current noise characteristics of a {T}-shape double-quantum-dot system},\ }\href {https://doi.org/10.1088/0953-8984/21/21/215604} {\bibfield  {journal} {\bibinfo  {journal} {J. Phys.: Condens. Matter}\ }\textbf {\bibinfo {volume} {21}},\ \bibinfo {pages} {215604} (\bibinfo {year} {2009})}\BibitemShut {NoStop}%
\bibitem [{\citenamefont {Karki}(2020)}]{Karki2020}%
  \BibitemOpen
  \bibfield  {author} {\bibinfo {author} {\bibfnamefont {D.~B.}\ \bibnamefont {Karki}},\ }\bibfield  {title} {\bibinfo {title} {Wiedemann-{F}ranz law in scattering theory revisited},\ }\href {https://doi.org/10.1103/PhysRevB.102.115423} {\bibfield  {journal} {\bibinfo  {journal} {Phys. Rev. B}\ }\textbf {\bibinfo {volume} {102}},\ \bibinfo {pages} {115423} (\bibinfo {year} {2020})}\BibitemShut {NoStop}%
\bibitem [{\citenamefont {Oguri}\ and\ \citenamefont {Hewson}(2018)}]{Oguri2018}%
  \BibitemOpen
  \bibfield  {author} {\bibinfo {author} {\bibfnamefont {A.}~\bibnamefont {Oguri}}\ and\ \bibinfo {author} {\bibfnamefont {A.~C.}\ \bibnamefont {Hewson}},\ }\bibfield  {title} {\bibinfo {title} {Higher-order {F}ermi-liquid corrections for an {A}nderson impurity away from half filling},\ }\href {https://doi.org/10.1103/PhysRevLett.120.126802} {\bibfield  {journal} {\bibinfo  {journal} {Phys. Rev. Lett.}\ }\textbf {\bibinfo {volume} {120}},\ \bibinfo {pages} {126802} (\bibinfo {year} {2018})}\BibitemShut {NoStop}%
\bibitem [{\citenamefont {Karki}\ and\ \citenamefont {Kiselev}(2017)}]{Karki2017}%
  \BibitemOpen
  \bibfield  {author} {\bibinfo {author} {\bibfnamefont {D.~B.}\ \bibnamefont {Karki}}\ and\ \bibinfo {author} {\bibfnamefont {M.~N.}\ \bibnamefont {Kiselev}},\ }\bibfield  {title} {\bibinfo {title} {Thermoelectric transport through a $\text{SU}({N})$ {K}ondo impurity},\ }\href {https://doi.org/10.1103/PhysRevB.96.121403} {\bibfield  {journal} {\bibinfo  {journal} {Phys. Rev. B}\ }\textbf {\bibinfo {volume} {96}},\ \bibinfo {pages} {121403} (\bibinfo {year} {2017})}\BibitemShut {NoStop}%
\end{thebibliography}%

\appendix
\onecolumngrid
\begin{center}
{\large \bf End Matter}\\
\end{center}
\twocolumngrid
\setcounter{equation}{0}
\renewcommand{\theequation}{A\arabic{equation}}
\paragraph*{Universal noise relations ---}
Within the universality regime, one can construct a set of ratios that connect all the transport and noise coefficients. Overall, there are five independent ratios (including the WF law) that relate all the coefficients between each other; other relations are redundant. For instance, one can define the ratio between the delta-T charge noise and the electric conductance. This is identical to the ratio of the following coefficients $\delta S_c^{\Delta T}/G=\frac{\mathcal{N}_1}{T\mathcal{L}_0}=L_1R^{\Delta T}_C$. Here, we denoted $L_1=1$ as the first extended Lorenz number for FL, and $R^{\Delta T}_C$ is the deviation of this ratio from the FL value:
\begin{align} \label{RdTC}
  R^{\Delta T}_C=2-\frac{2\int_{-\infty}^{\infty}dt\, \pi Tt\frac{\sinh\left(\pi T t\right)}{\cosh^{2}\left(\pi T t\right)}\mathcal{T}\left(\frac{1}{2T}+\textit{i}t\right)}{\int_{-\infty}^{\infty}dt \cosh^{-1}\left(\pi T t\right)\mathcal{T}\left(\frac{1}{2T}+\textit{i}t\right)}.  
\end{align}
This ratio has the same universality range and applicability conditions as $R_L$ of the WF law. Equation (\ref{RdTC}) is general for any form of $\mathcal{T}$. For the $\mathcal{T}\left(\frac{1}{2T}+\textit{i}t\right)\sim \cosh^{-\alpha}\left(\pi T t\right)$ transmission coefficient, it becomes
\begin{align} \label{RCT}
   R^{\Delta T}_C=\frac{2\alpha}{1+\alpha}. 
\end{align}
We further explore universality of the $R^{\Delta T}_C$ ratio for elastic and inelastic tunneling regimes, as well as breaking of this universality, for the SYK dot in \cite{Pavlov2025}. \\
We introduce further $\frac{\delta S_h^{\Delta T}}{T^2G}=\frac{\mathcal{N}_3}{T^3\mathcal{L}_0}=L_2R^{\Delta T}_H$, where $L_2=\pi^2$ is the second extended Lorenz number for FL. $R^{\Delta T}_H$ is the deviation of this ratio from the FL value:
\begin{align} \label{RdTH}
    R^{\Delta T}_H= \frac{2\alpha +6\alpha^2}{3+4\alpha+\alpha^2}.
\end{align}
With Eqs. (\ref{RLcosh}), (\ref{RdTC}), and (\ref{RdTH}), one can trivially find the three remaining ratios between the symmetric coefficients ($\mathcal{L}_2$, $\mathcal{N}_1$ and $\mathcal{N}_3$).\\
\begin{figure}[t!]
\vspace{-0.6cm}
\center
\hspace*{-.5cm}
\includegraphics[width=1.1\linewidth]{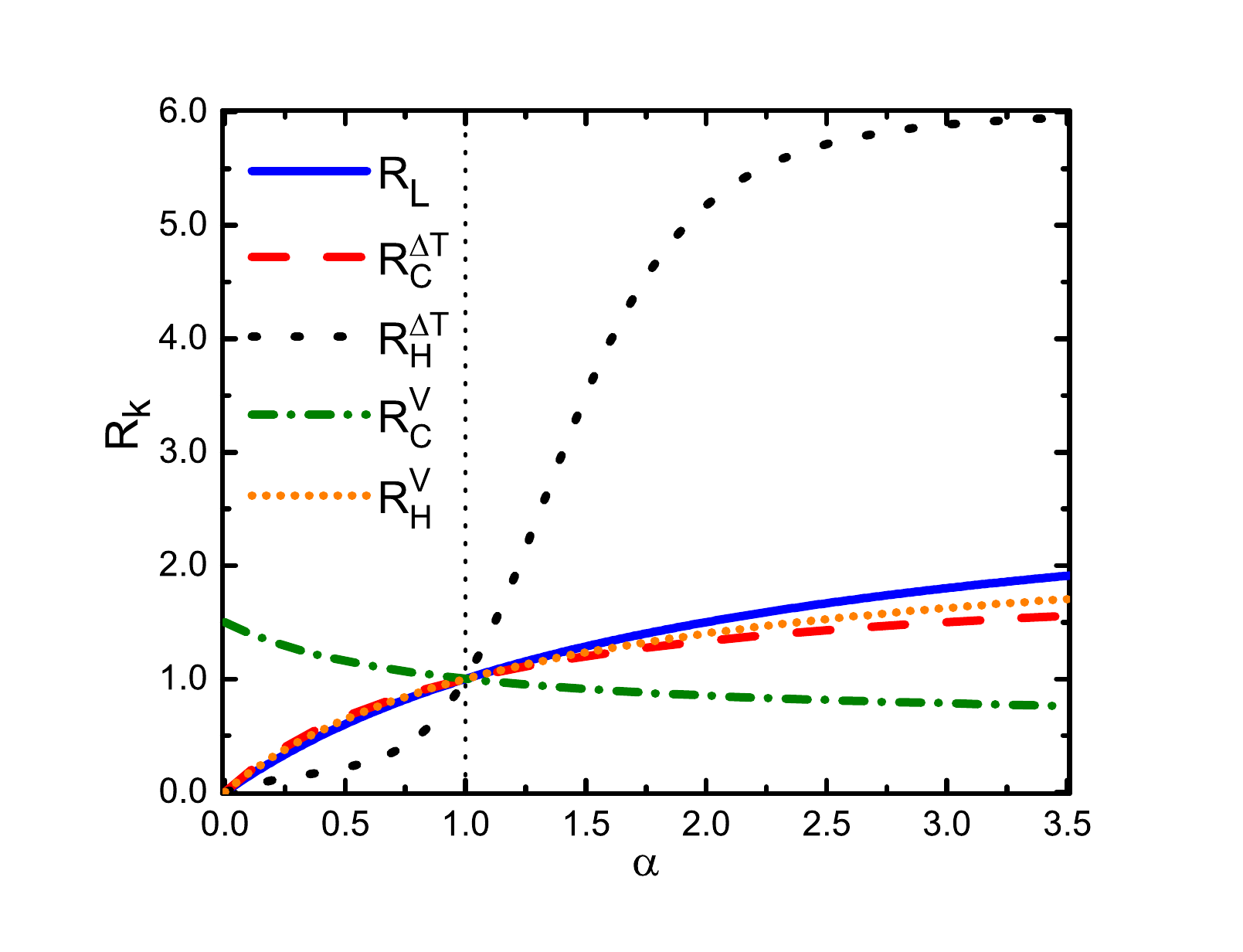} 
\vspace{-1cm}
\caption{Extended Lorenz ratios as functions of $\alpha$. $R_k$ stands for $R_L$ (solid blue line), $R^{\Delta T}_C$ (dashed red line), $R^{\Delta T}_H$ (dotted black line), $R^V_C$ (dash-dotted green line), $R^V_H$ (short-dotted orange line). All ratios are normalized by the extended Lorenz numbers $L_i$, $i=0,..,4$ such that $R_k=1$ at $\alpha=1$ (Fermi liquid regime, thin vertical dotted line).}
\label{fig:LorenzR} \vspace{-0.5cm}
\end{figure}
The same type of universal relations holds between the antisymmetric coefficients. Coefficients $\mathcal{L}_1$, $\mathcal{N}_0$ and $\mathcal{N}_2$ do not vanish only if the system is away from the hole-particle symmetric regime; this deviation can be characterized by the spectral asymmetry parameter $\mathcal{E}$ \cite{Pavlov2025}. In most cases, all the antisymmetric coefficients are proportional to $\sim \mathcal{E}$ for weak particle-hole asymmetry. Nevertheless, this proportionality cancels in their ratios (up to $O(\mathcal{E}^2)$ corrections), which remain finite even in the $\mathcal{E}\rightarrow 0$ limit. In some special cases, e.g., for the two-channel charge Kondo system \cite{Karki2020}, the prefactor at $\mathcal{E}$ for the antisymmetric coefficients is zero, so the leading contribution is $\sim \mathcal{E}^2$ (or $\mathcal{E}^n,\, n\in \mathbb{N}$ most generally), which again cancels in their ratios. These ratios can be found by introducing small spectral asymmetry into the transmission coefficient $\mathcal{T}\left(\frac{1}{2T}+\textit{i} t\right)$. This generates odd-in-$t$ terms in $\mathcal{T}$, so the antisymmetric coefficients are nonzero. Subsequently, one can take the $\mathcal{E}\rightarrow 0$ limit. In addition to the even part (which does not play a role for the antisymmetric coefficients), there is the odd component of the transmission coefficient that contributes to these ratios $\mathcal{T}_{odd}\left(\frac{1}{2T}+\textit{i}t\right)\sim \frac{\textit{i}\mathcal{E} t}{\cosh^{\alpha}\left(\pi T t\right)}$, and the vanishing $\mathcal{E}$ cancels in the nominator and denominator, providing universal asymptotic relation. Strictly speaking, the odd component of the transmission coefficient may have more complicated $t$ dependence, but this simple estimate captures its leading contribution at $\pi T t \lesssim 1$, while contributions from its larger values are exponentially suppressed due to $\cosh$ terms in the denominator. Nevertheless, one can calculate the odd component exactly by substituting the exact expression of Eq. (\ref{Tcoeff}) in the antisymmetric transport and noise integrals.\\
For the ratio $\frac{\delta S_c^{SN}}{G_T}=\frac{T\mathcal{N}_0}{\mathcal{L}_1}=L_3 R^V_C$, one has $L_3=\frac{1}{3}$,
\begin{align} \label{RCV}
     &R^V_C=\frac{6}{\pi^2}\frac{\int_{-\infty}^{\infty} dx\, x^2\cosh^{-1-\alpha}\left(x\right)}{\int_{-\infty}^{\infty} dx\, x\sinh\left(x\right)\cosh^{-2-\alpha}\left(x\right)}=&\\
 \nonumber    &\frac{6}{\pi^2}\frac{{_4F_3\left(\frac{1+\alpha}{2},\frac{1+\alpha}{2},\frac{1+\alpha}{2},1+\alpha;\frac{3+\alpha}{2},\frac{3+\alpha}{2},\frac{3+\alpha}{2};-1\right)}}{\sqrt{\pi}(1+\alpha)^{-1} 2^{-\alpha} \Gamma^{-2}\left(\frac{1+\alpha}{2}\right)\Gamma^3\left(\frac{3+\alpha}{2}\right)\Gamma^{-1}\left(\frac{2+\alpha}{2}\right)},&
    \end{align}
where $\Gamma\left(\cdot\right)$ is the Gamma function, and ${_4F_3}\left(\cdot;\cdot;\cdot\right)$ is the generalized hypergeometric function. \\
The ratio $\frac{\mathcal{N}_2}{T\mathcal{L}_1}=L_4 R^V_H$ contains $L_4=\frac{12+\pi^2}{9}\simeq 2.43$: 
\begin{align} \label{RHV}
 L_4R^V_H= \frac{4\int_{-\infty}^{\infty} dx\, x^2\cosh^{-3-\alpha}\left(x\right)}{\int_{-\infty}^{\infty} dx\, x\sinh\left(x\right)\cosh^{-2-\alpha}\left(x\right)}-\pi^2L_3R^V_C+4.
\end{align}
Using this expression, we can define $\frac{\delta S_h^{SN}}{T^2 G_T}=L_4R^V_H-4$.
The explicit form of $R^V_H$ is a cumbersome combination of the generalized hypergeometric functions, so we rather plot the exact value of $R^V_H$ in Fig. \ref{fig:LorenzR} along with all other extended Lorenz ratios. For $\alpha=\frac{1}{2}$, we reproduce the corresponding ratios numerically from the exact expressions in \cite{Pavlov2025}. The precision of such an estimate comparing to exact calculations is further analyzed in \cite{Suppl} for the two-channel charge Kondo system. We also analyze in \cite{Suppl} all the noise coefficients for the noninteracting two-stage Kondo problem, which has the FL set of the ratios at the first stage of the screening and its own unique set of the ratios at the second stage.

\onecolumngrid
\setcounter{equation}{0}
\setcounter{figure}{0}
\setcounter{table}{0}
\renewcommand{\theequation}{S\arabic{equation}}
\renewcommand{\thefigure}{S\arabic{figure}}
\renewcommand{\thetable}{S\arabic{table}}
\newpage
\begin{center}
{\large \bf Supplemental Material to ``Universal relations between thermoelectrics and noise in mesoscopic transport across a tunnel junction''}\\
\end{center}
\section{I. Connection between considered model and realistic experimental setups}
In the section, we provide a detailed explanation of the connection between our considered model and realistic experimental setups.\\
\begin{figure}[b!]
\center
\includegraphics[width=.5\columnwidth]{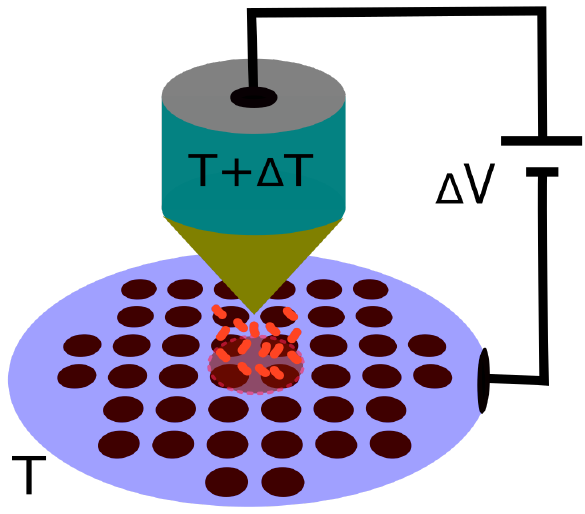} 
\caption{Considered setup: A system with some inner structure (light blue with dark dots) at temperature $T$ is coupled through a tunnel junction (red) to a metallic lead (teal and olive) at temperature $T+\Delta T$ serving as a probe. Additionally, there is a voltage bias $\Delta V$ induced between the system and the probe. The current and noise observables are governed by the local density of states in the probed system (red circle).}
\label{fig:setup}
\end{figure}

As an experimental setup for our theoretical analysis, we assume a well-established method of scanning tunnel microscope (STM) probe of the system where the probe's temperature can be different from the temperature of the reference system. The Johnson-Nyquist noise, shot noise, delta-T noise can be reliably measured through a tunnel junction, e. g., \cite{Spietz2003, Fevrier2018, Larocque2020}, including the STM break junction \cite{Tamir2022} (this work also discusses applications of this experimental technique for local measurements of shot noise in non-Fermi liquids).\\
We consider a lead of normal metal (Fermi liquid) serving as a probe, while the reference system (which is probed by the ``tip'' of the STM) is arbitrary (not necessarily a Fermi liquid). The important assumption is that the temperature and voltage drop occur across the tunnel junction. We are dealing with a weak out-of-equilibrium conditions (zero bias). Corresponding currents (charge and heat) are also vanishing. The metallic (macroscopic) lead is characterized by its chemical potential and temperature. This setup is illustrated in Fig. \ref{fig:setup}. The probed system may have some nontrivial internal structure, however, the tunnel junction transmission coefficient is determined solely by the local density of states at the tunnel junction. 

A paradigmatic example of such a system is a charge Kondo circuit \cite{Flensberg1993, Matveev1995, Furusaki1995, Andreev2002, Matveev2002}. In this mesoscopic device, a central island (a metallic quantum dot) is connected to one or several quantum point contacts (QPCs). This island exhibits a weak charge quantization operating in a mesoscopic Coulomb blockade regime. The charge quantization allows to fine tune the nano-device (by means of gate voltages applied to the QPCs) to a charge degeneracy point where the charge Kondo effect emerges.  Pronounce fingerprints of the non-Fermi liquid behavior associated with multi-channel charge
Kondo effect within the central island have been demonstrated experimentally \cite{Iftikhar2015, Iftikhar2018}. The design of these experiments assumes that the voltage probe is taken
far away from the QPC where the system already transitioned (through a crossover) to the Fermi liquid metallic regime, and the distribution function of the electrons is given by the Fermi distribution function. 
At the distances much larger than the size of the QD-QPC area, the temperature and chemical potential can be attributed to corresponding Fermi system, so the quantum dot in such a device has well-defined temperature and chemical potential. An STM probe can be applied to the central island of this mesoscopic device (or one of the QPCs in this setup can be pinched to make a tunnel junction, probing a QD connected to a single QPC, along the original proposals for the charge Kondo circuit \cite{Flensberg1993, Matveev1995, Furusaki1995, Andreev2002, Matveev2002}). The effects of the strong electron-electron correlations and resonance scattering occurring at the QD-QPC strongly renormalize the transmission coefficient across the tunnel junction, and as a result affect the quantum transport (namely, charge and heat currents and voltage and temperature driven charge, heat and mixed noises) through the junction.\\
\\
Recent STM experimental probes of the fractional quantum Hall states \cite{Farahi2023, Hu2025} are another example of the non-Fermi liquid investigation in the considered STM setup. Despite a complex structure of the investigated system, only the integrated density of states under the STM tip contributes to the measurements.\\
\\
The considered STM setup is also essential for observations of the transport signatures of the Sachdev-Ye-Kitaev (SYK) model in a proposed experimental realizations of this model in a disordered graphene flake \cite{Gnezdilov2018}.

\section{II. Resonant level model}
\begin{figure}[h!]
\vspace{-0.6cm}
\center
\hspace*{-.5cm}
\includegraphics[width=.7\linewidth]{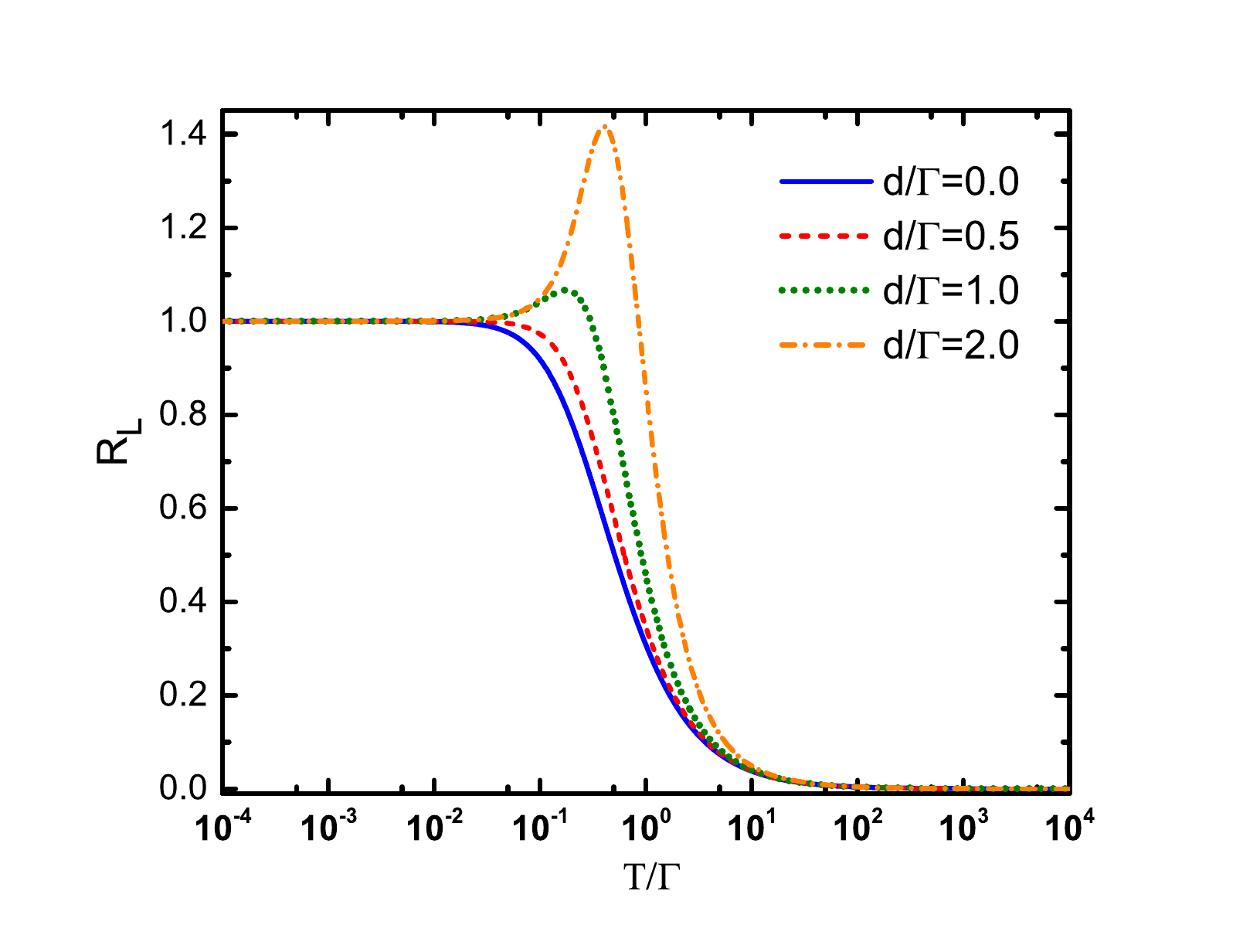} 
\vspace{-0cm}
\caption{Lorenz ratio $R_L$ of the resonant impurity model as function of temperature for various spectral asymmetries. Solid blue line: $d/\Gamma=0$; dashed red line: $d/\Gamma=0.5$, dotted green line: $d/\Gamma=1$; dot-dashed orange line: $d/\Gamma=2$.}
\label{fig:RLM} \vspace{-0.0cm}
\end{figure}
In this section, we provide analytical analysis of the transport coefficients for the resonant level model. 
The transmission coefficient of the resonant level model reads \cite{Manaparambil2025}
\begin{align} \label{TrCoefRL}
 T(\varepsilon)\sim\frac{\Gamma^2}{\Gamma^2+(\varepsilon-d)^2},
\end{align}
$\Gamma$ is the width of the resonant
level, $d$ is the energy shift inducing the spectral asymmetry. Using Eq. (4) of the main text, we can calculate the transport coefficients. The advantage of this integrable model is that it allows for analytical calculations in all regimes. By introducing $x=\Gamma/T$, $\tilde{d}=d/T$ we have 
\begin{align} \label{L0RLM}
&\mathcal{L}_0=\frac{1}{4}\int d z \frac{x^2}{x^2+(z-\tilde{d})^2}\frac{1}{\cosh^2(\frac{z}{2})}=\frac{\Gamma}{4\pi T}\left[\psi^{(1)}\left(\frac{1}{2}+\frac{\Gamma+\textit{i}d}{2\pi T}\right)+\psi^{(1)}\left(\frac{1}{2}+\frac{\Gamma-\textit{i} d}{2\pi T}\right)\right],&
\end{align}
where $\psi^{(1)}(y)=\sum_{n=0}^{\infty}\frac{1}{(n+y)^2}$ is the trigamma function. The detailed analytical evaluation of integrals of such structure can be found in \cite{Aristov2007, KiselevUn2}.
\begin{align} \label{L2RLM}
&\mathcal{L}_2=\frac{T^2}{4}\int d z \frac{x^2}{x^2+(z-\tilde{d})^2}\frac{z^2}{\cosh^2(\frac{z}{2})}=\frac{T^2x^2}{4}\int d z \left(1-\frac{x^2+\tilde{d}^2-2 z\tilde{d}}{x^2+(z-\tilde{d})^2}\right)\frac{1}{\cosh^2(\frac{z}{2})}=&\\
\nonumber &\Gamma^2-\frac{\Gamma}{4\pi T}\left[(\Gamma+\textit{i}d)^2\psi^{(1)}\left(\frac{1}{2}+\frac{\Gamma+\textit{i}d}{2\pi T}\right)+(\Gamma-\textit{i}d)^2\psi^{(1)}\left(\frac{1}{2}+\frac{\Gamma-\textit{i} d}{2\pi T}\right)\right].&
\end{align}

As is evident there, there are two regimes where the generalized WF law is obeyed. Namely, it is satisfied in the regime $T/\Gamma\ll 1$, with the Lorenz ratio reproducing the Fermi-liquid results $R_L=1$ (in this regime, only $\varepsilon\ll\Gamma$ energies contribute to $\mathcal{L}_0$ and $\mathcal{L}_2$, so the transmission coefficient Eq. (\ref{TrCoefRL}) can by approximated by a constant, corresponding to the $\alpha=1$ case of Eq. (\ref{RLcosh}) in the main text). The generalized WF law is further obeyed in the high temperature limit $T/\Gamma \gg 1$ with the Lorenz ratio converging to $R_L=0$. As expected, the generalized WL is violated at the intermediate temperature scales $T/\Gamma \sim 1$, where the integrals (\ref{L0RLM}), (\ref{L2RLM}) cannot be approximated as single-scale functions. 

For completeness, let's write here the exact expression for the $\mathcal{L}_1$ integral,
\begin{align}
&\mathcal{L}_1=\frac{T}{4}\int dz \frac{x^2}{x^2+(z-\tilde{d})^2}\frac{z}{\cosh^2\left(\frac{z}{2}\right)}=\frac{\Gamma}{4\pi T}\left[(d-\textit{i}\Gamma) \psi^{(1)}\left(\frac{1}{2}+\frac{\Gamma+\textit{i}d}{2\pi T}\right)+(d+\textit{i}\Gamma) \psi^{(1)}\left(\frac{1}{2}+\frac{\Gamma-\textit{i}d}{2\pi T}\right)\right].&
\end{align}
With this, one can explicitly calculate the experimentally motivated WF-law ratio $\kappa /(TG)$ (the WF law measured at the zero electric current condition). It has exactly the same validity range and high-$T$, low-$T$ Lorenz ratio values as $G_H/(TG)$ plotted in Fig. \ref{fig:RLM}, with only differences appearing in the non-universal regime $T\sim \Gamma$, where the generalized WF law is violated. 

\section{III. Competition between different energy scales}

While the transmission coefficient scaling $\mathcal{T}\left(\frac{1}{2T}+\textit{i}t\right)\sim \frac{1}{\cosh^{\alpha}\left(\pi T t\right)}$ captures limiting cases of a wide variety of quantum impurity systems discussed in the main text, the presented theory can be employed to calculate the noise components in arbitrary cases. Here, we demonstrate how the $\cosh$-scaling satisfies the limiting cases in the exact analysis of the two-channel charge Kondo system.

\begin{figure}[t!]
\vspace{-0.4cm}
\center
\hspace*{-.5cm}
\includegraphics[width=0.7\linewidth]{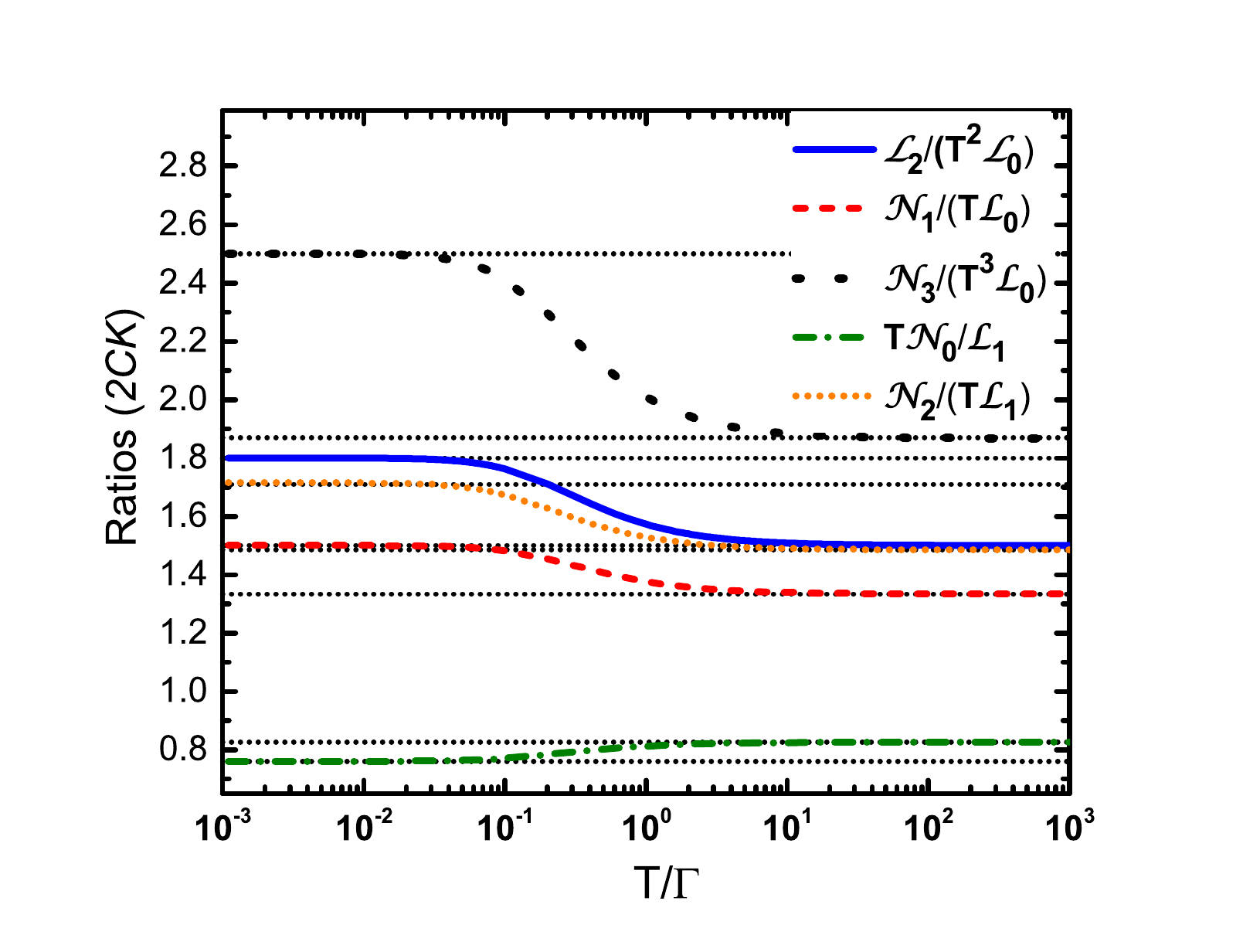} 
\vspace{-0cm}
\caption{Ratios between transport and noise integrals for a two-channel charge Kondo (2CK) system.}
\label{fig:ratios2CK} \vspace{-0.0cm}
\end{figure}

For the two-channel charge Kondo system (2CK), the transmission coefficient is known analytically \cite{Nguyen2010, KiselevUn2, Nguyen2025}. The density of states of the 2CK dot can be decomposed to even and odd components, $\nu(\varepsilon)=\nu^{e}(\varepsilon)+\nu^{o}(\varepsilon)$, these components are explicitly given in \cite{Nguyen2025}. Up to constant terms (that cancel out in the ratios of the noise integrals), these components are
\begin{align} \label{nuE}
    \nu^e(\varepsilon)\sim\int dx\frac{\cosh\left(\frac{\varepsilon}{2T}\right)}{\cosh\frac{x}{2}}\frac{p}{x^2+p^2}\frac{x+\varepsilon/T}{\sinh\left(\frac{x+\varepsilon/T}{2}\right)},\\ \label{nuO}
    \nu^o(\varepsilon)\sim\int dx \frac{\cosh\left(\frac{\varepsilon}{2T}\right)}{\cosh\frac{x}{2}}\frac{x}{x^2+p^2}\frac{x+\varepsilon/T}{\sinh\left(\frac{x+\varepsilon/T}{2}\right)},
\end{align}
where $p=\Gamma/T$, and $\Gamma$ is the Kondo-resonance width. For $T\gg \Gamma$, the system is in the non-Fermi liquid regime (corresponding to $\alpha=2$), while for $T\ll \Gamma$, the system exhibits the local Fermi-liquid behavior (the same as in the one-channel charge Kondo (1CK) $\alpha=3$). At the intermediate energy scale $T\simeq \Gamma$, there is a crossover between the two phases which has non-universal behavior. We use Eqs. (\ref{nuE}) and (\ref{nuO}) to calculate the symmetric and antisymmetric coefficients through Eqs. (\ref{Lint}), (\ref{Nint}) of the main text, and plot the ratios between different integrals in Fig. \ref{fig:ratios2CK}. As expected, the WF law and all the universal noise relations are obeyed in two limiting cases, $T\gg \Gamma$ (2CK) and $T \ll \Gamma$ (1CK), where the transmission coefficient scaling becomes a single-parameter function. In Table \ref{tab:ratios}, we provide the values of the corresponding ratios. For the symmetric coefficients, the agreement with the extended Lorenz numbers for symmetric coefficients from the main text is perfect. For the antisymmetric coefficients, there is a small discrepancy between these exact numerical results and the estimates given in the main text: $0.826$ vs $0.852$ for 2CK, $0.760$ vs $0.784$ for 1CK in $\frac{T\mathcal{N}_0}{\mathcal{L}_1}$; $1.485$ vs $1.400$ for 2CK, $1.715$ vs $1.625$ for 1CK in $\frac{\mathcal{N}_2}{T\mathcal{L}_1}$. The reason for that is the approximation used for the odd component of the transmission coefficient (which has more complicated structure, in general, than the even component \cite{Kiselev2023}). Nevertheless, this simple estimate provides us with the results within $\lesssim 5\%$ relative error for the numerical values between the universally related antisymmetric transport and noise integrals. At the intermediate energy scales, $T\simeq \Gamma$, the WF breaks, and all the noise relations break along with it too. This happens due to the energy scale $\Gamma$ which violates the single-parametric scaling of the transmission coefficient as elaborated in the main text.\\
A similar analysis for the SYK dot is provided in \cite{Pavlov2025}. The single-parametric scaling is broken in that case by the Coulomb blockade $E_C$, so the WF law and the noise relations break at $T\simeq E_C$, while they are obeyed in the $T\gg E_C$ and $T\ll E_C$ regimes, in accordance with our predictions.

\begin{table}[!t]
    \centering
    \resizebox{0.6\columnwidth}{!}{%
    \begin{tabular}{Sc| Sc | Sc | Sc | Sc | Sc}
   \,\,\,  & $\frac{\mathcal{L}_2}{T^2 \mathcal{L}_0}$ & $\frac{\mathcal{N}_1}{T\mathcal{L}_0}$ & $\frac{\mathcal{N}_3}{T^3\mathcal{L}_0}$ & $\frac{T \mathcal{N}_0}{\mathcal{L}_1}$ & $\frac{ \mathcal{N}_2}{T\mathcal{L}_1}$ \\ \hline
     2CK ($\alpha=2$) & $\frac{3}{2}$ &$\frac{4}{3}$ & $\frac{28}{15}$ & $0.826$ & $1.485$ \\ \hline
     1CK ($\alpha=3$) & $\frac{9}{5}$ &$\frac{3}{2}$ & $\frac{5}{2}$ & $0.760$ & $1.715$ \\ \hline     
    \end{tabular}}
    \caption{Extended Lorenz ratios for the 2CK, 1CK systems.} 
    \label{tab:ratios}
\end{table}

\section{IV. Two-stage Kondo effect}
\begin{figure}[!t]
\vspace{-0.6cm}
\center
\hspace*{-.5cm}
\includegraphics[width=.7\linewidth]{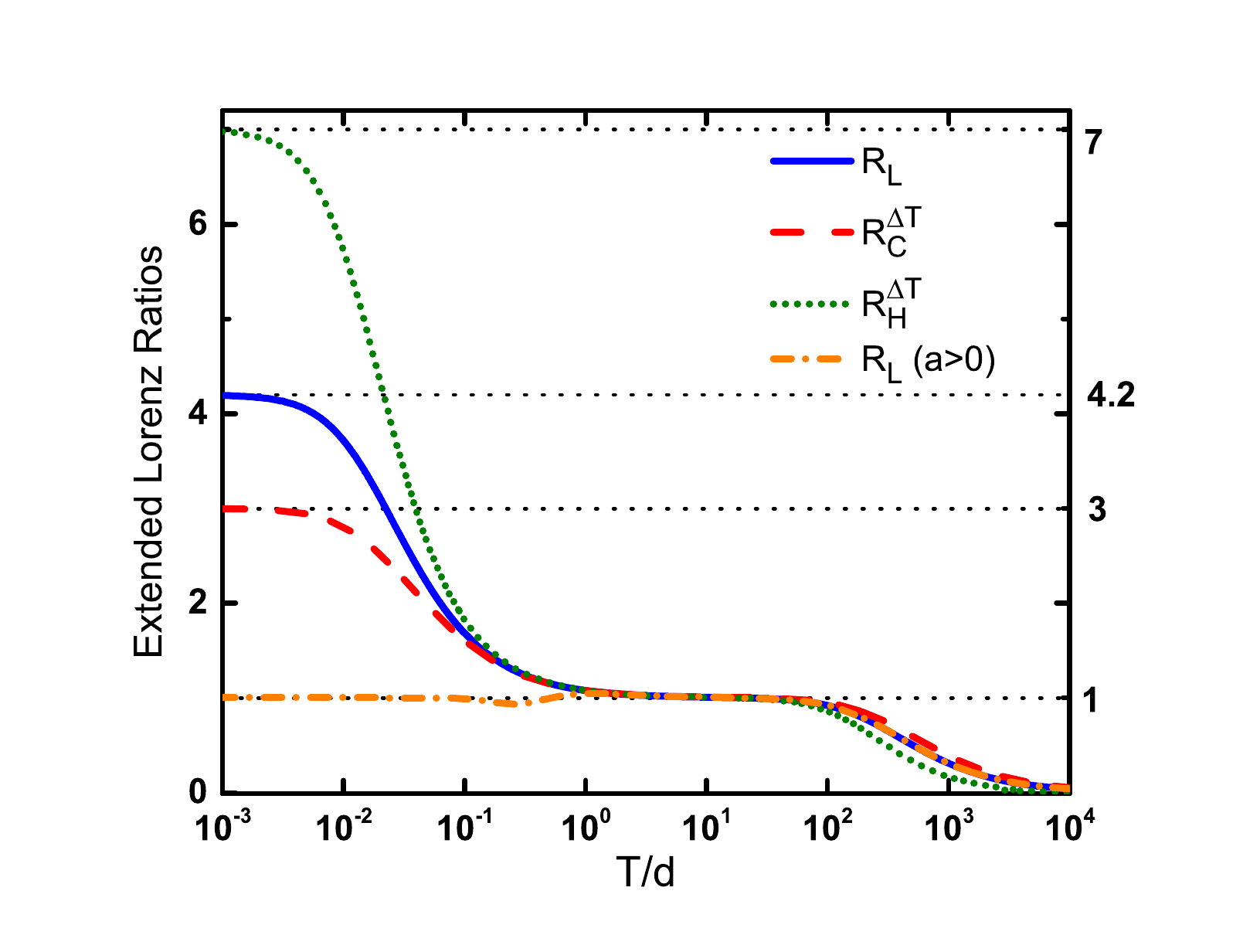} 
\vspace{-0cm}
\caption{Extended Lorenz ratios $R_L, R_C^{\Delta T}, R_H^{\Delta T}$ of the two-stage Kondo system as functions. Solid blue line: $R_L$; dashed red line: $R_C^{\Delta T}$; dotted green line: $R_H^{\Delta T}$. $\Gamma/d=10^3$, $g/d=10$, $a/d=0$. Dot-dashed range line: the Lorenz ratio $R_L$ for the same parameters except $a/d=1$.}
\label{fig:RL2SK} \vspace{-0.0cm}
\end{figure}
In this section, we demonstrate the Wiedemann-Franz (WF) law (and other universal ratios) for two different stages of the two-stage Kondo effect \cite{Pustilnik2001} recently observed experimentally \cite{Guo2021}, and the WF law violation on the intermediate scales in such a system. Generally, a two-level quantum dot may have the Coulomb interaction term $U$, so such a system needs to be treated numerically for obtaining quantitative results \cite{Wojcik2016}. For the $U=0$ case, however, the model is non-interacting and its transmission coefficient can be written exactly. In particular, let us consider two quantum dots coupled to metallic leads in a T-shaped configuration (the first dot is directly coupled to both leads, the second dot is coupled only to the first dot). The transmission coefficient for such a system at $U=0$ for both dots reads \cite{Brown2009}  
\begin{align} \label{T2SK}
\mathcal{T}(\varepsilon)=\frac{\Gamma^2}{\Gamma^2+(\varepsilon-d-\frac{g^2}{\omega-a})^2},
\end{align}
where $\Gamma$ is hybridization of the first dot with the leads, 
$d$ is the energy of the first dot, $a$ is the energy of the second dot (not coupled to the leads directly), $g$ is the hopping amplitude between the two dots. $g$ and $\Gamma$ are independent quantities that, in general, can have an arbitrary ratio. We assume $g\ll \Gamma$ for illustrative purposes.\\
The Kondo screening occurs in such a system in two stages. At the first stage, the spin of the first impurity (the one directly coupled to the leads) is screened, which happen at temperatures $g\ll T\ll \Gamma$. In this regime, the transmission coefficient (\ref{T2SK}) can be effectively approximated as a constant. This immediately provides us with the Lorenz ratio of the Fermi liquid $R_L=1$ (and the corresponding set of the extended Lorenz ratios). The second stage occurs at the low temperature regime $T\ll g$, where the second dot gets screened. In this regime, the transmission coefficient (\ref{T2SK}) behaves as $\mathcal{T}(\varepsilon)\sim \varepsilon^2$ if $a=0$ in Eq. (\ref{T2SK}), or $\mathcal{T}(\varepsilon)\sim const$ if $a\neq 0$ (see also \cite{Karki2020}). This effective transmission coefficient ensures that the transport integrals behave as single-scaling functions, so the generalized Wiedemann-Franz law is obeyed again. Due to the dip in the DoS in the $a=0$ case, the Lorenz ratio gets boosted, as discussed in the main text. Namely, $R_L=\frac{21}{5}$ was found in \cite{Karki2020}. Between these two universal regimes, the generalized WF law is violated, as expected from the fact that the transmission coefficient in this transitional regime depends on several energy parameters. In Fig. \ref{fig:RL2SK}, we demonstrate the extended Lorenz ratios between the symmetric transport coefficients. At the first stage of the screening (which takes place at intermediate temperatures), they all saturate to the Fermi-liquid value ($R_L=R_C^{\Delta T}=R_H^{\Delta T}=1$), while at the second stage they saturate to their specific values. In this regime, their limiting values are
\begin{align}
&R_L=\frac{3}{\pi^2}\frac{\int d z \frac{z^4}{\cosh\left(\frac{z}{2}\right)}}{\int dz\frac{z^2}{\cosh\left(\frac{z}{2}\right)}}=\frac{21}{5},&\\
&R_C^{\Delta T}=\frac{\int d z \frac{z^3\tanh\left(\frac{z}{2}\right)}{\cosh\left(\frac{z}{2}\right)}}{\int dz\frac{z^2}{\cosh\left(\frac{z}{2}\right)}}=3,&\\
&R_C^{\Delta T}=\frac{1}{\pi^2}\frac{\int d z \frac{z^5 \tanh\left(\frac{z}{2}\right)}{\cosh\left(\frac{z}{2}\right)}}{\int dz\frac{z^2}{\cosh\left(\frac{z}{2}\right)}}=7.&
\end{align}
As is evident from Fig. \ref{fig:RL2SK}, the extended Lorenz ratios indeed converge to these values in the zero-temperature limit. For completeness, we provide here also the universal ratios between the antisymmetric transport coefficients using the exact expression Eq. (\ref{T2SK}) (in the low-temparature limit, the leading contribution to the antisymmetric transmission coefficient at $a=0$ is $\mathcal{T}^{odd}(\varepsilon)\sim \varepsilon^3$):
\begin{align}
&L_3R^V_C=\frac{T\mathcal{N}_0}{\mathcal{L}_1}=\frac{15}{7\pi^2},&\\
&L_4R^V_H=\frac{\mathcal{N}_2}{T \mathcal{L}_1}=5.&
\end{align}

\section{V. Beyond linear response}
Here we discuss how the universal relations between currents and noise can be extended beyond the linear response regime. \\
We denote the Fermi-Dirac distribution function of the system and the lead, correspondingly, as $n_S=f(\varepsilon)$ and $n_L=f(\varepsilon)+\Delta f(\varepsilon)$.  The general expression for currents is
\begin{align}
I_{n}=\int d\varepsilon (\varepsilon-V)^{n}\nu_S(\varepsilon)\nu_L\Delta f(\varepsilon),
\end{align}
where $n=0,1$ for charge and heat current correspondingly. Using Eq. (2), and rewriting
\begin{align} \nonumber
n_L+n_S-2n_Ln_S=2f(\varepsilon)\left[1-f(\varepsilon)\right]+\Delta f(\varepsilon)\left[1-2f(\varepsilon) \right],
\end{align}
one gets the equilibrium noise (contribution from the first term) and the excess noise (the second term), which is
\begin{align}
\Delta S_l=\int d\varepsilon (\varepsilon-V)^l\nu_S(\varepsilon)\nu_L\tanh\left(\frac{\varepsilon}{2T}\right)\Delta f(\varepsilon),
\end{align}
$l=0,1,2$ for the charge, mixed and heat noise, and we used $1-2f(\varepsilon)=\tanh\frac{\varepsilon}{2T}$, $T$ is the temperature of the system.

Let us consider transport coefficients with respect to second derivatives. I. e., we analyze components of currents and noise in the second order of biases, $(\Delta V)^2, (\Delta T)^2, \Delta V\Delta T$ \cite{Mora2015, Oguri2018}. For the currents, all the components can be expressed in the closed form through the linear response transport and noise coefficients:
\begin{align}
&\frac{\partial^2 I_c}{\partial V^2}=\frac{\mathcal{N}_0}{T},&\\
&\frac{\partial^2 I_c}{\partial V \partial (\Delta T)}=\frac{\mathcal{N}_1}{T^2}-\frac{\mathcal{L}_0}{T},&\\
&\frac{\partial^2 I_c}{\partial (\Delta T)^2}=\frac{\mathcal{N}_2}{T^3}-2\frac{\mathcal{L}_1}{T^2},&\\
&\frac{\partial^2 I_h}{\partial V^2}=\frac{\mathcal{N}_1}{T}-2\mathcal{L}_0,&\\
&\frac{\partial^2 I_h}{\partial V \partial (\Delta T)}=\frac{\mathcal{N}_2}{T^2}-2\frac{\mathcal{L}_1}{T},&\\
&\frac{\partial^2 I_h}{\partial (\Delta T)^2}=\frac{\mathcal{N}_3}{T^3}-\frac{\mathcal{L}_2}{T^2}.&
\end{align}
We can further write expressions for the second order noise derivatives. For that purpose, it's convenient to introduce a new type of noise integrals (second order noise integrals)
\begin{align}
\mathcal{D}_n=\frac{1}{4T}\int d\varepsilon \mathcal{T}(\varepsilon)\frac{\varepsilon^n\tanh^2\left(\frac{\varepsilon}{2T}\right)}{\cosh^2\left(\frac{\varepsilon}{2T}\right)},\,\,\, n=0,..,4.
\end{align}
Note that, akin to the transport integrals $\mathcal{L}_n$, even integrals $\mathcal{D}_n$ are symmetric with respect to the contributions from the $\mathcal{T}(\varepsilon)$, while odd integrals are antisymmetric (for the $\mathcal{N}_0$ integrals, the situation is the opposite).
\begin{align} \label{nLinN}
&\frac{\partial^2 S_c}{\partial V^2}=\frac{1}{T}\mathcal{D}_0,& \\
&\frac{\partial^2 S_c}{\partial V\partial(\Delta T)}=\frac{1}{T^2}\mathcal{D}_1-\frac{1}{T}\mathcal{N}_0,& \\
&\frac{\partial^2 S_c}{\partial (\Delta T)^2}=\frac{1}{T^3}\mathcal{D}_2-\frac{2}{T^2}\mathcal{N}_1,&\\
&\frac{\partial^2 S_h}{\partial V^2}=\frac{1}{T}\mathcal{D}_2-4\mathcal{N}_1 ,&\\
&\frac{\partial^2 S_h}{\partial V\partial(\Delta T)}=\frac{1}{T^2}\mathcal{D}_3-\frac{3}{T}\mathcal{N}_2,&\\
&\frac{\partial^2 S_h}{\partial (\Delta T)^2}=\frac{1}{T^3}\mathcal{D}_4-\frac{2}{T^2}\mathcal{N}_3.&
\end{align}
Similar to the linear response regime, the mixed noise components do not bring us any new information about the system and can be fully expressed through components of charge and heat noise.
\begin{align} \label{nLinM}
&\frac{\partial^2 S_m}{\partial V^2}=\frac{1}{T}\mathcal{D}_1-2\mathcal{N}_0,& \\
&\frac{\partial^2 S_m}{\partial V\partial(\Delta T)}=\frac{1}{T^2}\mathcal{D}_2-\frac{2}{T}\mathcal{N}_1, &\\
&\frac{\partial^2 S_m}{\partial (\Delta T)^2}=\frac{1}{T^3}\mathcal{D}_3-\frac{2}{T^2}\mathcal{N}_2.&
\end{align}
From the expressions for non-linear current and noise components, one can construct multiple universal ratios and reciprocity relations which have the same applicability range as linear response results of the main text.
For instance, one can consider the generalized Fano factor \cite{Mora2009, Mora2015}, defined as the ratio between the corresponding components of the noise and currents with a subtraction of the linear response contributions. For the charge current noise due to voltage bias, we have
\begin{align}
\delta F^{SN}_c=\left.\frac{\Delta S_c-\Delta S_c^{lin}}{I_c-I_c^{lin}}\right\vert_{\Delta T=0}=\frac{\mathcal{D}_0}{\mathcal{N}_0},
\end{align} 
\begin{align}
\delta F^{SN}_c F^{SN}_c=\frac{\mathcal{D}_0}{\mathcal{N}_0}\frac{\mathcal{N}_0}{\mathcal{L}_0}=\frac{\mathcal{D}_0}{\mathcal{L}_0}.
\end{align}
Therefore, the product of the Fano factor and the generalized Fano factor becomes a universal constant. For instance, for the Fermi liquid ($\mathcal{T}(\varepsilon)=const$, or equivalently $\mathcal{T}\left(\frac{1}{2T}+\textit{i}t\right)\sim\cosh^{-1}\left(\pi T t\right)$), this product takes the value $\delta F^{SN}_c F^{SN}_c=\frac{\mathcal{D}_0}{\mathcal{L}_0}=\frac{1}{3}$.

\end{document}